\documentclass{aa}

\usepackage{longtable}
\usepackage[varg]{txfonts}
\usepackage{graphicx}
\usepackage{hyperref}
\usepackage{lscape}
\usepackage{subcaption}
\usepackage{mwe}
\usepackage{changepage}
\usepackage{booktabs}
\usepackage{multirow}
\usepackage{xcolor}

\begin{document} 

\title{\textbf{Cometary dust collected by MIDAS on board Rosetta\thanks{Table~\ref{table: MIDAS_particle_shape_descriptor_full_table} is only available in electronic form at the CDS via anonymous ftp to \url{cdsarc.cds.unistra.fr} (\url{130.79.128.5}) or via \url{https://cdsarc.cds.unistra.fr/cgi-bin/qcat?J/A+A/}. MIDAS dust coverage, clustering, and shape descriptor maps can be found \url{https://minjaegasparkim.github.io/Rosetta-MIDAS_dust_maps}}}}
\subtitle{II. Particle shape descriptors and pristineness evaluation}
\titlerunning{MIDAS particle shape descriptors}


\author{M.~Kim\inst{\ref{inst1}, \ref{inst2}, \ref{inst3}}
\and
T.~Mannel\inst{\ref{inst1}}
\and
J.~Lasue\inst{\ref{inst4}}
\and
A.~Longobardo\inst{\ref{inst5}}
\and
M.~S.~Bentley\inst{\ref{inst6}}
\and
R.~Moissl\inst{\ref{inst7}}
\and 
the MIDAS team
}
\institute{Space Research Institute of the Austrian Academy of Sciences, Schmiedlstrasse 6, 8042 Graz, Austria\\ \email{minjae.k.kim@warwick.ac.uk} \label{inst1} 
\and
Department of Physics, University of Warwick, Gibbet Hill Road, Coventry CV4 7AL, UK \label{inst2}
\and
Centre for Exoplanets and Habitability, University of Warwick, Gibbet Hill Road, Coventry CV4 7AL, UK \label{inst3}
\and
IRAP, Université de Toulouse, CNRS, CNES, UPS, 9 avenue Colonel Roche, FR-31400, Toulouse, France \label{inst4}
\and
Istituto Nazionale di Astrofisica, Istituto di Astrofisica e Planetologia Spaziali, via Fosso del Cavaliere 100, I-00133 Rome, Italy \label{inst5}
\and
European Space Astronomy Centre, Camino Bajo del Castillo, s/n., Urb. Villafranca del Castillo, 28692 Villanueva de la Cañada, Madrid, Spain \label{inst6}
\and
Scientific Support Office, Directorate of Science, European Space Research and Technology Centre, 2201 AZ Noordwijk, The Netherlands \label{inst7}
}

\date{Received 13 June, 2023 / Accepted 10 August, 2023}


\abstract
{The MIDAS (Micro-Imaging Dust Analysis System) atomic force microscope on board the Rosetta comet orbiter investigated and measured the 3D topography of a few hundred of nm to tens of $\mu$m sized dust particles of 67P/Churyumov-Gerasimenko with resolutions down to a few nanometers, giving insights into the physical processes of our early Solar System.}
{We analyze the shapes of the cometary dust particles collected by MIDAS on the basis of a recently updated particle catalog with the aim to determine which structural properties remained pristine.}
{We develop a set of shape descriptors and metrics such as aspect ratio, elongation, circularity, convexity, and particle surface and volume distribution, which can be used to describe the distribution of particle shapes.  
Furthermore, we compare the structure of the MIDAS dust particles and the clusters in which the particles were deposited to those found in previous laboratory experiments and by Rosetta/COSIMA. Finally, we combine our findings to calculate a pristineness score for MIDAS particles and determine the most pristine particles and their properties.}
{We find that the morphological properties of all cometary dust particles at the micrometer scale are surprisingly homogeneous despite originating from diverse cometary environments (e.g., different collection targets that are associated with cometary activities/source regions and collection velocities/periods). There is only a weak trend between shape descriptors and particle characteristics such as size, collection targets, and cluster morphology. We next find that the types of clusters found by MIDAS show good agreement with those defined by previous laboratory experiments, however, there are some differences to those found by Rosetta/COSIMA. Furthermore, our pristineness score shows that almost half of MIDAS particles suffered severe alteration by impact, which indicates structural modification by impact (e.g., flattening and/or fragmentation) is inevitable despite the very low collection speeds (i.e., $\sim$ 3 - 7 m s$^{-1}$). Based on our result, we rate 19 out of 1082 MIDAS particles at least moderately pristine that is they are not substantially flattened by impact, not fragmented, and/or not part of a fragmentation cluster.}
{}


\maketitle

	\section{Introduction}\label{sec:introduction}
	
Since the formation of comets, asteroids, planetesimals, and planets is expected to start with the aggregation of the smallest interstellar dust particles \citep{tielens_dust_2005, li_dust_2003, weidenschilling_formation_1993, blum_growth_2008}, the properties of the smallest dust grains in the early solar system are of great importance. These have been the focus of various studies, for example, the sample return mission flying by comet 81P/Wild~2, Chondritic Porous Interplanetary Dust Particles collected in the stratosphere, as well as the ultracarbonaceous Antarctic micrometeorites (\citealp{brownlee_comet_2006, flynn_organic_2013, noguchi_dust_2015, duprat_ucamms_2010}). In particular, cometary material is the result of aggregation processes known to take place in extremely low gravity conditions that can result in aggregates with low tensile strength \citep{Sierks_2015, Attree_2018}. These fragile structures are suspected to have been stored in comets and to be lifted into the coma by cometary activity (e.g., \citealp{blum_evidence_2017}). However, these investigations are hampered by the severe alteration of the dust particles between their time of release from the comet and their investigation, for example, due to heating to temperatures above the melting point of silica as well as due to the capturing process \citep{hoerz_impact_2006}, and the passage through Earth’s atmosphere \citep{flynn_organic_2013}. 

To obtain structural information on pristine cometary dust particles, a unique opportunity to sample the dust and gas environment of the inner coma of comet 67P/Churyumov-Gerasimenko (hereafter 67P) was pioneered by the Rosetta mission between 2014 and 2016. The Micro-Imaging Dust Analysis System (MIDAS; ~\citealt{riedler_midas_2007, bentley_lessons_2016}), one of three in-situ dust instruments on board, is the first generation Atomic Force Microscope (AFM) that was launched into space to collect the smallest cometary dust particles with sizes from hundreds of nanometers to tens of micrometers. It recorded the 3D topographic/textural information with nanometer resolution and collected statistical parameters (\citealp{Kim_Mannel_MIDAS_catalog}). Most importantly, dust particles were collected close to the comet (less than $\sim$ 30 km afar) with low relative velocities between spacecraft and nucleus, roughly between some cm\,s$^{-1}$ and 10 m\,s$^{-1}$ (e.g., \citealp{dellacorte_giada_2015}), which enables to collect dust particles with only a small degree of alteration.  In particular, small particles are assumed to be aggregates of subunits \citep{bentley_morphology_2016} that have a higher internal strength so their sizes and shapes might be possibly still pristine \citep{hornung_assessment_2016}.

Depending on the particle strength and the collection velocity, dust particles hitting a collection target either undergo alteration such as break up or compaction, or they stay intact \citep{Ellerbroek_labstudy_2017, hornung_assessment_2016}. 
The secondary ion mass spectrometer COSIMA (Cometary Secondary Ion Mass Analyser; \citealp{langevin_typology_2016}), using a similar ‘hit and stick' collection technique to MIDAS but for particles one to two orders of magnitude larger, observed that the majority of dust particles would fragment upon collection and create large scatter fields of fragments \citep{langevin_typology_2016}. This finding was confirmed by a laboratory experiment for particles larger than about 80 micrometers \citep{Ellerbroek_labstudy_2017}. It is, however, still an open question to understand to what extent the one order of magnitude smaller MIDAS particles were altered during the collection process, and which of the MIDAS particles show pristine structures. The straightforward dust collection strategy with MIDAS (i.e., the dust particles from 67P entered the MIDAS instrument via a funnel and simply hit the targets; \citealp{bentley_lessons_2016}) could lead to an unknown degree of collection alteration. Consequently, dust alteration during the collection process has to be understood to retrieve the properties of pristine and more realistic cometary particles. Furthermore, the flow properties of bulk materials (e.g., flowability) are influenced by particle shape and particle size, amongst others (\citealp{Mellmann2013}). Thus, the investigation of unaltered and thus more realistic particles could open up new possibilities for unraveling unresolved questions in space science, particularly in gas flow experiments in cometary environments (\citealp{Laddha2023}). The importance of shape descriptors of the non-spherical MIDAS particles (\citealp{bentley_morphology_2016, Kim_Mannel_MIDAS_catalog}) is further enhanced as traditional models using purely spherical particles might not be sufficient to significantly improve gas flow models further.

In this paper, a ‘pristine' particle is a particle that was not altered by the collection that is a particle as it was moving through the cometary coma. Depending on its history between its formation in the early solar system and its collection, this particle may or may not be modified by its flight through the coma, ejection from the cometary surface, storage, or incorporation into the comet. Thus, it has to be judged case by case in which sense a particle's properties reflect properties of cometary dust in the coma, the comet, or even in the early solar system. 

Using the final MIDAS particle catalog that contains all cometary dust particles collected by MIDAS together with their basic properties, called particle catalog V6\footnote{ESA planetary science archive of MIDAS/Rosetta (PSA 6.2.4), dataset identifier: RO-C-MIDAS-5-PRL-TO-EXT3-V3.0, \newline
\url{https://archives.esac.esa.int/psa/\#!Table\%20View/MIDAS=instrument}\label{ESA/PSA address}} (\citealt{Kim_Mannel_MIDAS_catalog}), we focus on the development of MIDAS particle shape descriptors to understand the degree of dust alteration and we determine the structural properties of the MIDAS dust particles that remained pristine. This paper\footnote{The present study follows a dust classification scheme (i.e., the definition of language for all the different dust pieces) that has already been unified with a clearly defined vocabulary within the Rosetta dust studies (\citealt{Mannel_classification_2019, Guettler_morphology_2019, Kim_Mannel_MIDAS_catalog}).} is organized as follows: We provide a description of the MIDAS instrument and its collection procedure in Sect.~\ref{sec:methods}. Next, we discuss the development of MIDAS particle shape descriptors and the MIDAS cluster classifications. We describe the evaluation of MIDAS particles and clusters based on MIDAS shape descriptors and compare them to previous studies in Sect.~\ref{sec:results}. Based on these results, we investigate the development of a pristineness descriptor of MIDAS particles. We summarize our findings and address the outlook of this study in Sect.~\ref{sec:summary and outlooks}.

	\section{Methods}\label{sec:methods}

\subsection{The MIDAS AFM and its dust collection}\label{method: MIDAS_AFM}

MIDAS was a unique instrument designed to collect and investigate the size, shape, texture, and microstructure of cometary dust particles down to the 10s of nanometers scale \citep{riedler_midas_2007, bentley_lessons_2016}. The MIDAS dust intake system consisted of a funnel and a shutter that could be opened to allow dust particles to fall into the collection targets mounted onto a wheel. After the collection of a dust particle, the wheel was rotated to transport the dust particle to the AFM, which physically scanned over the particles. 

A scientific sequence of MIDAS was started with a prescanning of an area of a target, exposing it to the dust flux of the comet, and rescanning of the same area with low resolution. If a dust particle was detected, it was then scanned with optimized parameters (\citealp{bentley_lessons_2016}). There are two operating modes (i.e., amplitude-modulated `dynamical mode' and `contact mode') in MIDAS, while MIDAS mainly utilized an amplitude-modulated mode, which effectively avoided feedback issues (e.g., most AFMs that operate in a vacuum use frequency modulation, potentially causing damage and fragmentation, especially with large particles having high Z gradients; \citealp{Hecht2008, McSween2010}) by performing software-controlled point approaches to each pixel of the image (\citealp{bentley_lessons_2016}). This approach is slower than traditional AFMs, but minimizes damage at each pixel position of particles, making it ideal for studying delicate or sensitive samples. 

The force arising from the tip-sample interaction in MIDAS causes the deflection of the piezo-resistive cantilevers. These cantilevers, attached to the sharp tip, can electrically detect their own deflection without the need for additional sensing elements. Specifically, when the tip-sample separation is small (around 5–10 nm), the electron orbits interact, resulting in a weak attractive force. Consequently, the resonance frequency of the cantilever changes due to a virtual increase in its spring constant. When excited close to their resonant frequency ($\sim$ 80 - 100 kHz) the cantilevers have an amplitude of cantilever vibration on the order of 40 - 100 nm. The tip-sample interaction arises from a combination of short-range, repulsive atomic forces, and long-range Van der Waals force (\citealp{riedler_midas_2007}). When operated in the contact mode, which operates close to the repulsive force regime where the very sharp tip (i.e., tip radius $\sim$ 10 to 15 nm) physically touches the surface of particles, the force applied by the tip to the sample is approximately 10$^{-7}$ – 10$^{-6}$ N (\citealp{riedler_midas_2007}), translating to a pressure range during measurements spanning approximately 100 MPa to 2500 MPa. However, when operated in dynamic mode, the expected force is reduced to 10$^{-8}$ N (\citealp{riedler_midas_2007}), corresponding to the pressure values during the measurements that lie within the range of around 10 MPa to 25 MPa. This relatively small force and the absence of lateral forces minimize the likelihood of damaging the tip or sample. In addition, the approach (force-distance) curve of the MIDAS instrument can be recorded throughout an image, providing valuable information about the physics of the tip-sample interaction and the physical properties of the sample.

Rastering over the particles' surface, a 3D image of cometary dust was created, which enabled the determination of the particle structure, and thus an insight into the properties of micro- to nanometer-sized dust particles. In particular, the high-resolution MIDAS images allow us to determine particle shape factors and descriptors. We note that the volume estimate in MIDAS may be susceptible to over-estimation due to several factors (e.g., tip dilation and concave surfaces; see Sect.~\ref{result: MIDAS_particle_shape_descriptor}), which can lead to incomplete imaging of certain particles, potentially resulting in inaccuracies in the volume measurements. In the retrieved scans identifying dust grains and marking them with a mask is performed by hand with the open-source package Gwyddion (\citealp{Gwyddion_paper}). More detailed information about MIDAS and its dataset can be found in previous MIDAS studies (e.g., \citealp{riedler_midas_2007, bentley_lessons_2016, mannel_fractal_2016, Mannel_classification_2019, Kim_Mannel_MIDAS_catalog}). 

MIDAS collected nano- to micrometer-sized cometary dust particles on 4 different small (1.4 mm x 2.4 mm) and sticky targets (i.e., target 11, target 13, target 14, and target 15)\footnote{Note that previous studies (e.g., \citealp{Longobardo_2022_MNRAS, Kim_Mannel_2022EGU}) adopt different indexing, which comes from the number of telemetries counted from 1 to 64. However, the MIDAS particle catalog (\citealp{Kim_Mannel_MIDAS_catalog}) and browse images count the targets from 0 to 63. Thus, target numbers 10, 12, 13, and 14 in the previous studies correspond to target numbers 11, 13, 14, and 15, respectively, in this study.}, where each target collected dust in different periods. For example, the dust particles on Target 11, Target 13, and Target 15 were collected before perihelion and the ones on Target 14 were collected in an outburst after perihelion. Furthermore, the average compact dust speed was estimated for exposition periods of each MIDAS target via speed measurements of GIADA (\citealp{Longobardo_2022_MNRAS}). In particular, the mean velocities for the particle collections on targets 11 and 13 are slower (i.e., 3.1 $\pm$ 1.5 and  2.7 $\pm$ 0.3 m\,s$^{-1}$, respectively), while the velocities of the particles collected on target 14 and 15 are faster (i.e., 7.21 $\pm$ 0.09 and 7.2 $\pm$ 0.2 m\,s$^{-1}$, respectively; \citealp{Longobardo_2022_MNRAS}). The standard deviation range corresponds to each MIDAS target 1-2 m s$^-1$. Furthermore, the standard deviation of the entire GIADA-MIDAS dataset is 6 m s$^-1$. The detailed list of target exposure history and duration can be found in Appendix Table B.1. in \citet{Kim_Mannel_MIDAS_catalog}.

Assuming the same source region of particles detected by GIADA and MIDAS in the same period, \citet{Longobardo_2022_MNRAS} determined which MIDAS particles stem from which source area of the comet based on a traceback algorithm applied to GIADA data (\citealp{Longobardo_2020a_Merging_data}). In particular, MIDAS particles on target 14 originated from rough terrains (or deeper layers), implying that these particles are expected to be more pristine. On the other hand, particles on target 15 were ejected from smooth terrains, which are thought to originate from dust particles falling back to the cometary surface after ejection, creating smooth plains that can be as much as a few meters thick (\citealp{Longobardo_2020a_Merging_data}). Since ejection, travel through the coma, backfall, and re-ejection are likely to result in particle alteration, particles on target 15 may not be as pristine as those on target 14. Furthermore, particles collected on targets 11 and 13 originated from both types of terrain.

\subsection{Development of MIDAS particle shape descriptors}\label{method: MIDAS_particle_shape descriptors}

Since particle shape descriptors represent a range of properties, for instance, a spatial variation on a large scale or irregularity/surface textures on a small scale (\citealt{shape_descriptors_Berrezueta_2019}), they are well suited to evaluate and classify our particle collection. Thus, we develop MIDAS particle shape descriptors to determine particles preserving their pristineness. 

We start from 3523 MIDAS particle identifications based on the updated MIDAS particle catalog V6 (\citealt{Kim_Mannel_MIDAS_catalog}). For further investigation, we select particles only usable for further investigation (e.g., only particles of cometary origin, particles with a preferably low amount of alteration, particles not shown to be fragments of particles, particles were not susceptible to being multiple fragments, particles not suspected to have fallen off a tip), which is necessary to ensure the integrity and reliability of the dataset. Additionally, we remove duplicates as many MIDAS dust particles were scanned several times and thus select the best representation of particles (\citealt{Kim_Mannel_MIDAS_catalog}). Next, we again exclude particles with a very small number of pixels in images (e.g., < 10) due to the physical difficulties of further analysis with a small number of pixel particles. After applying these selection criteria, 1082 MIDAS particles are used for the analysis of shape descriptors in total. These selected particles will serve as the basis for our in-depth analysis and study of cometary particles and their properties including shape descriptors. More detailed information about MIDAS particle selection for further investigation can be found in previous MIDAS studies (e.g., \citealp{Kim_Mannel_MIDAS_catalog}).

For the development of MIDAS particle shape descriptors, we first apply several pre-processing procedures, most importantly, rotation of the data to align the scan direction to the horizontal direction, correction of horizontal scars in scan files, alignment of rows in scan files using the median of differences algorithm, leveling of the data by means of a plane subtraction, and the rotation of scan files in the opposite direction of the first rotation with the Gwyddion Python module called Pygwy\footnote{\url{http://gwyddion.net/documentation/user-guide-en/pygwy.html}} to all scan files to clean up the MIDAS data. This is necessary to avoid falsification of the result by AFM image artifacts. 

We consider several independent 2D or 3D shape descriptors and shape/morphological parameters, for example, aspect ratio (\citealt{Lasue_simulation_2019}), elongation (\citealt{bentley_morphology_2016}), circularity (often called 2D sphericity; \citealt{shape_descriptors_Berrezueta_2019}), and convexity (often called solidity parameter or 2D roughness; \citealt{shape_descriptors_Berrezueta_2019}). In the present section, we introduce the shape descriptors and correlate them with already derived particle properties. We note that there are limitations (e.g., resolution) and possible sources of uncertainty that must be considered when performing particle shape analysis as with all imaging/measurement techniques. Furthermore, AFM measurements provide only a 2D projection of the particle's surface. As a result, the actual 3D shape is not directly determined by the applied method; rather, only the 3D topology is obtained. Therefore, this limitation should be considered in the following analyses.\newline

\noindent\textbf{Aspect ratio}: We calculate the aspect ratio $\frac{H}{\sqrt{A}}$, where $H$ is the height (i.e., subtraction of the median height value to all pixels surrounding a particle from the maximal z value of that particle) and $A$ is the projected 2D area of the dust particle/cluster on the target. The aspect ratio is used as a tool to assess the degree of flattening of MIDAS dust particles (\citealt{Lasue_simulation_2019}). For example, higher aspect ratio particles indicate that particles are less flattened. In particular, we assume particles with ratios over about 0.8 are unflattened based on the aspect ratio distribution of MIDAS particles (see Sect. \ref{result: Relation_shape_descriptors_ohter_parameters}). 
    
In the present work, we determine the aspect ratio for MIDAS particles and clusters (\citealp{Kim_Mannel_MIDAS_catalog}), compare them with those found for COSIMA particles and clusters in \citet{Lasue_simulation_2019}, and check if the aspect ratio can be used as a proxy for particle pristineness (i.e., investigate whether results from \citet{Lasue_simulation_2019} are also valid for MIDAS particles).\newline

\noindent\textbf{Elongation}: We calculate two elongations of MIDAS particles to determine how much the shape deviates from a sphere: i) the major axis to the height $\frac{a}{h}$, where $a$ is the major axis of the 2D projection and $h$ is the height of the particle, called ‘3D elongation’ and ii) the major axis to the minor axis $\frac{a}{b}$, where $b$ is the minor axis of 2D projection, called ‘2D elongation’, where all axes and heights have to be in 90 degrees angle to each other in the coordinate system based on the target after leveling work (\citealp{bentley_morphology_2016}). An elongation value larger than 1 indicates that the shapes of the particles deviate from a spherical form. The final elongation value is chosen as the bigger elongation between the 2D and 3D elongation.\newline
  
\noindent\textbf{Circularity}: We calculate a circularity $\frac{4\,\pi\,A}{P^{\rm{\,2}}}$, where $A$ is the area of the particle of the 2D projection and $P$ is the 2D projection perimeter of the particle, to quantify roundness. Physically, a circularity value of 1 corresponds to a circle.\newline
    
\begin{figure}
\includegraphics[width=9cm, height= 6cm]{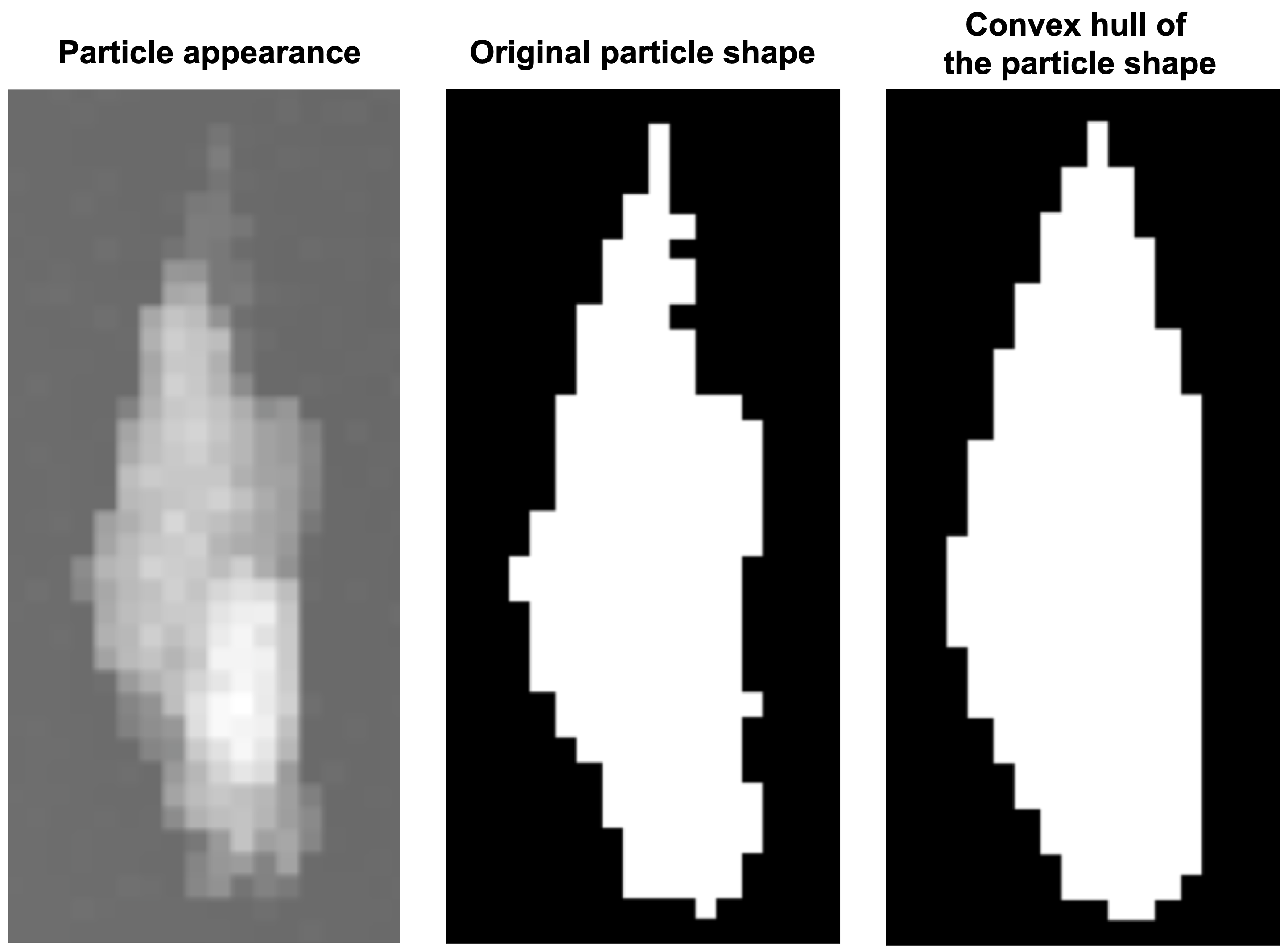}
\caption{Brief sketch of the comparison between the area of the silhouette and the convex hull of the silhouette on the example of  an original MIDAS particle (Particle ID: 2015-11-22T165903$\_$P01$\_$T15; \citealp{Kim_Mannel_MIDAS_catalog}). The value of the convexity of this particle shows 0.456.}
\label{fig:ex_convexity}
\end{figure}

\noindent\textbf{Convexity}: We calculate a convexity $\frac{A}{A_{\rm{convex}}}$, where $A$ is the area of the particle and $A_{\rm{convex}}$ is the area defined by the convexity produced by the irregularity of the edge of the particle, to describe how fringed the rims are. This parameter is defined as the ratio between the area of the silhouette and the convex hull of the silhouette. Thus, it represents the 2D projection roundness of particles. For example, a convexity value of 0.5 corresponds to 100$\%$ of the convex hull area occupied by perimeter concavities. A brief sketch of an example is given in Fig.~\ref{fig:ex_convexity}.\newline

\noindent\textbf{Particle surface and volume distribution}: Since the impact residue strongly depends on the size of the initial particle and velocity \citep{Ellerbroek_labstudy_2017}, the deposit left by such an impact can be investigated to understand how the impact on the target modified the particles and thus determine the degree of alteration of its structure. We aim to link the amount of dust deposited to the distance from the center of the deposit by refining previous ideas \citep{Guarino_2019} and applying them to MIDAS particles. This allows an understanding of the shape of a deposit (e.g., pyramidal or rectangular) and the nature of its rims (e.g., steep cliff-like or shallow sloped, asymmetric curved rim or straight edge). 

The residue from the impact covered by dust is computed as the number of pixels with dust as the volume (or height) associated with those pixels. The algorithm of the two developed deposit studies is described below:\newline

\noindent - Particle surface distribution: We investigate the area of the particle on the 2D projection. For each particle, the image is divided into N rings (where N is the total number of rings from the center to the outermost possible ring with each ring being 1 pixel thick) from the center of the deposit. This center is computed as the mean value of the x and y coordinates of the pixels of the mask (i.e., left \& right and up \& down, Sect.~\ref{method: MIDAS_AFM}). We compute the area of the particle laying inside each i-th ring (i.e., summing the number of pixels with dust). Finally, we calculate for each ring the percentage of pixels with dust (i.e., that belonging to the particle) in that ring.

\noindent - Particle volume distribution: We investigate the volume of the particle by weighing the particle surface distribution by the height of each pixel. Note that the value of the particle volume distribution plot can be smaller than the one of particle surface distribution because the unit of height values is in micrometer (e.g., in case the height value is below 1~$\mu$m). \newline

\begin{figure}
\includegraphics[width=9cm, height= 7cm]{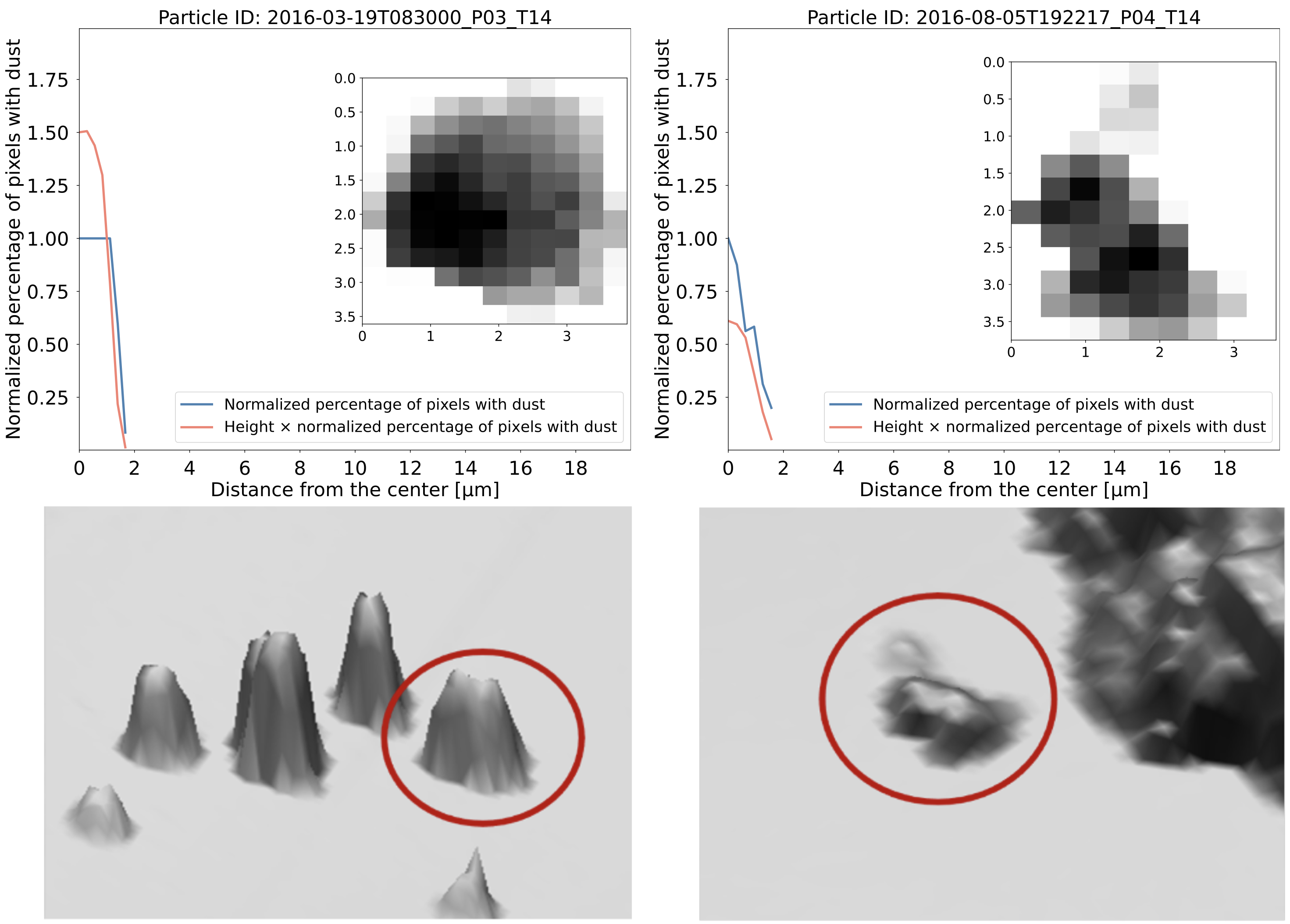}
\caption{Example of particle surface and volume distribution of ‘MIDAS Single' (e.g., typically a solid, roundish bulk and shows a plateau in the particle surface distribution; left column) and ‘MIDAS Pile' (e.g., elongated shapes,  fringed rims, and no plateau visible in the surface distribution and thus less steep slope than one for the MIDAS Single particles; right column) particles. 3D images of the MIDAS Single and MIDAS Pile particles are marked with red circles in the left and right bottom images.}
\label{fig:ex_particle_surface/volume_distribution}
\end{figure}

Based on the particle surface and volume distribution, a clear distinction between solid and roundish particle bulks with defined rims (the so-called ‘MIDAS Single' particles) and other shapes (e.g., elongated shapes with fringed rims; so-called ‘MIDAS Pile’) become apparent. The name convention is adopted from the work of \citet{Ellerbroek_labstudy_2017} because ‘MIDAS Single' particles look like ‘single morphology' from \citet{Ellerbroek_labstudy_2017}. Examples of the results from the two algorithms (particle surface and volume distribution) along with 3D images of the investigated particles are given in Fig.~\ref{fig:ex_particle_surface/volume_distribution}.

\subsection{MIDAS cluster/particle classification}\label{method: MIDAS cluster classification}

We investigate possible similarities between the MIDAS dataset and previous studies that classified shapes of cometary dust particles (e.g., \citealt{Ellerbroek_labstudy_2017, langevin_typology_2016}). This classification work is a cornerstone to determine particle pristineness because particle properties that were already connected to alteration or pristineness in other studies could be identified in MIDAS data as well and be similarly interpreted.

\citet{Ellerbroek_labstudy_2017} investigated the footprint of cometary dust analogs in laboratory experiments using low-velocity impacts of analog materials and compared these with COSIMA and MIDAS data. They found that the degree of alteration is strongly determined by the initial particle size and the impact velocity. In particular, for certain combinations of size and velocity, only a part of the dust particle is left on the target and the amount of mass transferred increases with the impact velocity. Thus, the velocity is the main driver of the appearance of the resulting deposits for most collisions. For example, for velocities $v$ below  $\sim$ 2 m\,s$^{-1}$, initial parent particles either stick without visible alteration and leave a ‘single' deposit on the target plate, or the parent particles bounce, leaving a shallow ‘footprint' that is a flat layer of monomers. 
At velocities $v$ > 2 m\,s$^{-1}$ and initial parent particle sizes > 80 $\mu$m, particles fragment upon collision, transferring up to 50 percent of their mass in a rubble-pile-like deposit named ‘pyramid' morphology on the target.

In the case of MIDAS particle collection, we investigate possible fragmentation of the collected particles using a sophisticated algorithm (the mean-shift algorithm method; \citealt{Kim_Mannel_MIDAS_catalog}) to identify dust pieces belonging together, so-called MIDAS clusters. The algorithm was cautiously tested in \citet{Kim_Mannel_MIDAS_catalog}, for example, the expected size of the clusters for each target and thus the optimal cluster number was validated by the k-nearest neighbors algorithm (\citealp{Cover1967}) and silhouette analysis (\citealp{Rousseeuw1987}), the assignment of particles into clusters was shown to be statistically significant, similar results were obtained by a different algorithm described in \citet{Longobardo_2022_MNRAS}, and the identified cluster morphologies are comparable to those found in other experiments (\citealp{Ellerbroek_labstudy_2017, langevin_typology_2016}).

\begin{table*}[]
\caption{Mean velocity, orbital epoch of collection, type of terrains, and the number of the different cluster types (deposits) listed for each MIDAS target.} 
\label{table:MIDAS_Cluster_Ellerbroek}
\renewcommand{\arraystretch}{1.4}
\centering\begin{tabular}{ccccccc}
\toprule
\multirow{3}{*}{\textbf{Target}} &  \textbf{Mean}   &\textbf{Orbital} &  \textbf{Type} &\textbf{MIDAS} & \textbf{MIDAS}         & \textbf{MIDAS} \\
 &  \textbf{velocity} &\textbf{epoch of}  &\textbf{of} & \textbf{Single}  & \textbf{Footprint} & \textbf{Pyramid}\\
& [m s $^{-1}$] &\textbf{collection} &\textbf{terrains} & \textbf{Cluster}  & \textbf{Cluster} & \textbf{Cluster}\\
\hline
11  & 3.1 $\pm$ 1.5 &  Pre-perihelion  &  Rough + Smooth      & 31                      & 14            & 7    \\
13  & 2.7 $\pm$ 0.3 &  Pre-perihelion  &  Rough + Smooth      & 10                      & 4             & 0    \\
14  & 7.21 $\pm$ 0.09 & Post-perihelion  &  Rough & 48                      & 61            & 17   \\
15  & 7.2 $\pm$ 0.2 &  Pre-perihelion  &   Smooth     & 3                       & 0             & 1    \\
\hline
Total       &    & &  & 92                     & 79            & 25    \\                  
\bottomrule
\end{tabular}
\end{table*}

Using this identification, we investigate whether the morphologies of MIDAS deposits are qualitatively similar to those found by previous laboratory studies \citep{Ellerbroek_labstudy_2017}. The automatic algorithm determined clusters with characteristics that are strongly aligned with the three morphological groups defined in \citet{Ellerbroek_labstudy_2017}. In particular, we consider the number of all fragments in one cluster (MIDAS single cluster vs. MIDAS footprint and MIDAS pyramid clusters), the existence of a dominant central fragment (MIDAS footprint cluster vs. MIDAS pyramid cluster), and the diameter of the scattering field compared to the size of the central fragment or deposit (MIDAS footprint cluster vs. MIDAS pyramid cluster). In conclusion, we summarize the MIDAS classification into three different morphological types defined based on \citet{Ellerbroek_labstudy_2017} as follows:\newline

\noindent -\textbf{MIDAS single cluster\footnote{We note that the ‘MIDAS single cluster' term should not be confused with ‘MIDAS Single' from particle surface and volume distribution in Sect.~\ref{method: MIDAS_particle_shape descriptors}}}: Deposits should have a clearly defined central component with only a few ($\leq$ 2) or no scattered fragments, resulting from the sticking of the entire aggregate upon impact with little or no fragmentation. 
    
\noindent -\textbf{MIDAS footprint cluster}: The most important morphological characteristic is that there is a collection of monomers with no dominant central fragment, resulting from the bouncing of the parent particle upon impact. The scattered field is about twice the size of the left central fragment due to the bouncing effect.
    
\noindent -\textbf{MIDAS pyramid cluster}: Deposits should have a clearly defined central fragment with a lot of scattered debris around it, resulting from the fragmentation upon impact. The diameter of the scattering field is about twice the central deposit. \newline

\begin{figure*}
\includegraphics[width=6.1cm, height= 10cm]{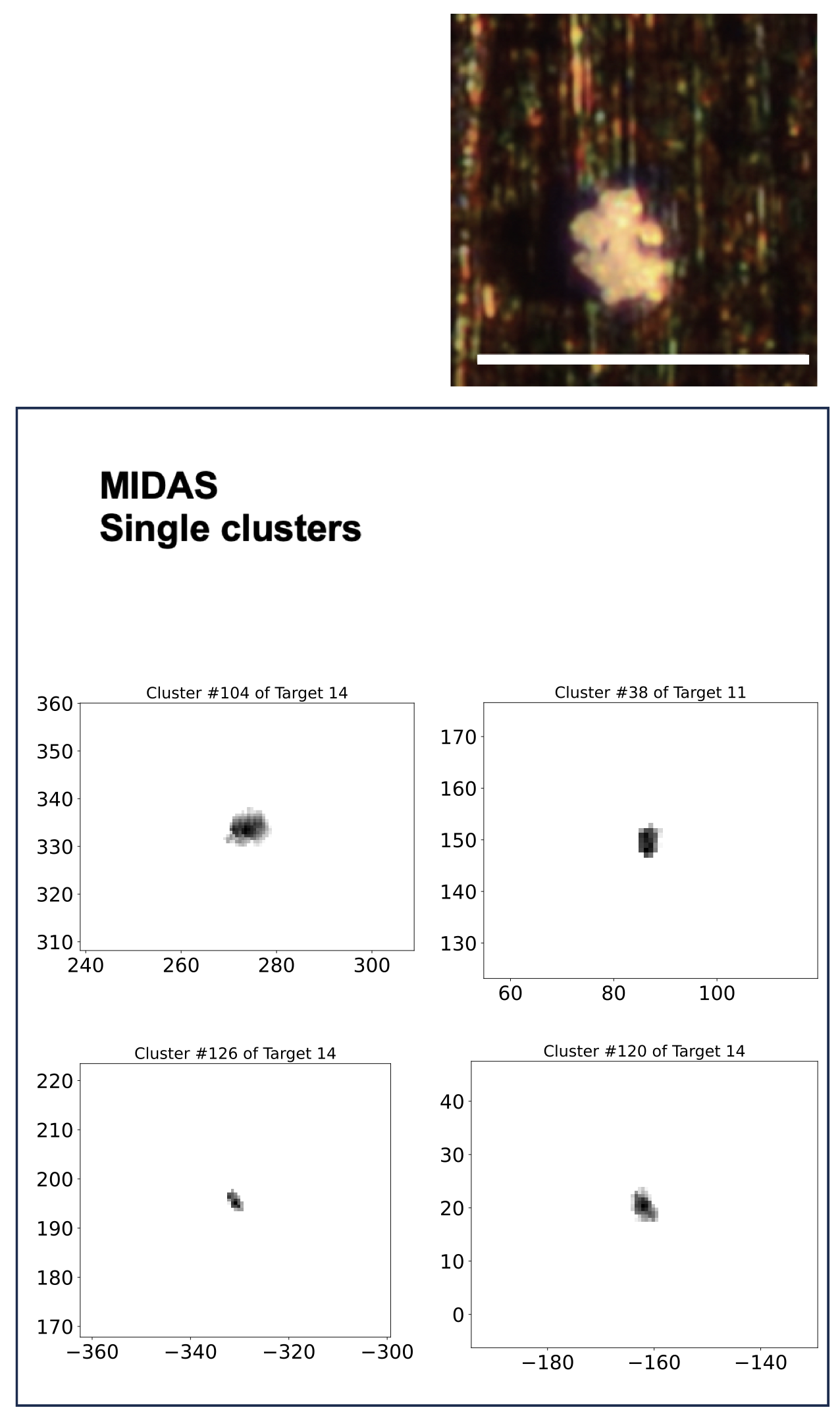}
\includegraphics[width=6.1cm, height= 10cm]{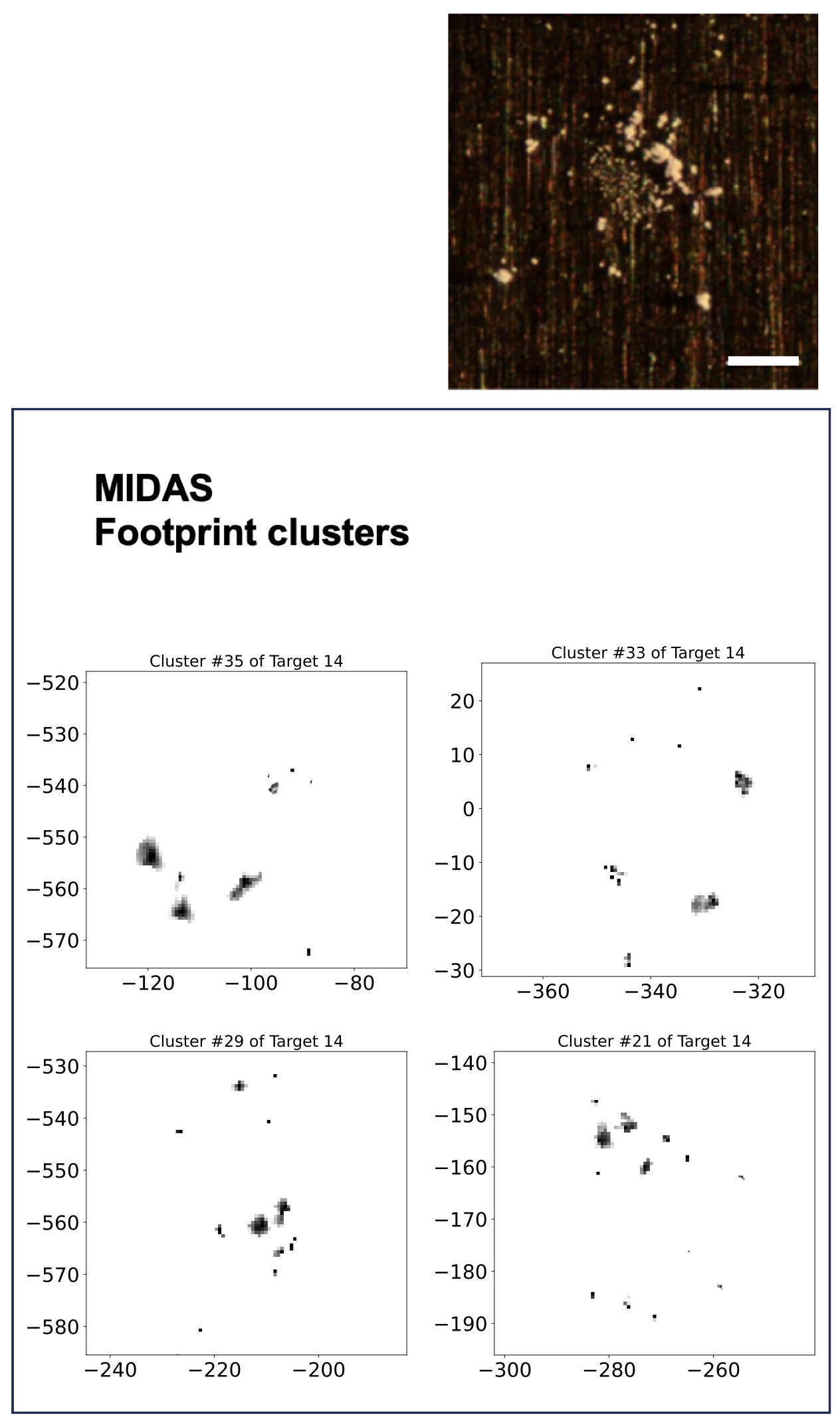}
\includegraphics[width=6.1cm, height= 10cm]{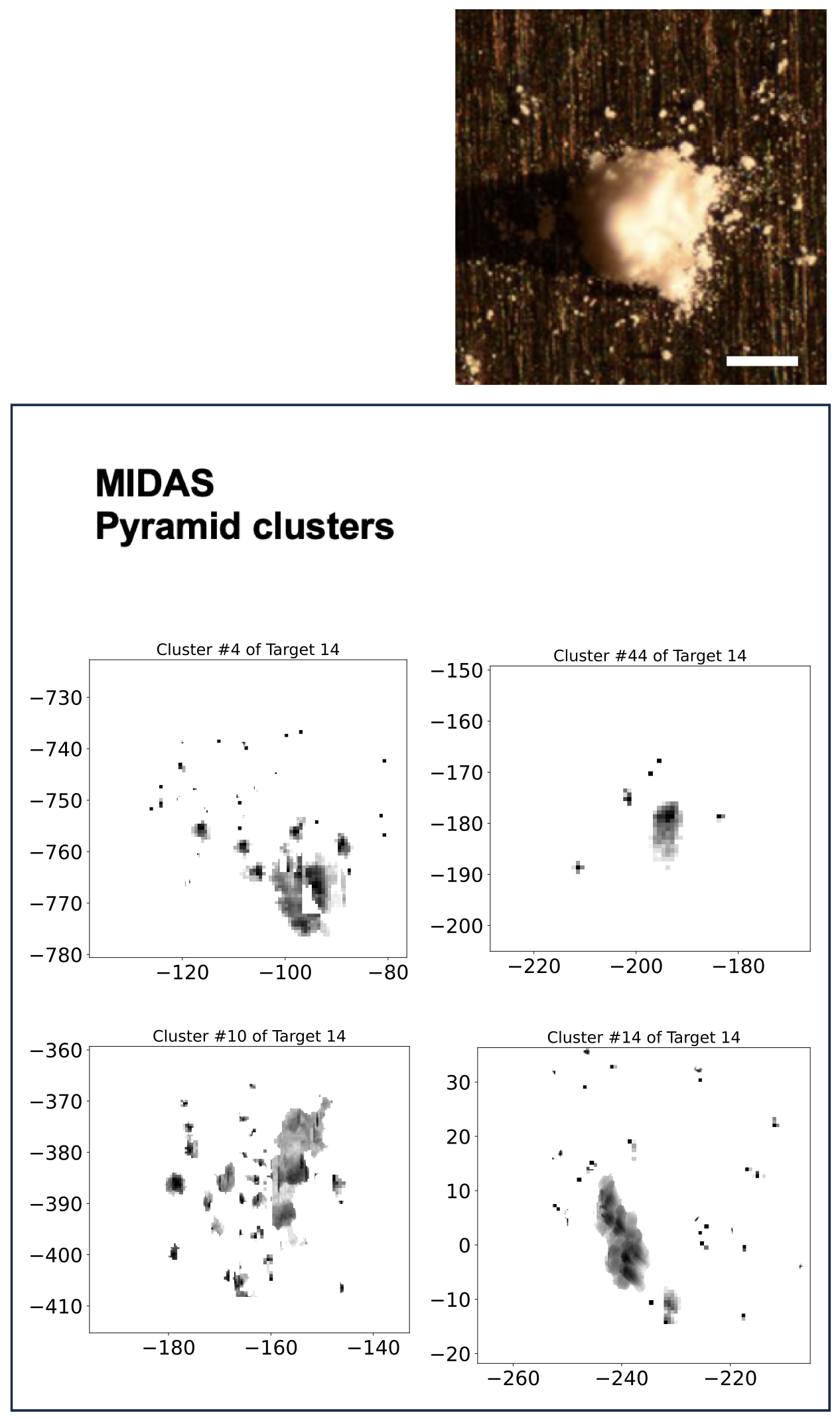}
\caption{An example showcasing the classification of MIDAS clusters is presented, highlighting three distinct morphological classes: MIDAS single, MIDAS footprint, and MIDAS pyramid clusters. Each class is depicted through 2D representations featuring four MIDAS morphologies (clusters). For comparison purposes, a confocal microscope image showing a dust deposit, which exhibits corresponding morphological features observed by~\citet{Ellerbroek_labstudy_2017}, is displayed in the bottom-right corner of each sub-figure. The white scale bar accompanying the image indicates a length of 100 $\mu$m, providing a reference for size.}
\label{fig:ellerbroek_MIDAS}
\end{figure*}

An example of the classification of MIDAS clusters, e.g., a MIDAS single cluster, a MIDAS footprint cluster, and a MIDAS pyramid cluster, and the appearance of different morphological types from \citealt{Ellerbroek_labstudy_2017} (right top in each sub-figure) are shown in Fig.~\ref{fig:ellerbroek_MIDAS}.

We find that the majority of morphological types of individual clusters in all targets are single or footprint deposits. According to \citet{Ellerbroek_labstudy_2017}, small (< 80 $\mu\rm{m}$) particles leave single deposits, thus a large number of MIDAS Single deposits underline the small parent particle size. Additionally, the dominance of MIDAS footprints suggests that if larger particles arrived they had slow impact velocities (\citealp{Ellerbroek_labstudy_2017}). A summary of a number of different morphological types of all targets based on the mean shift method (\citealp{Kim_Mannel_MIDAS_catalog}) and the target mean velocity (\citealp{Longobardo_2022_MNRAS}) is given in Table~\ref{table:MIDAS_Cluster_Ellerbroek}.

\indent Furthermore, we compare MIDAS clusters to COSIMA studies \citep{langevin_typology_2016, Merouane_COSIMA_size_D_2016, hornung_assessment_2016}, which showed the typology and optical properties of cometary particles collected by the COSIMA mass spectrometer in the inner coma of 67P. They found four different morphological classes of particles: ‘compact', ‘shattered clusters', ‘glued clusters', and ‘rubble piles'. However, we find that only the compact particles, which are particles with well-defined boundaries with only a very small number of satellite particles, are relatively easy to identify in the MIDAS data. The lack of a numerical definition of the different shapes found by COSIMA and the different particle sizes (COSIMA in the hundreds of micrometer range, MIDAS in the tens of micrometer range) precludes the application of comparative algorithms. Consequently, we only use the MIDAS cluster classification based on \citet{Ellerbroek_labstudy_2017} for further investigations in the present study. 

We note that all these morphological classes of the particles detected by MIDAS and COSIMA belong to the GIADA ‘compact' particles, although GIADA also defined a second category called ‘fluffy' that was not found by COSIMA and only once by MIDAS (\citealp{Longobardo_2020a_Merging_data, Longobardo_2022_MNRAS, mannel_fractal_2016}).

	
	\section{Results and discussion}\label{sec:results}

\subsection{Evaluation of MIDAS particles}\label{result: Evaluation_MIDAS_particles}

\subsubsection{MIDAS particle shape descriptor}\label{result: MIDAS_particle_shape_descriptor}

We calculate the shape descriptors described in Sect.~\ref{method: MIDAS_particle_shape descriptors} to all MIDAS particles usable for analysis. A summary of the statistics of the MIDAS particle shape descriptors is given in Table~\ref{table: MIDAS_particle_shape_descriptor}. The detailed information on all MIDAS particle characteristics and shape descriptors can be found in Table~\ref{table: MIDAS_particle_shape_descriptor_full_table} of Appendix \ref{appendix: Shape descriptor of MIDAS particles}.

\begin{table*}[]
\caption{Overview of the mean and standard deviation of the shape descriptors of MIDAS particles for each target.} 
\label{table: MIDAS_particle_shape_descriptor}
\renewcommand{\arraystretch}{1.4}
\centering\begin{tabular}{cccccc}
\toprule
\textbf{Target} & \textbf{Number of particles} &  \textbf{Aspect ratio} &  \textbf{Elongation} &  \textbf{Circularity} &  \textbf{Convexity}\\
\hline
11  & 250 & 0.47 $\pm$ 0.27  & 3.68 $\pm$ 1.95 & 0.60 $\pm$ 0.17 & 0.47 $\pm$ 0.02 \\
13  & 40 & 0.39 $\pm$ 0.13  & 4.54 $\pm$ 2.31 & 0.53 $\pm$ 1.86 & 0.47 $\pm$ 0.03 \\
14  & 785 & 0.57 $\pm$ 0.44  & 4.00 $\pm$ 2.41 & 0.57 $\pm$ 0.20 & 0.47 $\pm$ 0.03 \\
15  & 7 & 0.45 $\pm$ 0.09  & 2.83 $\pm$ 1.02 & 0.67 $\pm$ 0.18 & 0.47 $\pm$ 0.01 \\
\hline
MIDAS Total  & 1082 & 0.54 $\pm$ 0.40  & 3.93 $\pm$ 2.31 & 0.58 $\pm$ 0.19 & 0.47 $\pm$ 0.03 \\
\bottomrule
\end{tabular}
\end{table*}

First, we find that the value of the aspect ratio of all the MIDAS particles is found around 0.53 $\pm$ 0.40, which is smaller than expected since one would expect the value of 1 for unflattened particles. According to the finding from \citealp{Ellerbroek_labstudy_2017}, smaller and slower arriving particles (e.g., MIDAS particles) seem to show almost no alteration. Thus, we conclude that MIDAS particles seem to be flattened upon the impact, indicating not enough internal strength to stay unaltered based on our results. Additionally, we find that the value of the aspect ratio of all the MIDAS particles is higher than the one of COSIMA particles showing 0.36 $\pm$ 0.21 (\citealp{langevin_typology_2016, Lasue_simulation_2019}, which indicates higher strength of smaller particles.

Furthermore, we find that the majority of MIDAS particles have an elongation larger than 1 with the value of elongation 3.93 $\pm$ 2.31 (i.e., the largest axis is about 4 times longer than the smallest one). This means that their shapes highly deviate from a sphere, which shows good agreement with previous studies (e.g., \citealt{fulle_dust_2017}). We note that the elongation of the particles on target 15 appears to be significantly smaller compared to the other targets. However, considering the uncertainty of one of target 15, this difference becomes less significant. 

The link between interstellar dust grains and MIDAS dust particle can be found via their approximate shapes. For example, linear polarisation of the light (\citealp{Hiltner_polarization_1949, Mathis1990}) suggests that the interstellar dust grains should be both elongated and partially aligned. Furthermore, the presence of elongated grains is also considered in various cometary dust models (e.g., \citealp{Greenberg_Gustafson_1981, ACLR2007, Lasue2006}). In particular, \citet{Greenberg_Gustafson_1981} proposed that cometary dust comprises agglomerations of elongated submicrometer-size interstellar dust with aspect ratios of 2-4, which showed a satisfactory agreement between light scattering experiments and observations.

We also find that 3D elongation is bigger than 2D elongation in most cases, meaning that the shortest axis between the height and the semi-minor axis is the height. These results show consistency with \citet{bentley_morphology_2016}, where most particles had an elongation of about $\sim$ 3, which mostly means that the height of the particle is three times smaller than the maximum diameter. Both results may imply either the particles had a natural elongation and are more probable to stick on MIDAS targets when arriving with the flattest side up-wards (i.e., having the largest cross-section sticking to the target), they collided with the target in an arbitrary alignment but turned their shortest axis upwards upon impact, or the particles are flattened upon impact. This result is also consistent with the statistics of the MIDAS particle catalog (\citealp{Kim_Mannel_MIDAS_catalog}), with the bigger mean value of equivalent radius on the 2D projection compared to the radius derived from the volume of the particle. On the assumption that the fractal particle E \citep{mannel_fractal_2016} is not heavily altered by the impact since it preserved its fractal structure, it is suggested that the elongation of its sub-units forming particles is natural and not created by the impact \citep{bentley_morphology_2016}. However, sub-units of particle E mostly show an elongation around 3 with their height as the shortest axis \citep{bentley_morphology_2016}, suggesting that the sub-units may have suffered flattening upon impact. If a rotation of the sub-units cannot happen to keep the fractal structure intact, the aspect ratio (and/or 3D elongation) could be a reasonable parameter to find particles with the most pristine morphology. 

Additionally, we find that the value of the circularity of MIDAS particles is 0.58 $\pm$ 0.19, indicating MIDAS particles deviate from being circular. On the other hand, we find that the value of the convexity of MIDAS particles showed  0.47 $\pm$ 0.19, meaning that most of the convex hull area of MIDAS particles is occupied by their perimeter concavities and thus showed the rather smooth particle shape on 2D projection view. In particular, all MIDAS particles show a similar value of convexity. 

We note that these shape descriptors are particularly tip convolution dependent, which is an inherent feature of all AFMs (e.g., \citealp{Vijendran2007}) and hard to be fully removed as all AFM images are the convolution of the tip shape and the particle shape, particularly for rough surfaces. For example, multiple points on the tip may be simultaneously in contact with the surface and so the shape and geometry of the tip has an effect on the accuracy of the imaging. This has the effect of making protruding features appear wide, while holes appear smaller with a broader tip. More specifically, the variation in particle height leads to considerable uncertainties in tip convolution (\citealp{Mannel_classification_2019}), with a calculated uncertainty ranging from 0.4 to 178 percent (using the equation in \citealt{Mannel_classification_2019}) for the dataset of all MIDAS particles. The smallest and largest uncertainties are associated with the flattest and roughest particles, respectively. Therefore, due to the broad tip used in the MIDAS instrument (with a half-opening angle of 25$^{\circ}$ $\pm$ 5$^{\circ}$), very small fringes on the particles may not be effectively imaged.

\subsubsection{Relation between shape descriptors and particle characteristics, collection target, and cluster morphology, and MIDAS shape descriptor maps}\label{result: Relation_shape_descriptors_ohter_parameters}

In the present section, we investigate possible relationships between the discussed shape descriptors and particle characteristics (e.g., size), MIDAS collection targets (i.e., here related to collection velocities/periods and cometary source regions with dust activities; see Sect.~\ref{method: MIDAS_AFM}), and cluster morphology derived from laboratory experiments \citep{Ellerbroek_labstudy_2017} to find similarities/differences between the particles. In particular, we aim to find particles that preserved at least some properties pristinely (i.e., as they were when stored in the comet and/or as they formed in the early solar system). \newline

\noindent \textbf{Shape descriptors vs. particle size} \hspace{2mm} We first investigate the relation between MIDAS shape descriptors and particle size. Furthermore, we further investigate the correlation between large particles (e.g., r > 8 $\mu$m) and small particles (e.g., r < 8 $\mu$m) as indicated by the size distribution of MIDAS particles. For example, particles larger than 8 $\mu$m in radius show lower aspect ratios. In particular, we find no large particles with an aspect ratio larger than 0.8. 

\begin{figure}
\includegraphics[width=9cm, height= 6.5cm]{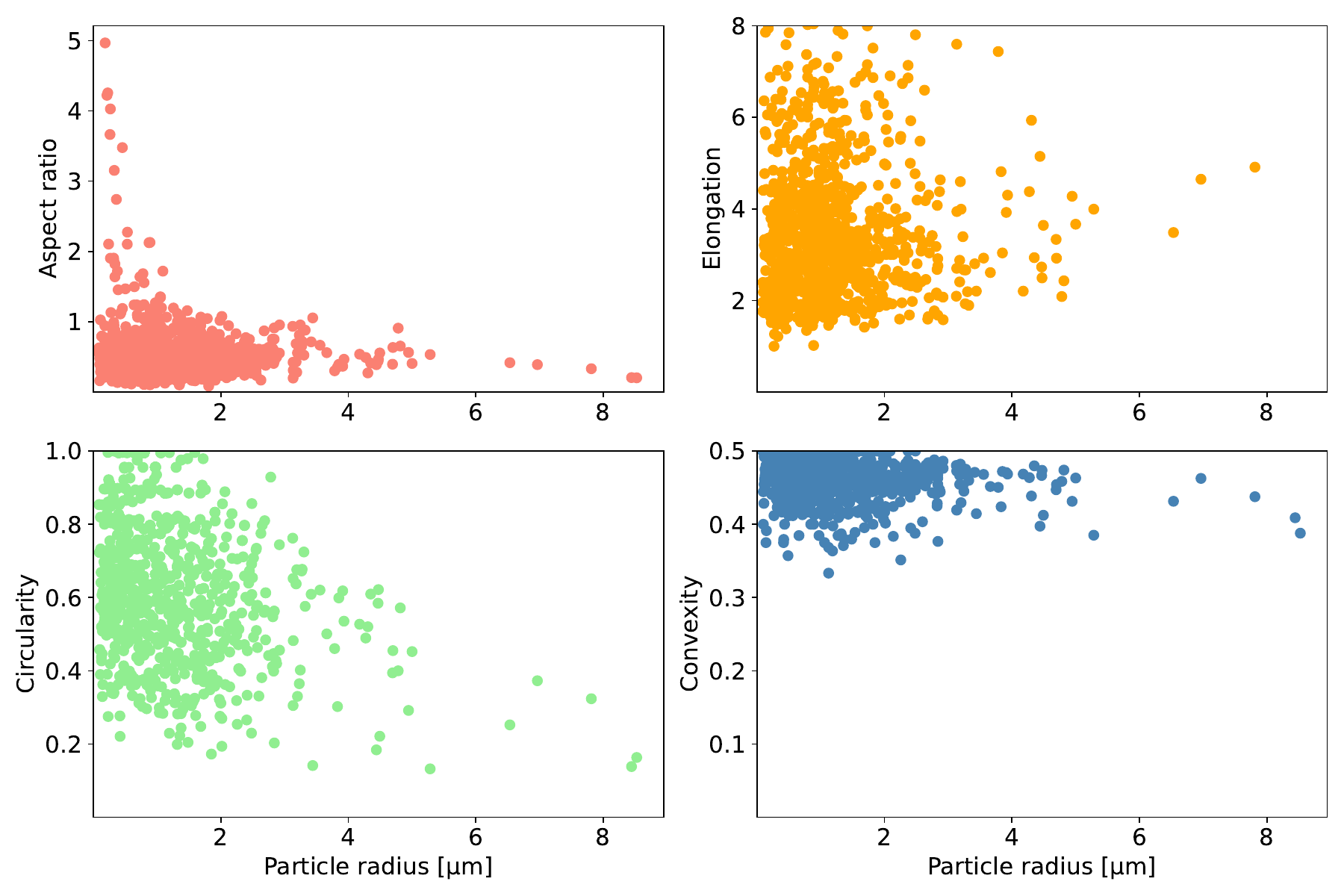}
\caption{Relations between MIDAS particle shape descriptors and sizes of particles.}
\label{fig:Relations_MIDAS_particle_shape_descriptors_and_size}
\end{figure}

The scattered plot of the relation between all MIDAS shape descriptors and particle size can be found in Fig.~\ref{fig:Relations_MIDAS_particle_shape_descriptors_and_size}. We find no clear trend between shape descriptors and particle size, in particular, only weak correlations between shape descriptors and size.  

We first find that larger MIDAS particles within low standard deviation (i.e., 0.26 $\pm$ 0.09) show a tendency to lower aspect ratios, which can also be explained by their larger size and therefore expected weaker strength. We also find that the aspect ratio of larger MIDAS particles shows a rather similar value of COSIMA particles (see Sect. \ref{result: MIDAS_particle_shape_descriptor}), compared to the one of smaller MIDAS particles. On the other hand, small particles are the only candidates with extremely large aspect ratios, however, they show a large variance as well (i.e., 0.49 $\pm$ 0.41). Based on this finding, it is reasonable to conclude that particles with a high aspect ratio have undergone less alteration or deformation upon impact and thus have retained their original morphology compared to particles with lower aspect ratios.

We next find that larger MIDAS particles show relatively smaller elongation values, showing again the degree of flattening as the shortest axis between the height and the semi-minor axis in the particle is mostly the height in our analysis (see Sect. \ref{result: MIDAS_particle_shape_descriptor}). 

Furthermore, we find that larger MIDAS particles show smaller values of circularity and convexity. As \citealp{Kimura2020} indicated that dust aggregates in the early stage of comet formation may grow under BCCA (ballistic cluster-cluster aggregation), corresponding to the decreasing density as the particle
size increases (e.g., \citealp{Kozasa1992, Mukai1992, Kataoka2013}) up to scales of 10 $\mu$m (i.e., corresponding to MIDAS size scale) due to the motion of dust particles that are controlled by Brownian motion, where dust particles of similar size hit and stick to each other (\citealp{Weidenschilling_1989, weidenschilling_formation_1993}), and then by BPCA (ballistic particle-cluster aggregation) once aggregate particles grow to the size of 10 to 100 $\mu$m. Based on this finding, MIDAS particles would show the change in different shape descriptors (e.g., circularity, convexity, and elongation) for the larger particles, if larger MIDAS particles consist of a few agglomerates (e.g., relatively few large sub-particles) of the next smaller hierarchical step. For example, larger MIDAS particles would naturally deviate from a spherical shape and lead to more fringed rims. Consequently, we suggest that our findings with smaller values of circularity and convexity for larger MIDAS particles are possibly explained by a hierarchical growth process and would show consistency with the fractal growth process as it may be expected by models of early cometary dust accretion (e.g., \citealp{Weidenschilling_1989, Kimura2020}). This finding also aligns with the previous finding of a higher fragmentation or alteration for larger MIDAS particles (\citealp{Mannel_classification_2019}), indicating a relatively lower internal strength keeping the larger particles together. During the particle selection for further investigation (see Sect. \ref{method: MIDAS_particle_shape descriptors}), we also find that out of 135 particles larger than 10 $\mu$m in total, only 22 particles remained unfragmented, indicating around 85\% of larger particles underwent fragmentation during the collection and/or scanning process. This indicates that fragmentation of the particle in subunits is easier than fragmentation of the subunits in grains, which also underlines the suggested hierarchical dust structure with characteristic subunit size regimes (\citealp{Mannel_classification_2019}). \newline

\begin{figure}
\centering
\renewcommand{\arraystretch}{0}
\setlength{\tabcolsep}{0pt}
\begin{tabular}{cc}
\includegraphics[width=0.5\linewidth]{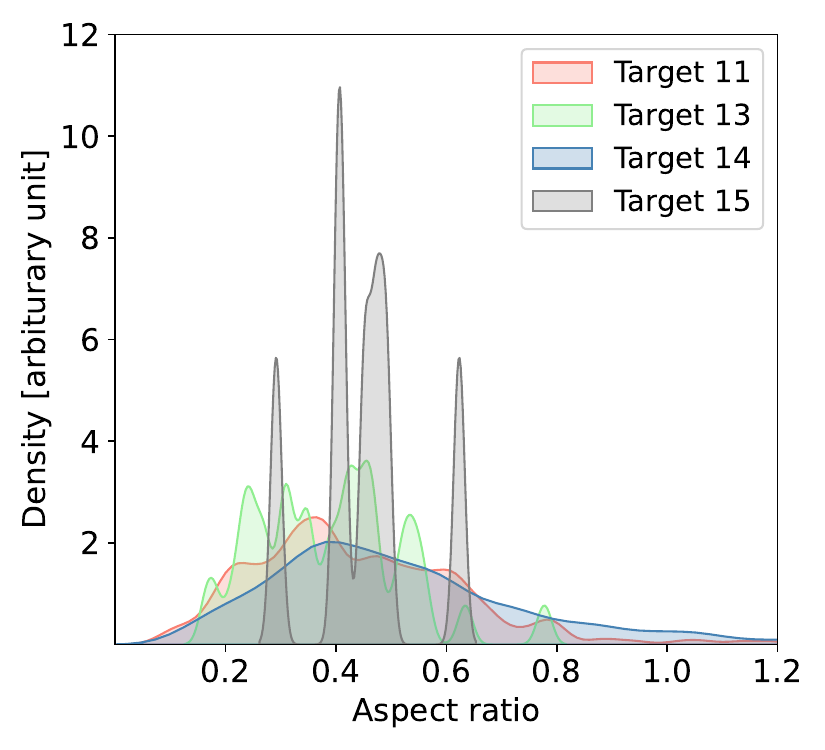} &
\includegraphics[width=0.5\linewidth]{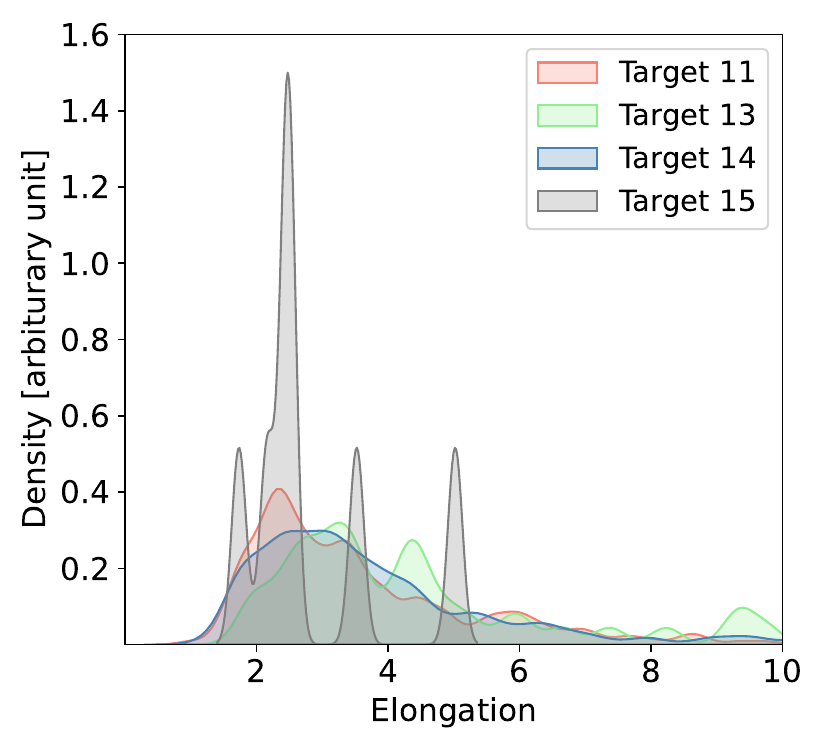} \\
\includegraphics[width=0.5\linewidth]{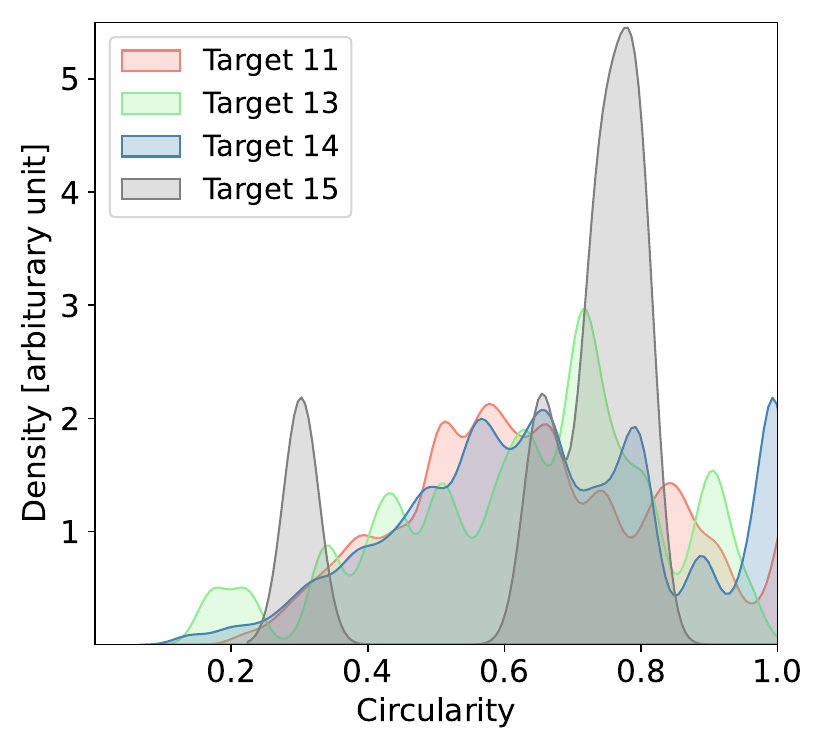} &
\includegraphics[width=0.5\linewidth]{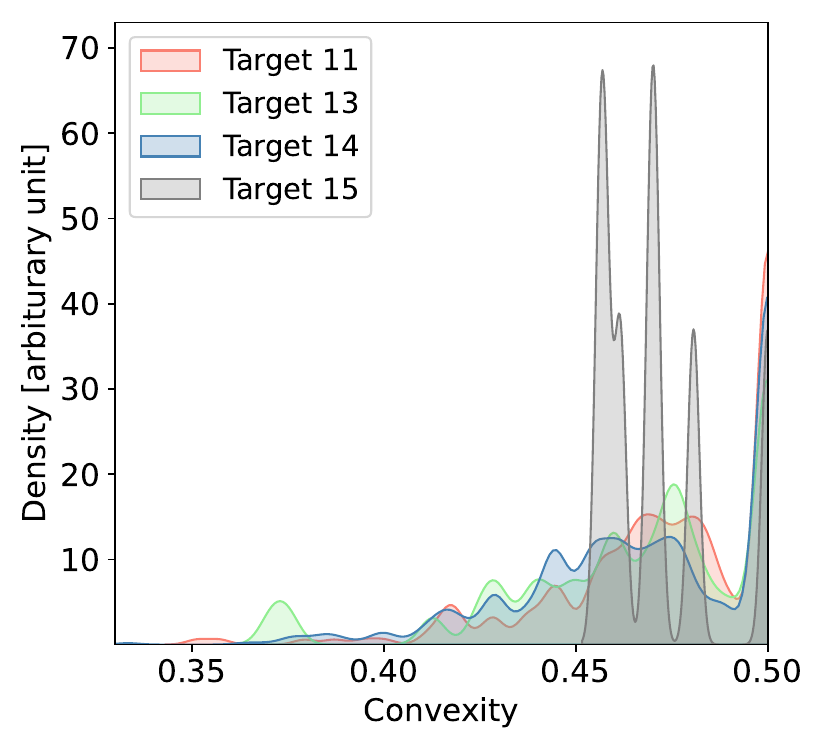}
\end{tabular}
\caption{Density plots of MIDAS particle shape descriptors for each dust collection target. The graph of target 15 shows narrow high spikes due to low statistics, contrary to the smooth graph of target 14 resulting from a high number of data points. Overall, the particle shapes seem to be similarly distributed independently from the collection target, i.e., the collection velocity, the cometary source region, or the period of dust emission from the comet.}
\label{fig:Relations_MIDAS_particle_shape_descriptors_and_target}
\end{figure}

\noindent \textbf{Shape descriptors vs. collection targets} \hspace{2mm} Next, we investigate the relation between MIDAS shape descriptors and collection targets. The distributions of the four MIDAS shape descriptors for each dust collection target, which is here a proxy for the source region (i.e., particles on target 14 originated from rough terrains/deeper layers, particles on target 15 originated from smooth terrains, and particles on targets 11 and 13 originated from both types of terrains; see Sect.~\ref{method: MIDAS_AFM}; \citealp{Longobardo_2022_MNRAS}), the collection velocity (i.e., target number 11 equates to a mean collection velocity of 3.1 $\pm$ 1.5 m\,s$^{-1}$, target number 13 to 2.7 $\pm$ 0.3 m\,s$^{-1}$, target number 14 to 7.21 $\pm$ 0.09 m\,s$^{-1}$, and target number 15 to 7.2 $\pm$ 0.2 m\,s$^{-1}$ (\citealp{Longobardo_2020a_Merging_data}), and/or the period of dust emission from the comet (i.e., particles on target 11, 13, 15 were collected when comet showed a nominal activity, while particles on target 14 were collected when comet showed an outburst), can be found in Fig.~\ref{fig:Relations_MIDAS_particle_shape_descriptors_and_target}.

We find that there is no clear trend between target and shape descriptors even for the unflattened (thus potentially less altered) particles (see Sect.~\ref{method: MIDAS_particle_shape descriptors}). Although the statistics on target 15 are very poor, the data still shows similar behavior in this investigation. Thus, we conclude that particles coming from different terrain and activity are similar in their investigated shape properties. This means dust activity such as dust ejection, partial dry-out (loss of volatiles), and recycling of dust material (backfall and subsequent ejection) did not alter the structure of particles on the micrometer scale. These results indicate that dust parameters associated with the collection target, such as dust activities, cometary source regions, and collection velocities and periods, do not play a significant role in the alteration of particle structure at the micrometer scale. We note that our finding about the homogeneity of MIDAS particles in their properties may show the consistency with globally uniform Martian soil all but the coarsest components (e.g., larger fragments, clasts, and spherules) at widely separated sites (\citealp{McSween2010}).\newline

\begin{figure}
\centering
\renewcommand{\arraystretch}{0}
\setlength{\tabcolsep}{0pt}

\begin{tabular}{cc}
\includegraphics[width=0.5\linewidth]{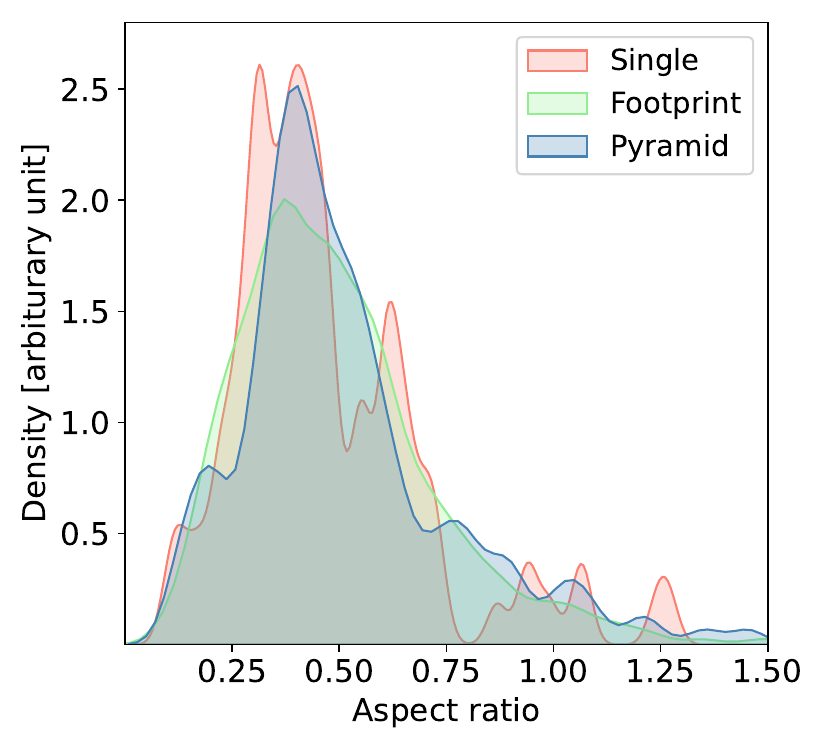} &
\includegraphics[width=0.5\linewidth]{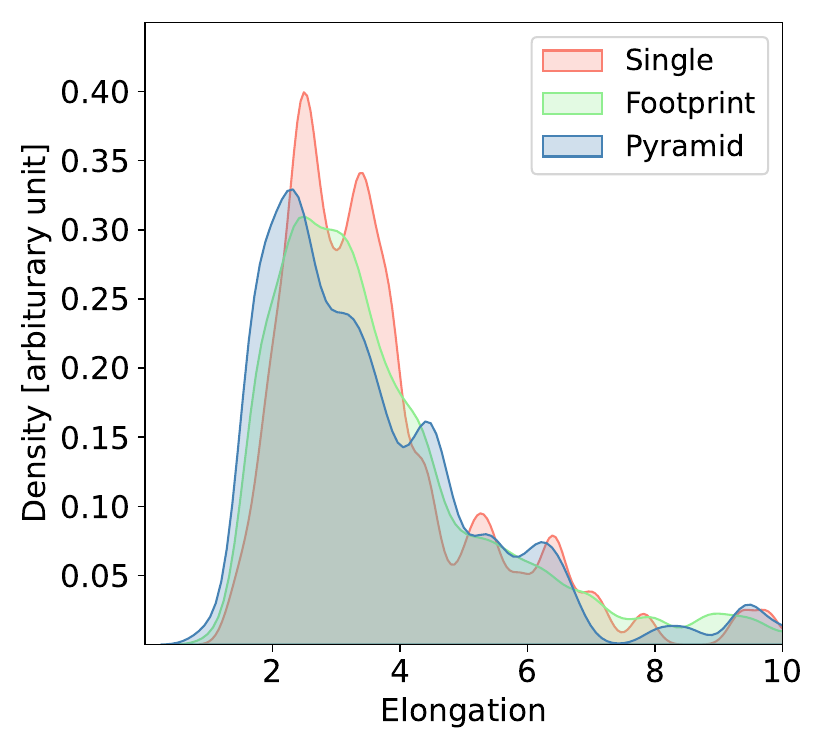} \\
\includegraphics[width=0.5\linewidth]{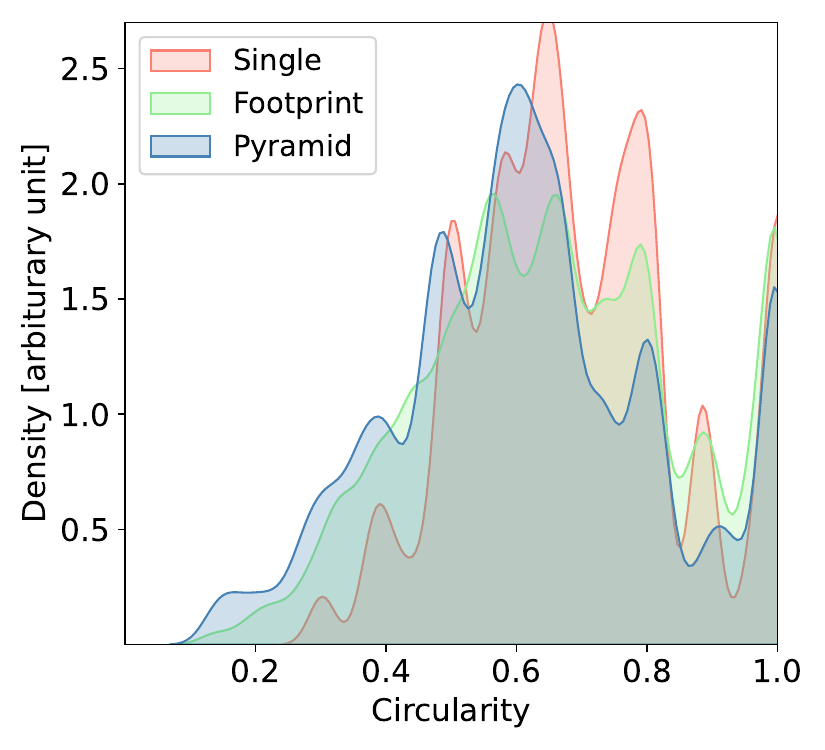} &
\includegraphics[width=0.5\linewidth]{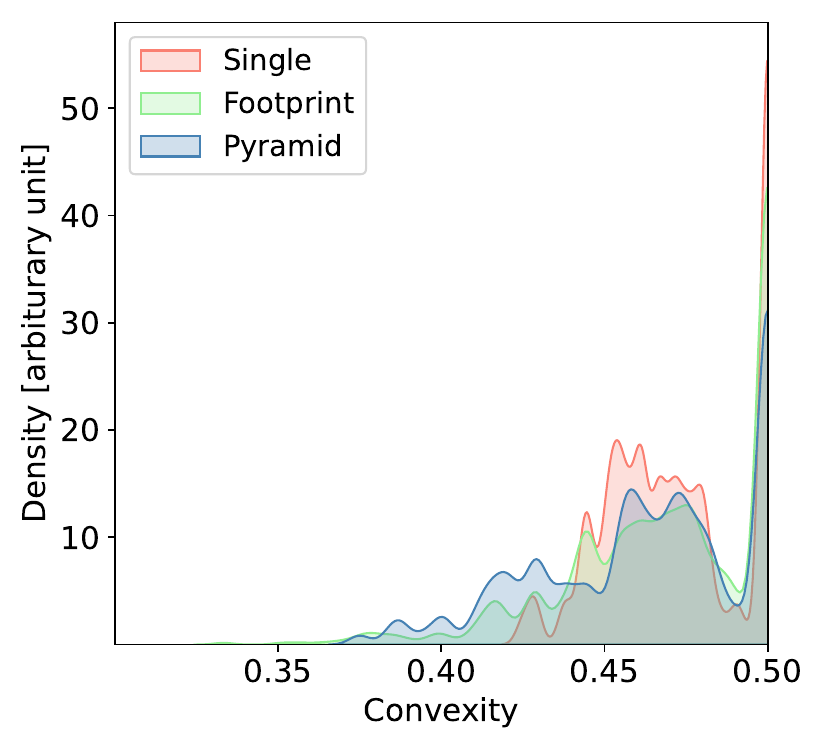}
\end{tabular}
\caption{Density plots of MIDAS particle shape descriptors depending on the cluster morphology. The similar particle shapes independent from the cluster morphology suggest that the collected parent particles were homogeneous in particle properties and consisted of similar subunits.}
\label{fig:Relations_MIDAS_particle_shape_descriptors_and_morphology}

\end{figure}

\noindent \textbf{Shape descriptors vs. cluster morphologies} \hspace{2mm} We finally investigate the relation between MIDAS particle shape descriptors and cluster morphologies according to the definition of \citet{Ellerbroek_labstudy_2017} and Sect.~\ref{method: MIDAS cluster classification}.  The distributions of the four MIDAS shape descriptors for each cluster morphology can be found in Fig.~\ref{fig:Relations_MIDAS_particle_shape_descriptors_and_morphology}.

We find that cometary dust particles of different cluster morphology are rather homogeneous in properties. More specifically, we find that the aspect ratio distributions of the particles are very similar and independent of the cluster type (see the left top panel of Fig.~\ref{fig:Relations_MIDAS_particle_shape_descriptors_and_morphology}). This may imply that the sub-units of different cluster types are very similar in their shape and composition. On this assumption, different cluster types may not be created by differently composed parent particles. A similar situation of surprisingly homogeneous properties of the sub-units of cometary dust particles down to the sub-micrometer scale was found for IDPs \citep{Bradley_IDP_2014} and material collected by Stardust \citep{hoerz_impact_2006}. Following the results from \citet{Lasue_simulation_2019}, the different cluster morphologies detected by MIDAS, COSIMA, or in simulations are not only created by a change in sub-unit properties (e.g., different particle structure and/or composition) but also by different impact velocities. Thus, based on our findings, which indicate the similar composition and internal cohesive strength of the parent particles, the variations in impact velocities during the collection process might lead to distinct cluster morphologies. This result is in good agreement with the laboratory experiments of \citet{Ellerbroek_labstudy_2017}, where MIDAS (and in a wider sense also COSIMA) cluster types were reproducible by using one simple cometary analog material (SiO$_{\rm2}$ spheres with a polydisperse size distribution) with different impact velocity.

However, we find that the different mean velocities of the MIDAS dust collection (approximately between 3 and 7m\,s$^{-1}$; \citealt{Longobardo_2020a_Merging_data}; see also Sect.~\ref{method: MIDAS_AFM}) did not alter the mean deposit morphology (see Fig.~\ref{fig:Relations_MIDAS_particle_shape_descriptors_and_target}). This discrepancy with \citet{Ellerbroek_labstudy_2017} can be interpreted as a result of different material compositions that, in particular, lead to different cohesive strengths and thus different threshold velocities for the creation of different cluster morphologies. This is not surprising as the material used by \citet{Ellerbroek_labstudy_2017} was extremely simple SiO$_{\rm2}$ aggregates. On the contrary, compositional analysis of dust of comet 67P (e.g., by the COSIMA instrument) indicated a more varied composition, in particular with a share of about 45 mass percent of the particles consisting of a long-chained organic material \citep{bardyn_dust_2017}. The expected different cohesiveness of SiO$_{\rm2}$ aggregates and dust of comet 67P might be confirmed by the MIDAS results.\newline

\noindent Based on the investigation in Sect.~\ref{method: MIDAS_particle_shape descriptors}, we create maps combining dust clustering (\citealp{Kim_Mannel_MIDAS_catalog}) and MIDAS shape descriptors, which can be found in Appendix \ref{appendix_shape_descriptor_maps}. MIDAS shape descriptor maps allow a quick overview of dust distribution and properties.

We find that the least altered particles (e.g., higher aspect ratio particles) can occur even in large clusters (e.g., MIDAS footprint cluster or MIDAS pyramid cluster). This contradicts the finding from \citet{Ellerbroek_labstudy_2017}, which indicated that the smaller and slower a particle arrives at the target the higher the chance that it sticks without alteration. No alteration would mean the particle does not fragment, making it a cluster with only a single particle inside, a so-called single-particle-cluster (SPC). Thus, the aspect ratio of SPCs is expected to be higher. However, high aspect ratio MIDAS particles seem to be randomly scattered over the target and mostly be part of clusters (see Figs.~\ref{fig:2D_dust_clustering_map_10_aspect_ratio}, \ref{fig:2D_dust_clustering_map_12_aspect_ratio},  \ref{fig:2D_dust_clustering_map_13_aspect_ratio}, and 
\ref{fig:2D_dust_clustering_map_14_aspect_ratio}). As \citet{hornung_assessment_2016} suggested, unflattened particles are possibly fragments that are pre-existing in parent particles. To be specific, on the assumption that the force that holds the grains in the subunits together is stronger than the force that holds the subunits of the particle together (\citealp{Mannel_classification_2019}), the impact energy may first be used for fragmentation (i.e., breaking the bonds between the pre-existing sub-units) when the parent particle hit the target, and if all impact energy is used up before compaction of the sub-units could happen some sub-units are set free unaltered. In other words, the impact energy at dust collection did only suffice to destroy the bonds between the sub-units in the parent particle but did not alter the internal structure of the sub-units (i.e., easier fragmentation of the particle in subunits compared to one of the subunits in grains). This may lead to MIDAS particles with high aspect ratios so thus MIDAS particles with unaltered structures can be found in clusters as well. This finding again aligns well with the hierarchical agglomerate structure for cometary dust as the tensile strength of the porous dust layer would decrease with increasing aggregate size (see also discussion in the relation between shape descriptors and particle characteristics such as size; \citealp{skorov_dust_2012, Kimura2020}) and could lead to differences in the cohesive strength of particle units and subunits at MIDAS scales (e.g., 10 micrometers or less) if the cometary dust is characterized by a hierarchical structure of fractal aggregates derived from the coagulation process in the solar nebula. Furthermore, \citet{Kimura2020} indicated an outer layer of organic material that may be of interest for the bond strength between particle subunits, implying that different processes of aggregation could lead to more differences in the cohesive strength of particle units at these different scales, supporting the MIDAS measurements.

\subsubsection{MIDAS cluster morphology – aspect ratio classification compared to COSIMA results}\label{result: MIDAS_Morphology_COSIMA}

\citet{Lasue_simulation_2019} conducted a numerical simulation of aggregate impact flattening to interpret the initial properties of particles/clusters collected by COSIMA. They found that the particle morphologies observed by COSIMA and those generated by the laboratory experiments of \citet{Ellerbroek_labstudy_2017} are consistent. Thus, it is of interest to compare MIDAS cluster structures to those found by COSIMA and laboratory experiments, and check if the same structures are present, although the size of COSIMA particles is one to two orders of magnitude larger. Furthermore, the laboratory study and the COSIMA images yield insight into parent particle alteration, which might help to determine the least altered MIDAS particles.

Following the approach taken in \citet{Lasue_simulation_2019} allows a quantitative comparison of the morphological cluster classification between COSIMA and MIDAS in terms of aspect ratio calculations. As discussed in Sect.~\ref{method: MIDAS cluster classification}, the MIDAS cluster morphology classification appears to follow \citet{Ellerbroek_labstudy_2017} such that the dust can be categorized into one of three classes: MIDAS single, MIDAS footprint, or MIDAS pyramid cluster morphology. 

\begin{figure*}
\includegraphics[width=9cm, height= 6.45cm]{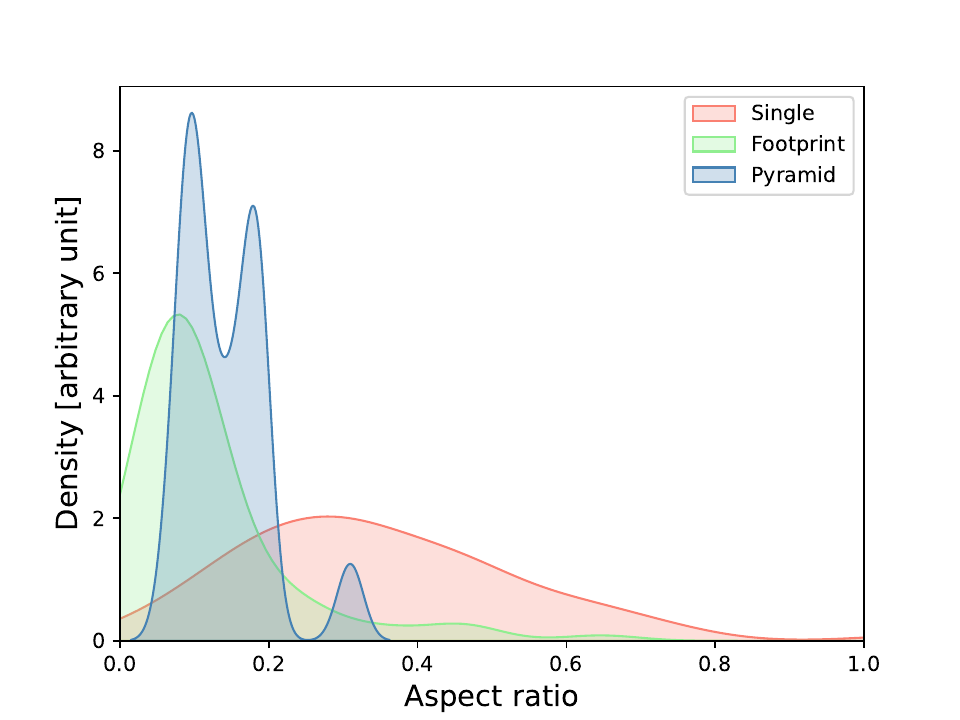}
\includegraphics[width=9cm, height= 6.45cm]{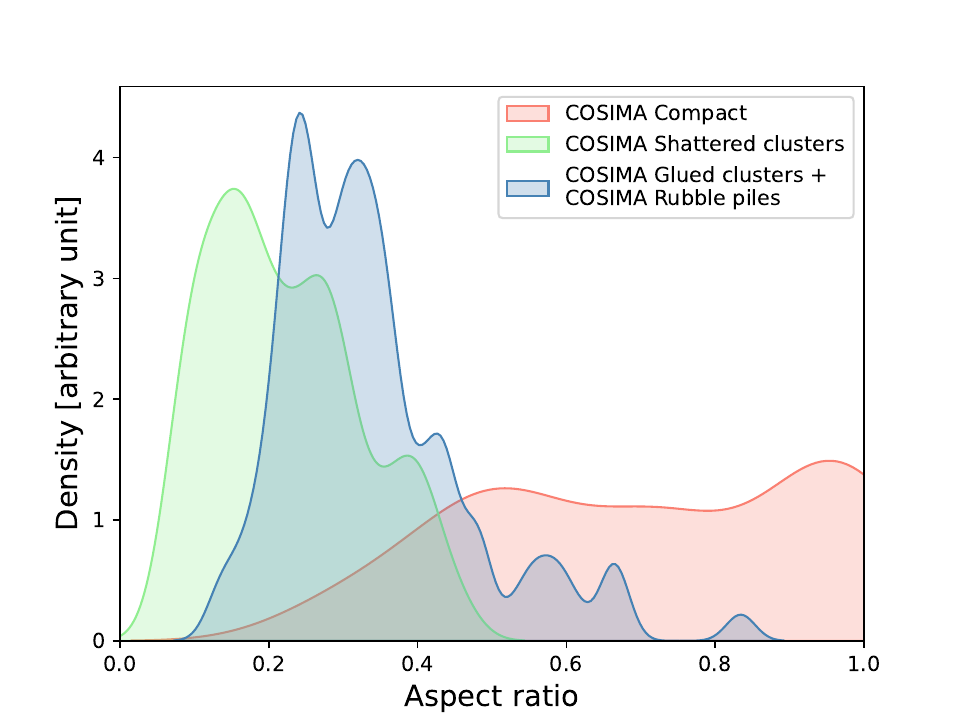}
\caption{Comparison between calculated aspect ratios of MIDAS clusters (left) and the COSIMA clusters (right; \citealt{Lasue_simulation_2019}). Both graphs show a peak at smaller ratios and a long shoulder towards higher ratios, but overall the COSIMA data show higher ratios. }
\label{fig:MIDAS_cluster_COSIMA_cluster}
\end{figure*}

The radius of clusters $r_{\rm{cluster}}$ is calculated by taking half of the mean value between the length in the x-direction (right position of the right-most particles in a cluster – left position of the left-most particle in a cluster) and length in the y-direction (upwards position of the uppermost particles in the cluster - downwards position of the lowermost particles in the cluster) on the assumption that clusters are circular/spherical shapes. Furthermore, the height of the cluster $h_{\rm{cluster}}$ is chosen as the highest height of the particles within the cluster. Calculated aspect ratios (i.e., $\frac{h_{\rm{\,cluster}}}{\sqrt{\,\pi\, r_{\rm{\,cluster}}^2}}$) and their density plots of all MIDAS clusters and of COSIMA particles \citep{Lasue_simulation_2019} can be found in Table~\ref{table:MIDAS_Cluster_aspect_ratio} and Fig.~\ref{fig:MIDAS_cluster_COSIMA_cluster}.\newline

\begin{table}[]
\caption{The mean value and standard deviation of calculated aspect ratios of all MIDAS clusters} 
\label{table:MIDAS_Cluster_aspect_ratio}
\renewcommand{\arraystretch}{1.3}
\centering\begin{tabular}{cc}
\toprule
\multirow{2}{*}{\textbf{MIDAS Cluster}}                  & \multirow{2}{*}{\textbf{Aspect ratio $\frac{h_{\rm{\,cluster}}}{\sqrt{\pi\, r_{\rm{\,cluster}}^2}}$}} \\
& \\ \hline
Single cluster                & 0.38 $\pm$ 0.25\\
Footprint cluster                & 0.13 $\pm$ 0.17\\
Pyramid cluster                &  0.14 $\pm$ 0.05\\
\hline
Total             & 0.22 $\pm$ 0.16  \\                  
\bottomrule
\end{tabular}
\end{table}

\noindent\textbf{MIDAS single cluster} \hspace{2mm} We find that the aspect ratio distributions of COSIMA and MIDAS are not similar although the MIDAS single cluster morphology looks very similar to the COSIMA compact morphology (i.e., particles with well-defined boundaries). The values of the aspect ratio of MIDAS single and COSIMA compact morphologies are 0.38 $\pm$ 0.25 (see Table \ref{table:MIDAS_Cluster_aspect_ratio}) and 0.73 $\pm$ 0.24 (\citealp{langevin_typology_2016}), respectively. This discrepancy is basically definition-dependent that is COSIMA and MIDAS classification used different definitions for compact and single morphology. The COSIMA compact particles have, by definition, a high aspect ratio and present only well-defined boundaries without smaller satellite particles and with an apparent height above the collecting plane of the same order of magnitude as their x and y dimensions \citep{langevin_typology_2016}. However, we find that MIDAS cluster analysis shows that several high aspect ratio MIDAS particles are contained in MIDAS footprint and even MIDAS pyramid clusters (see Sect.~\ref{result: Relation_shape_descriptors_ohter_parameters}). Thus, we suggest that contrary to the COSIMA classification high aspect ratio particles should not be treated as a unique cluster class. Furthermore, the MIDAS classification strictly follows the definition from \citet{Ellerbroek_labstudy_2017}, resulting in only particles with clearly defined central components or no scattered fragments being labeled as single (see Sect.~\ref{method: MIDAS cluster classification}).

We also find a slight deficit of high aspect ratio particles in the MIDAS Single morphology collection compared to COSIMA's compact collection (see  Fig.~\ref{fig:MIDAS_cluster_COSIMA_cluster}). There is no apparent reason for this behavior at the present state of the work, thus this may be an effect due to the low statistics of COSIMA and MIDAS (167 COSIMA clusters of over $\sim$ 30,000 COSIMA detections were analyzed, compared with 196 MIDAS clusters). Consequently, it would be of great interest to include the rest of the COSIMA data set, possibly even re-classified following the \citet{Ellerbroek_labstudy_2017} definition, and to check how the density plot of COSIMA is changed.\newline

\noindent\textbf{MIDAS footprint cluster} \hspace{2mm} According to the aspect ratio calculation, COSIMA shattered clusters and MIDAS footprints show similar behavior, for example, a strong overlap to other cluster types and a wide distribution of ratios covering the smallest up to values of $\sim$ 0.5. The mean ratio of the COSIMA shattered cluster and MIDAS footprints show similar behavior as well (i.e., 0.22 $\pm$ 0.10; \citealp{langevin_typology_2016}) for COSIMA shattered cluster (see Fig.~\ref{fig:MIDAS_cluster_COSIMA_cluster}) and 0.13 $\pm$ 0.17 for MIDAS footprints (see Table~\ref{table:MIDAS_Cluster_aspect_ratio}). In particular, images of COSIMA shattered clusters show a depression in the central part of the deposit, similar to those found in scans of MIDAS footprint clusters. Thus, we may conclude that the closest resemblance to the MIDAS footprints can be found in COSIMA shattered clusters, although the morphology of MIDAS footprints does not have a direct counterpart in the COSIMA data set.\newline

\noindent\textbf{MIDAS pyramid cluster} \hspace{2mm} MIDAS pyramids may show similar behavior to COSIMA rubble piles (i.e., the value of the aspect ratio is 0.37 $\pm$ 0.13; \citealp{langevin_typology_2016}) and glued clusters (i.e., the value of aspect ratio is 0.27$\pm$ 0.09; \citealp{langevin_typology_2016}), with the difference that MIDAS pyramids show a lower aspect ratio (i.e., the value of aspect ratio is 0.14 $\pm$ 0.05; see Table \ref{table:MIDAS_Cluster_aspect_ratio}). This can be explained by the fact that MIDAS had a 10-micrometer absolute height detection limit, and in most practical cases the limit is less than 5 micrometers. If a pyramidal cluster has a diameter over 10 micrometers, a ratio 1 pyramidal cluster could not be detectable by MIDAS. Therefore, pyramids are biased towards flattened deposits resulting in artificially low ratios. We note that this is a limitation common to all AFMs of current in-situ planetary exploration instruments, more so those capable of higher resolutions. For example, AFM investigation of Martian soil FAMARS was able to provide data on the parameters including particle heights up to 6 - 12 micrometers. To address this limitation, future developments in AFM technology or the use of complementary imaging techniques are required to obtain a more comprehensive analysis of particles with diverse surface features, including those exceeding the current height limitations.

Furthermore, we find that the class of MIDAS pyramids might contain some borderline cases due to the complex appearance of clusters (except for the clearly detectable ones with compact morphology). The boundary between MIDAS footprints and pyramids is fluid. As a result, the MIDAS pyramid class contains COSIMA's glued clusters and rubble piles. We note that glued clusters may be the outcome of particles with slightly different material composition \citep{langevin_typology_2016}. Furthermore, a slight angle during deposition might also help to create slightly asymmetric features such as those found for glued clusters or rubble piles. \newline

\noindent According to the results from the comparison of MIDAS parent particle structures to those detected by COSIMA and laboratory experiments, we find that MIDAS morphologies and aspect ratio show both similarities and differences to those found by COSIMA. Classification differences can be understood when taking into account the difference in resolution and imaging technique. \citet{Ellerbroek_labstudy_2017} showed that the single morphology may contain the particles to be least altered, which implies particles found in MIDAS single cluster morphology can also be the candidates for least altered particles. Furthermore, using the similarities between MIDAS clusters and COSIMA clusters as well as the aspect ratio calculation of each cluster, we suggest that the MIDAS single clusters are the least altered among MIDAS morphologies. This conclusion shows agreement with the results of laboratory experiments \citet{Ellerbroek_labstudy_2017} and the observations of COSIMA.

\subsection{Pristineness descriptor}\label{pristineness_descriptor}

Current gas flow models in experiments and simulations (e.g., \citealp{Laddha2023}) have an indication that cometary activity can be strongly associated with the cometary surface and microphysical properties (e.g., structures). Thus, less altered and thus more pristine shape descriptors that are used for particle classification of dust in space science will yield valuable clues on the effect of more realistic grain shape on the gas flow in the cometary environment.

For this purpose, we now combine shape descriptors and knowledge about seemingly pristine morphologies to determine their level of alteration, ultimately deriving a final pristineness descriptor based on the previous findings on the deposition of dust on MIDAS targets by the identification of dust clusters and their morphology. Thus, the aim of the present section is to identify the particles that have undergone the least amount of alteration and provide a comprehensive description of their properties.

\subsubsection{Development of MIDAS particle pristineness descriptors}\label{Development of MIDAS particle pristineness descriptors}

\begin{figure}
\includegraphics[width=9cm, height= 9cm]{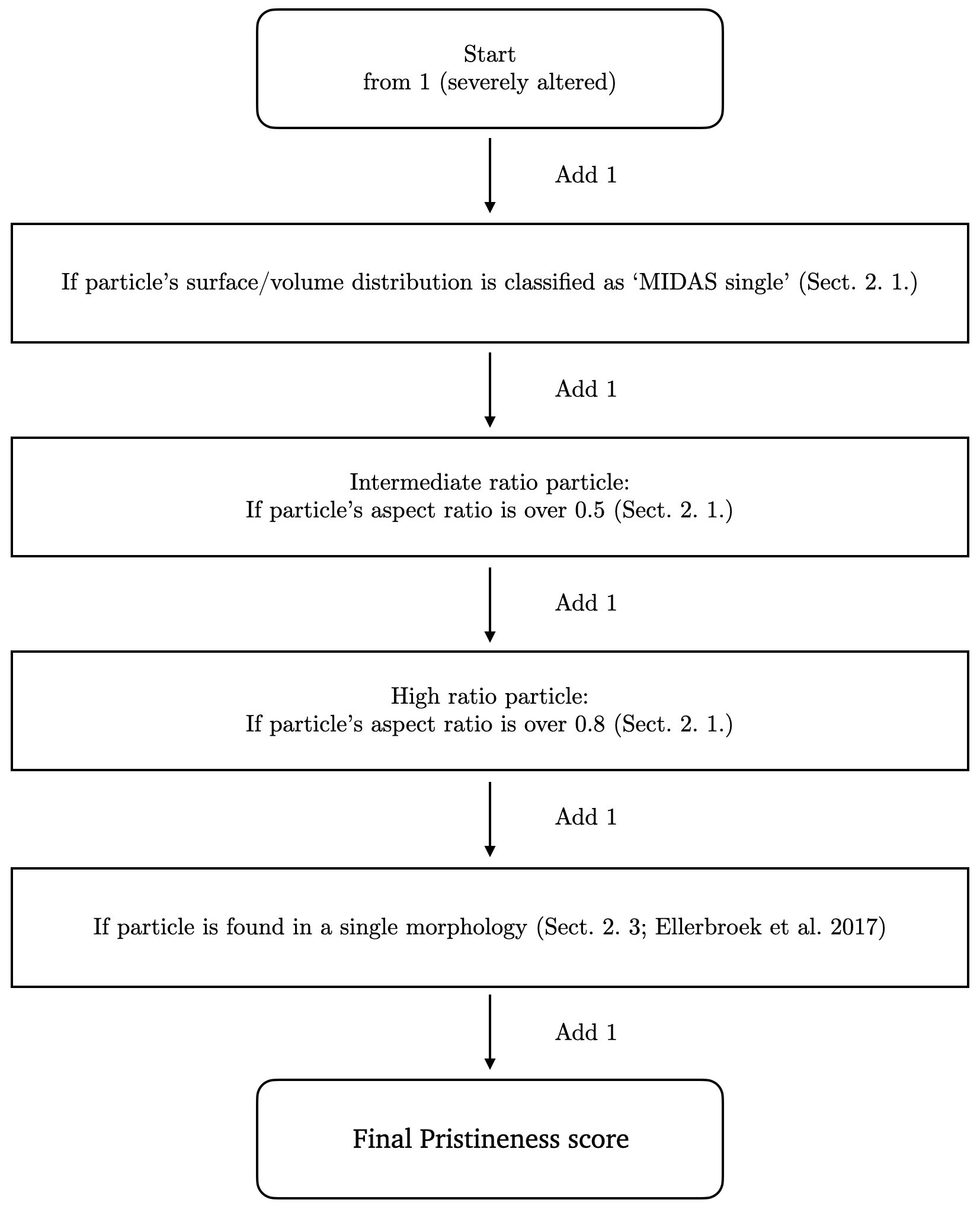}
\caption{Flow chart on how to calculate dust particle pristineness.}
\label{fig:scaling_system_of_pristineness}
\end{figure}

We calculate a final cumulative pristineness score by weighting one point if an individual particle satisfies each parameter that indicates pristineness that is if the particle surface and volume distribution of MIDAS particle classified as ‘MIDAS Single' (see Sect.~\ref{method: MIDAS_particle_shape descriptors}), a particle shows a moderately high aspect ratio (> 0.5; motivated by $\sim$ mean value of aspect ratio of whole MIDAS particles; see Sect.~\ref{result: MIDAS_particle_shape_descriptor}), a particle shows a higher aspect ratio (> 0.8; see Sect.~\ref{result: MIDAS_particle_shape_descriptor}), or if a particle is found in the MIDAS single cluster (i.e., SPC; see Sect.~\ref{result: MIDAS_Morphology_COSIMA}; \citealp{Ellerbroek_labstudy_2017}). Consequently, the final pristineness score starts from 1 (‘severely altered’), 2 (‘altered’), 3 (‘possibly altered’), 4 (‘moderately pristine’), and 5 (‘highly pristine’). The flow chart on how to calculate dust particle pristineness is described in Fig.~ \ref{fig:scaling_system_of_pristineness}. This approach allows us to assess and describe the properties of the particles that have experienced minimal alteration. 

Based on the result from pristineness calculator, we create maps combining dust clustering (\citealp{Kim_Mannel_MIDAS_catalog}) and the pristineness score of MIDAS particles, which can be found in Appendix \ref{appendix_pristineness_maps}. 

\subsubsection{Statistics of the pristineness descriptor}\label{Statistics of pristineness descriptor}

\begin{figure}
\includegraphics[width=9cm, height= 7.1cm]{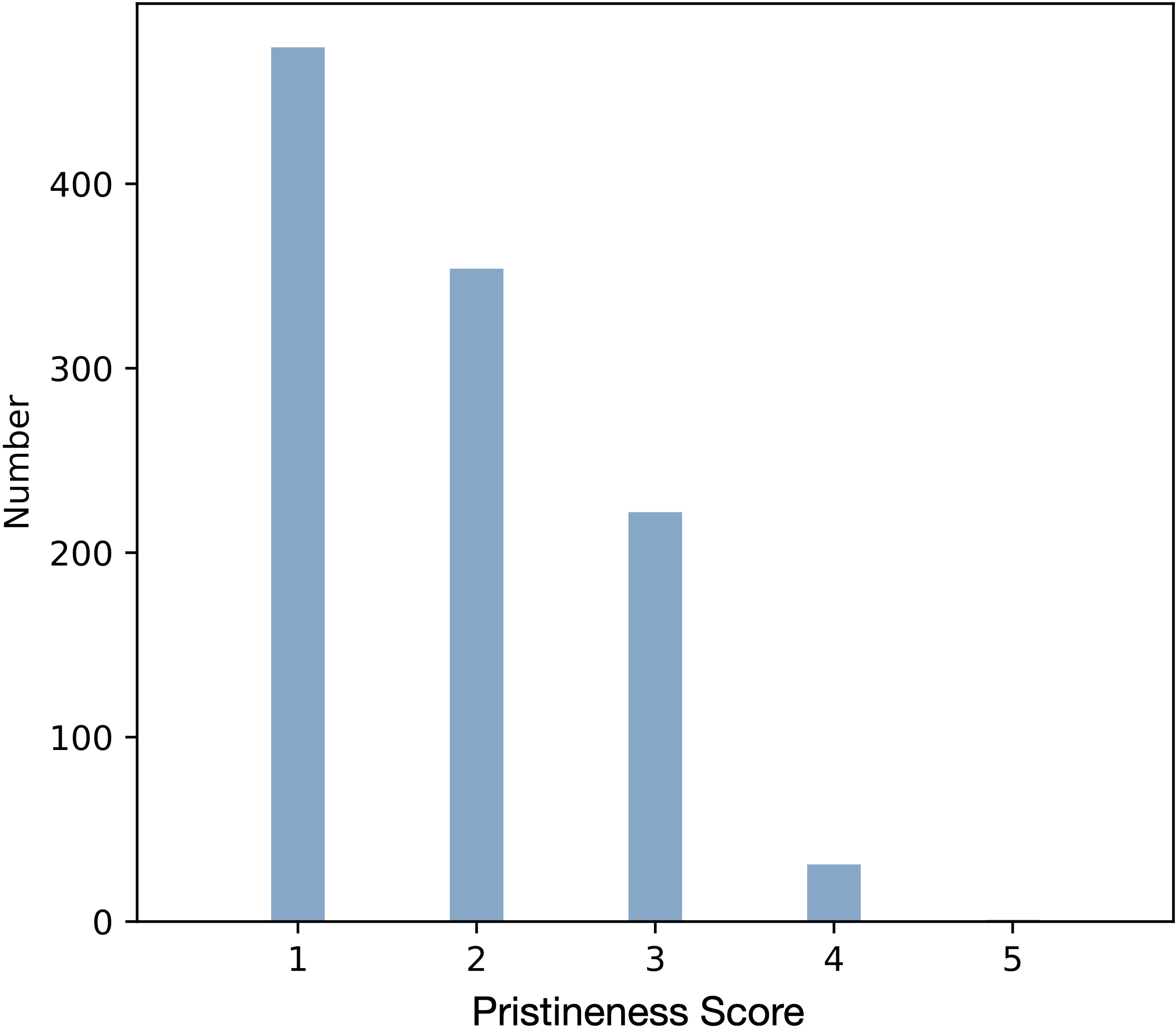}
\caption{Distribution of the final pristineness score of MIDAS particles, where 1 stand for ‘severely altered’ and 5 for ‘highly pristine’. Only 1 particle is found to be highly pristine and 29 candidates are rated with 4 ‘moderately pristine’.}
\label{fig:pristineness_score_histogram}
\end{figure}

The final pristineness scores and the related histogram can be found in Fig.~\ref{fig:pristineness_score_histogram}. 

We find that out of 1082 investigated particles, there is only one particle rated as highly pristine with a score of ‘5'. This result is mainly because particles with very high aspect ratios are predominantly found in MIDAS footprints and pyramid clusters, rather than in MIDAS single clusters (SPCs: single-particle-cluster; see Sect. \ref{result: Relation_shape_descriptors_ohter_parameters}). Additionally, several particles with intermediate aspect ratios were detected as SPCs. The findings from Sect.~\ref{result: MIDAS_particle_shape_descriptor}, where flattened sub-units were either compressed upon impact or turned into the target direction on impact despite the parent particle not being fragmented, support that those SPCs with low aspect ratios have likely undergone some alteration, such as flattening due to the impact. This suggests that SPCs could be either subunits of the MIDAS footprint cluster or particles that have remained pristine, which differs from the expectations proposed by \citet{Ellerbroek_labstudy_2017}, where SPCs were presumed to be pristine. However, we note that the SPCs in \citet{Ellerbroek_labstudy_2017} are at the edge of their resolution capability, and thus an alteration such as flattening is likely not identified. Based on the current data analysis, we conclude that SPCs are susceptible to flattening, including even high degrees of alteration. 

Furthermore, we find that there are 29 particles rated as ‘4' in terms of pristineness calculation and thus 30 particles in total as the most pristine particles (particles rated ‘4' or ‘5'). Interestingly, no single particle rated ‘4' or ‘5' is found on target 13, where particles were collected at nominal activity with slower speed from smooth and rough terrains, and target 15, where particles were collected at nominal activity with faster speed from smooth terrain (see Sect. \ref{method: MIDAS_AFM}). Most importantly, we find that 11 ‘MIDAS Pile’ particles were included in moderately pristine particles (pristineness rated ‘4'). Given that the appearance of MIDAS Pile particles is expected to result from the target impact, we thus exclude 11 ‘MIDAS Pile’ particles from the final list of most pristine particles (particle rated ‘4' and ‘5') and thus reasonably rescale the pristineness score (i.e., 19 most pristine MIDAS particles in total). On the other hand, we find that there are 474 particles rated ‘1', indicating that nearly half of the MIDAS particles have undergone significant alteration due to impact, while 222 and 354 particles were rated ‘3' and ‘2', respectively.

Overall, our findings reveal the presence of only one highly pristine particle, a small number of moderately pristine particles, and a substantial proportion of altered particles within the MIDAS dataset. These results emphasize that dust alteration is an unavoidable process, even at the relatively slow collection speeds achieved by the Rosetta spacecraft.

\subsubsection{Characteristics of pristine particles}\label{Characteristics of pristine particles}

\begin{figure}
\centering
\includegraphics[width=9cm, height=7cm]{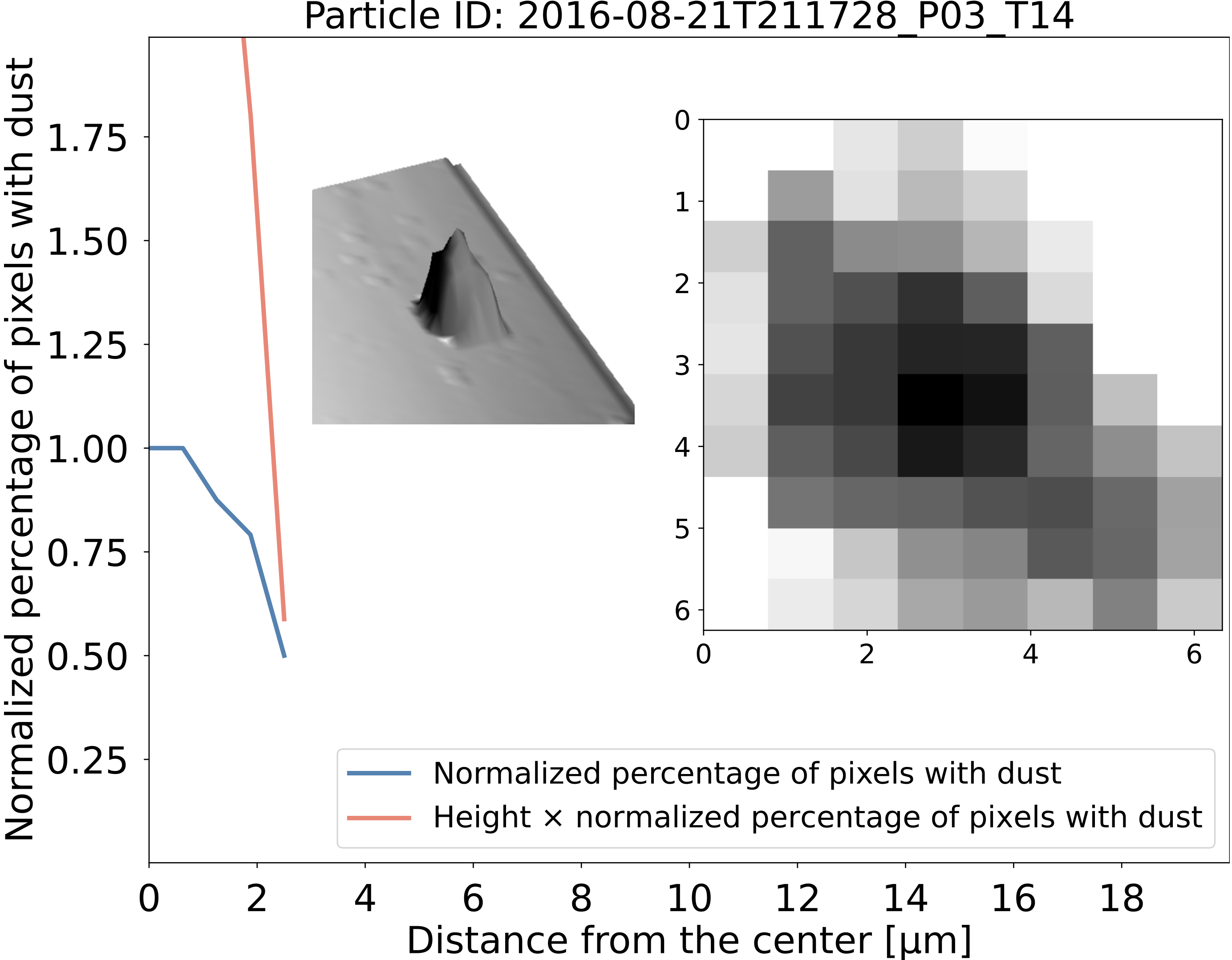}
\caption{Appearance, surface and volume distribution, and characteristics of the most pristine particle (particle rated 5, Particle ID: 2016-08-21T211728\_P03\_T14).}
\label{fig:particle_pristineness_5}
\end{figure}

\begin{figure*}
\centering

\includegraphics[width=\textwidth, height=15cm]{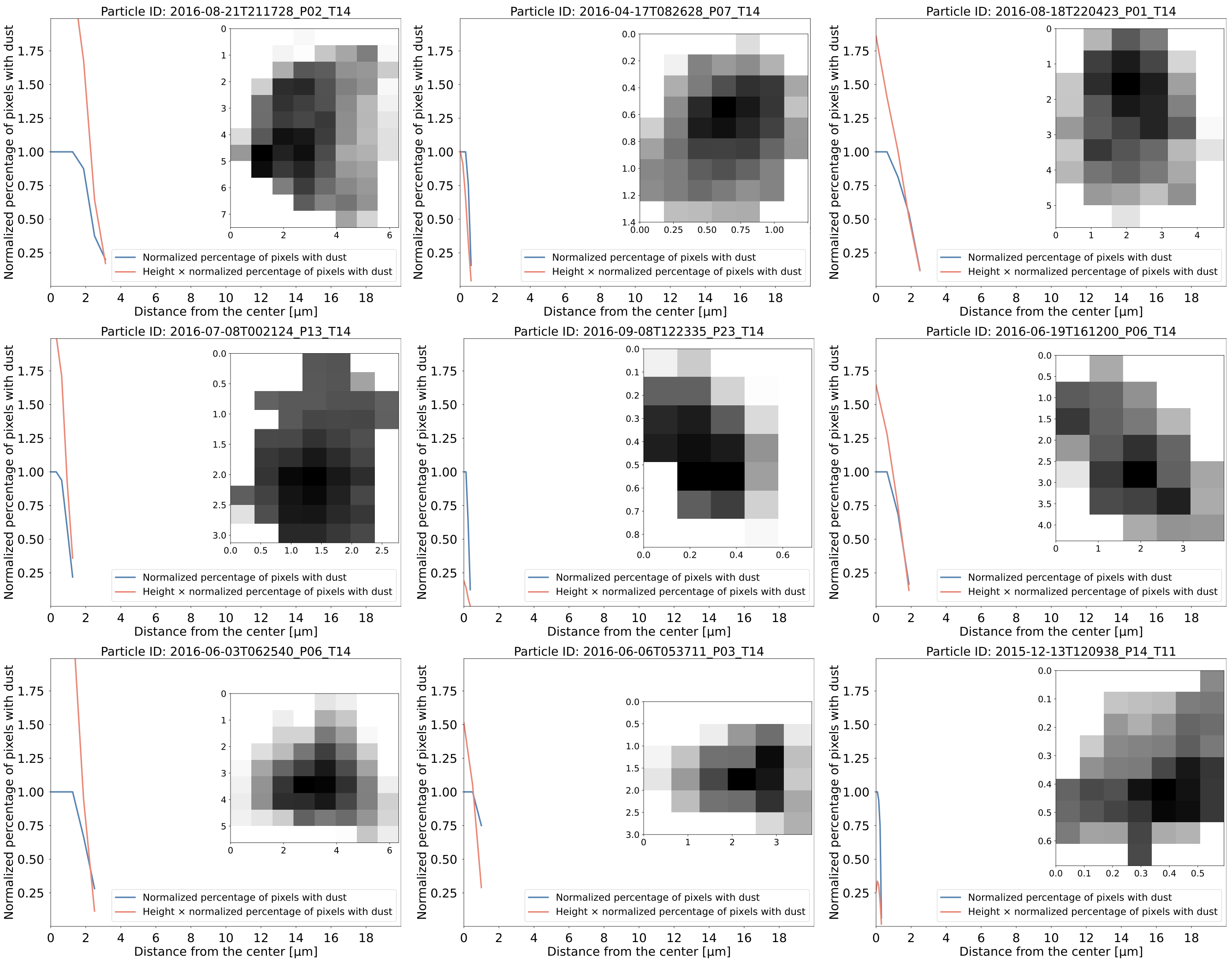}
\caption{Selected examples of appearance, surface and volume distribution, and their characteristics of particles rated 4, i.e. moderately pristine, on the pristineness scaling.}
\label{fig:particle_pristineness_4}
\end{figure*}

A summary of characteristics and shape descriptors of the 19 particles rated ‘4' or ‘5', sorted by the number of pixels on the pristineness descriptor, is given in Appendix \ref{appendix_pristineness_of_MIDAS_particles}. Furthermore, the 3D appearance, surface and volume distribution, and characteristics of the selected examples of most pristine particles (particle rated ‘4' and ‘5') can be found in Figs.~\ref{fig:particle_pristineness_5} and  \ref{fig:particle_pristineness_4}.

We find that $\sim$ 40 \% of most pristine MIDAS particles (7 of 19 particles) that are rated 4 and 5 are found in MIDAS single clusters (i.e., SPCs). This implies again that moderately pristine particles can be found even in footprints or pyramids (resembling COSIMA glued cluster, shattered cluster, or large rubble pile morphology), which is thought to be initial sub-clusters of the particles that did not get modified by the collection. We also find these candidates for pristine particles are not that elongated, showing the mean value of elongation $\sim$ 1.8. This value is close to the one from observations and simulation of ISM grains \citep{Hiltner_polarization_1949, Voshchinnikov_elongated_ISM_1999}, compared to the average elongation of all MIDAS particles, showing the mean value of elongation $\sim$ 3.93 (see Table~\ref{table: MIDAS_particle_shape_descriptor} and Sect.~\ref{result: MIDAS_particle_shape_descriptor}). This finding may indicate that a relatively higher value of elongation of all MIDAS particles is possibly created by impact rather than subunits turning in one direction when deposited (i.e., particles have a natural elongation and preferentially stick with their longer side on the targets; \citealp{bentley_morphology_2016}). 


	\section {Summary and outlook}\label{sec:summary and outlooks}
	
The MIDAS AFM on board the Rosetta comet orbiter investigated the 3D topography of about 3523 cometary dust particles with sizes between a few hundreds of nm to tens of $\mu$m based on images with resolutions down to a few nanometers. Based on the updated version of the MIDAS particle catalog (\citealp{Kim_Mannel_MIDAS_catalog}), we analyzed the shapes of the cometary dust particles. In particular, we developed MIDAS particle shape descriptors such as aspect ratio, elongation, circularity, convexity, and particle surface and volume distribution. Next, we compared the structure of the MIDAS dust particles and clusters (i.e. fragments of a parent particle that broke up upon collection) to those found in laboratory experiments \citep{Ellerbroek_labstudy_2017} and by COSIMA onboard Rosetta \citep{Lasue_simulation_2019}. Finally, we calculated a pristineness score for the MIDAS particles and determined the most pristine particles and their properties. Based on the particle shape descriptors and pristineness calculator, we created dust maps combining clustering and the derived parameters, as well as the pristineness score. We finally established the microphysical properties of pristine cometary materials in the present study, which have drawn on a wide range of shape descriptors. The key results are as follows:

\begin{itemize}
    \item We found that the mean value of the aspect ratio of all the MIDAS particles is found around 0.53 $\pm$ 0.40, which is implying the degree of flattening and is smaller than expected since one would expect the value of 1 for unflattened particles. We also found that the mean value of all MIDAS particles’ elongation is 3.93 $\pm$ 2.31, meaning that their shapes highly deviate from a sphere. Additionally, we found that the value of the circularity of MIDAS particles is 0.58 $\pm$ 0.19, meaning that MIDAS particles deviate from being circular. Finally, we found that the value of the convexity of MIDAS particles showed  0.47 $\pm$ 0.19, meaning that most of the convex hull area of MIDAS particles is occupied by their perimeter concavities.

    \item We found no clear trend between shape descriptors and particle size, only weak correlations between aspect ratio and size, for instance, larger particles within low standard deviation (i.e., 0.26 $\pm$ 0.09) showing a tendency to lower aspect ratios and small particles being the only candidates with extremely large aspect ratios. However, small particles show a large variance as well (i.e., 0.49 $\pm$ 0.41). We found that large particles show relatively smaller elongation values, implying again the degree of flattening. Large particles show smaller circularity and convexity, which may possibly be explained by a hierarchical growth process.
    
    \item We found that there is no clear trend between shape descriptors and the properties related to the target (and thus dust activities/cometary source region). These results imply that various dust activities depending on the cometary source region and different velocities of the MIDAS dust collection do not play a role in the structural alteration of the particles on the micrometer scale.
    
    \item We found that cometary dust particles of different cluster morphology (i.e., MIDAS single, footprints, and pyramid clusters) are rather homogeneous in properties. We found that the aspect ratio distributions of the particles are very similar and independent of the cluster type, implying that the sub-units of different cluster types are very similar in their shape and composition. 
    
    \item We made a quantitative comparison between the data of COSIMA and MIDAS clusters. We found that MIDAS clusters and their aspect ratio distribution show both similarities and differences to those found by COSIMA. Based on the finding and results from \citet{Ellerbroek_labstudy_2017}, we concluded that the MIDAS Single cluster may contain the particles to be least altered. 
    
    \item We calculated a final pristineness score by weighting based on combined shape descriptors and knowledge about seemingly pristine morphologies. We found that there is only 1 particle rated as highly pristine and 18 particles rated as moderately pristine, while 222 and 354 particles are rated as possibly altered and altered. We also found that there are 474 particles rated as severely altered. This finding indicates dust alteration is inevitable even at the slow collection speeds that were available at the Rosetta spacecraft. \newline

\end{itemize}

\noindent The particle catalog (\citealt{Kim_Mannel_MIDAS_catalog}) and shape descriptor in the present study will serve as references for future scientific projects such as the development of more realistic analog materials, and the design of dust particles in simulations for comets and early solar system studies. The database will also be a landmark for any future cometary or interplanetary dust investigation, be it a re-consideration of Stardust data, the analysis of returned material of asteroids, or even - in the far future - of a comet. Gas flow properties are significantly influenced by the shape of the grains. Therefore, adopting a more quantitative approach that correlates with shape descriptors, particularly for realistic (and thus pristine) cometary dust particles used in particle classification, can provide valuable insights into how grain shape affects the characteristics of gas flow through the medium (e.g., packed beds).

There are still possibilities to improve the particle’s structural description, for instance, an investigation of 3D shape descriptors such as roughness calculation (Kim et al. in preparation) can be applied to increase our understanding of the microphysical properties of pristine cometary materials. Furthermore, subunit size investigation could be taken into account to further understand particle growth (e.g., \citealt{Mannel_classification_2019}). Finally, an estimation of particle strength similar to that given in \cite{hornung_assessment_2016} can be studied to verify the microphysical properties of cometary materials. These studies are in preparation and will further increase our knowledge about cometary dust.


\begin{acknowledgements}
This paper is dedicated to the memory of MIDAS Co-I Prof. Anny-Chantal Levasseur-Regourd, a true Renaissance person. The authors gratefully thank our referee and editor for the constructive comments and recommendations which definitely help to improve the readability and quality of the paper. M. Kim and T. Mannel acknowledge funding by the ESA project "Primitiveness of cometary dust collected by MIDAS on-board Rosetta" (Contract No. 4000129476). M. Kim also acknowledges the funding from the Royal Society. J. Lasue acknowledges funding by CNES and the PNP INSU CNRS for working on the Rosetta data. This research was supported by the International Space Science Institute (ISSI) through the ISSI International Team “Characterization of cometary activity of 67P/Churyumov-Gerasimenko comet”.
\end{acknowledgements}

\bibliographystyle{aa} 
\bibliography{AA_2023_47173.bib} 

\begin{appendix}

\section{Shape descriptor of MIDAS particles}\label{appendix: Shape descriptor of MIDAS particles}

The MIDAS particle shape descriptor presented in this paper contains in total of 1082 scans of particles (see Sect.~\ref{result: MIDAS_particle_shape_descriptor}). This table is an excerpt only and the full table of the ultimately chosen 1082 particles is available at the CDS.\newline

\begin{landscape}
\begin{table}
\caption{Summary of characteristics and shape descriptors of all MIDAS particles.}
\renewcommand{\arraystretch}{1.3}
\begin{minipage}{\linewidth}
\scriptsize

\begin{tabular}{llllllllllllll}

\textbf{Particle ID}                & \textbf{Target} & \textbf{Area} & \textbf{Radius} & \textbf{Height} & \textbf{Aspect}      & \textbf{Elongation}      & \textbf{Circularity}  & \textbf{Convexity} & \textbf{PSVD\footnote{Particle surface and volume distribution}}         & \textbf{Cluster} & \textbf{Ellerbroek\footnote{MIDAS Cluster classification according to \citealp{Ellerbroek_labstudy_2017}}} & \textbf{COSIMA\footnote{MIDAS Cluster classification according to the COSIMA classification (e.g., \citealp{langevin_typology_2016})}}            & \textbf{Pristineness} \\

 & & [$\mu{m}^2$] & [$\mu{m}$] & [$\mu{m}$]& \textbf{Ratio}      &      &   &  &  & \textbf{Number} &  &           & \textbf{score} \\ \hline

2014-11-18T032520\_P01\_T11  & 11     & 1.991                & 0.796             & 0.870  & 0.617         & 1.900       & 0.796      & 0.481               & MIDAS Single & 44                & Single     & Compact           & 4                 \\
2014-11-18T131924\_P01\_T11  & 11     & 0.664                & 0.460             & 0.320  & 0.393         & 2.732       & 0.871      & 0.480               & MIDAS Pile   & 8                 & Footprint  & Shattered cluster & 1                 \\
2014-11-18T131924\_P02\_T11  & 11     & 2.757                & 0.937             & 1.190  & 0.717         & 1.653       & 0.838      & 0.461               & MIDAS Single & 8                 & Footprint  & Shattered cluster & 3                 \\
2014-11-18T131924\_P03\_T11  & 11     & 0.357                & 0.337             & 0.110  & 0.184         & 5.963       & 0.739      & 0.500               & MIDAS Pile   & 8                 & Footprint  & Shattered cluster & 1                 \\
2014-11-18T131924\_P04\_T11  & 11     & 0.408                & 0.361             & 0.230  & 0.360         & 3.648       & 0.553      & 0.500               & MIDAS Pile   & 37                & Single     & Compact           & 2                 \\
2014-11-18T131924\_P05\_T11  & 11     & 1.225                & 0.625             & 0.530  & 0.479         & 2.662       & 0.650      & 0.479               & MIDAS Single & 5                 & Pyramid    & Large Rubble pile & 2                 \\
2014-11-18T131924\_P06\_T11  & 11     & 2.195                & 0.836             & 0.760  & 0.513         & 2.539       & 0.839      & 0.420               & MIDAS Pile   & 5                 & Pyramid    & Large Rubble pile & 2                 \\
2014-11-18T131924\_P07\_T11  & 11     & 0.204                & 0.255             & 0.100  & 0.221         & 4.844       & 0.577      & 0.500               & MIDAS Pile   & 8                 & Footprint  & Shattered cluster & 1                 \\
2014-11-18T131924\_P10\_T11  & 11     & 0.306                & 0.312             & 0.110  & 0.199         & 5.908       & 0.577      & 0.500               & MIDAS Pile   & 8                 & Footprint  & Shattered cluster & 1                 \\
2014-11-27T193208\_P02\_T11  & 11     & 0.141                & 0.212             & 0.170  & 0.453         & 2.921       & 0.706      & 0.492               & MIDAS Single & 39                & Single     & Compact           & 3                 \\
2014-11-28T034542\_P02\_T11  & 11     & 0.197                & 0.250             & 0.100  & 0.225         & 5.305       & 0.886      & 0.471               & MIDAS Single & 27                & Single     & Compact           & 3                 \\
2014-11-28T034542\_P03\_T11  & 11     & 0.150                & 0.218             & 0.140  & 0.362         & 3.791       & 0.694      & 0.443               & MIDAS Single & 27                & Single     & Compact           & 3                 \\
2014-11-29T222402\_P06\_T11  & 11     & 0.313                & 0.315             & 0.440  & 0.787         & 1.967       & 0.568      & 0.460               & MIDAS Single & 6                 & Footprint  & Compact           & 3                 \\
2014-11-29T222402\_P07\_T11  & 11     & 0.221                & 0.266             & 0.290  & 0.616         & 1.883       & 0.988      & 0.464               & MIDAS Single & 6                 & Footprint  & Compact           & 3                 \\
2014-11-29T222402\_P08\_T11  & 11     & 0.049                & 0.125             & 0.110  & 0.496         & 2.990       & 0.558      & 0.475               & MIDAS Single & 6                 & Footprint  & Compact           & 2                 \\
2014-11-29T222402\_P09\_T11  & 11     & 0.049                & 0.125             & 0.120  & 0.541         & 3.162       & 0.442      & 0.450               & MIDAS Single & 6                 & Footprint  & Compact           & 3                 \\
2014-11-29T222402\_P10\_T11  & 11     & 0.081                & 0.161             & 0.170  & 0.597         & 2.393       & 0.629      & 0.457               & MIDAS Single & 6                 & Footprint  & Compact           & 3                 \\
2014-11-29T222402\_P11\_T11  & 11     & 0.256                & 0.285             & 0.280  & 0.553         & 2.293       & 0.811      & 0.464               & MIDAS Single & 6                 & Footprint  & Compact           & 3                 \\
2014-11-30T141103\_P02\_T11  & 11     & 0.059                & 0.137             & 0.120  & 0.494         & 2.652       & 0.749      & 0.460               & MIDAS Single & 6                 & Footprint  & Compact           & 2                 \\
2014-12-02T113032\_P02\_T11  & 11     & 0.532                & 0.411             & 0.310  & 0.425         & 2.875       & 0.874      & 0.466               & MIDAS Single & 6                 & Footprint  & Compact           & 2                 \\
2014-12-02T113032\_P03\_T11  & 11     & 0.037                & 0.108             & 0.120  & 0.625         & 2.313       & 0.619      & 0.429               & MIDAS Single & 6                 & Footprint  & Compact           & 3                 \\
2015-01-24T051029\_P07\_T11  & 11     & 0.588                & 0.433             & 0.870  & 1.134         & 3.337       & 0.300      & 0.400               & MIDAS Pile   & 14                & Footprint  & Shattered cluster & 3                 \\
2015-01-24T051029\_P08\_T11  & 11     & 0.706                & 0.474             & 0.190  & 0.226         & 5.752       & 0.382      & 0.455               & MIDAS Pile   & 14                & Footprint  & Shattered cluster & 1                 \\\hline
    \multicolumn{14}{c}{...\footnote{This table is an excerpt only and} the full table is available at the CDS.} \\

\bottomrule
\label{table: MIDAS_particle_shape_descriptor_full_table}
\end{tabular}
\end{minipage}
\end{table}
\end{landscape}

\section{MIDAS shape descriptor maps}\label{appendix_shape_descriptor_maps}

MIDAS dust maps combining shape descriptors that is aspect ratio, elongation, circularity, and convexity, are in general 2D images (e.g., white: lower value, red: intermediate value, and black: higher value) showing their shape descriptors at the respective locations on the individual target. Furthermore, particle clustering with a sophisticated clustering algorithm  (i.e., mean shift method) is described to identify fragments with a common origin (\citealp{Kim_Mannel_MIDAS_catalog}).

\subsection{MIDAS shape descriptor maps - aspect ratio}\label{appendix_shape_descriptor_maps_aspect_ratio}

\begin{figure*}
\centering\includegraphics[width=17cm, height=23cm]{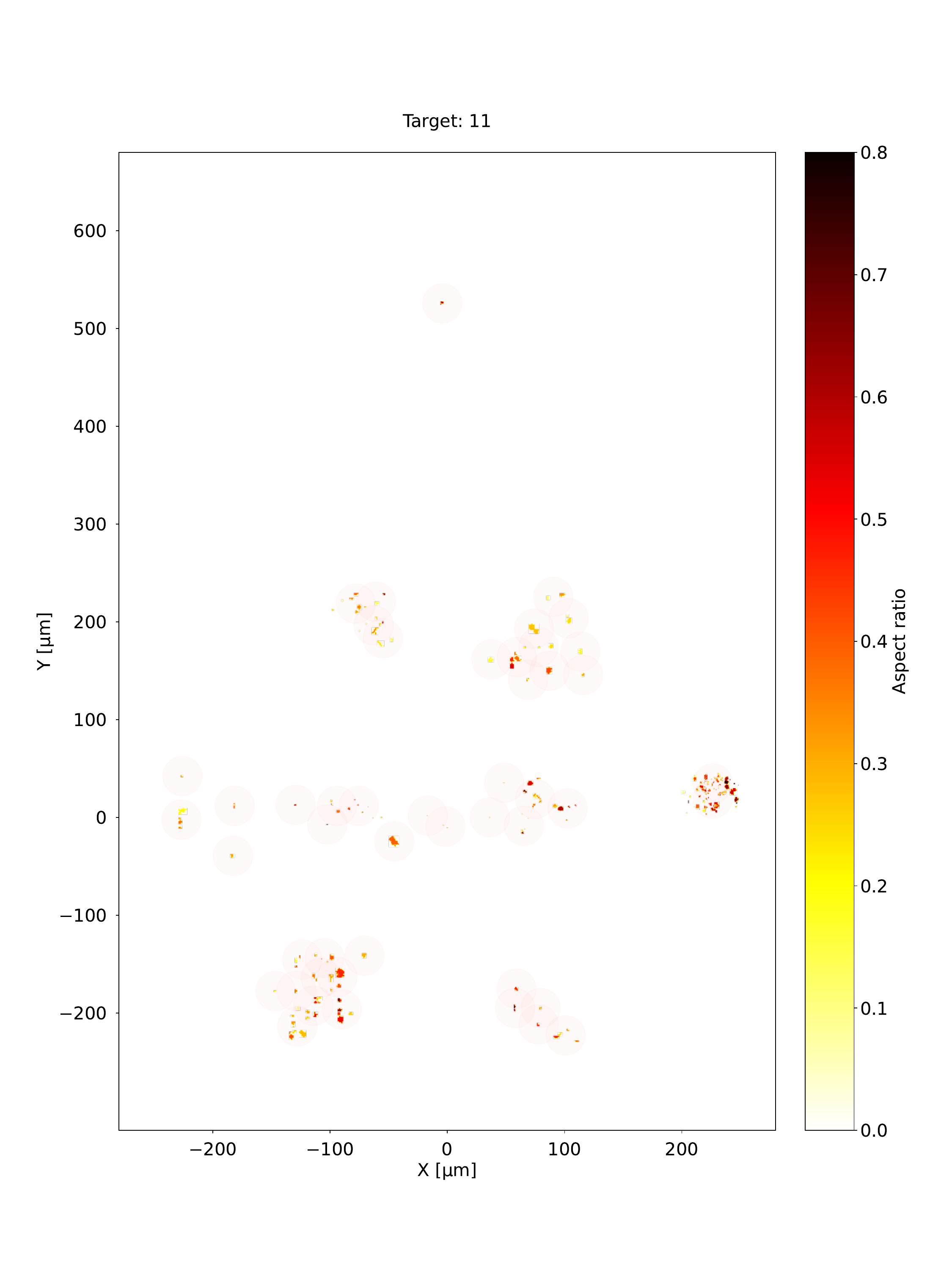}
\caption{Map showing the aspect ratio distribution of dust particles and the particle clustering of target 11. The pink circles represent an approximate size estimation of individual MIDAS clusters.}
\label{fig:2D_dust_clustering_map_10_aspect_ratio}
\end{figure*}

\begin{figure*}
\centering
\includegraphics[width=13.3cm, height=20cm]{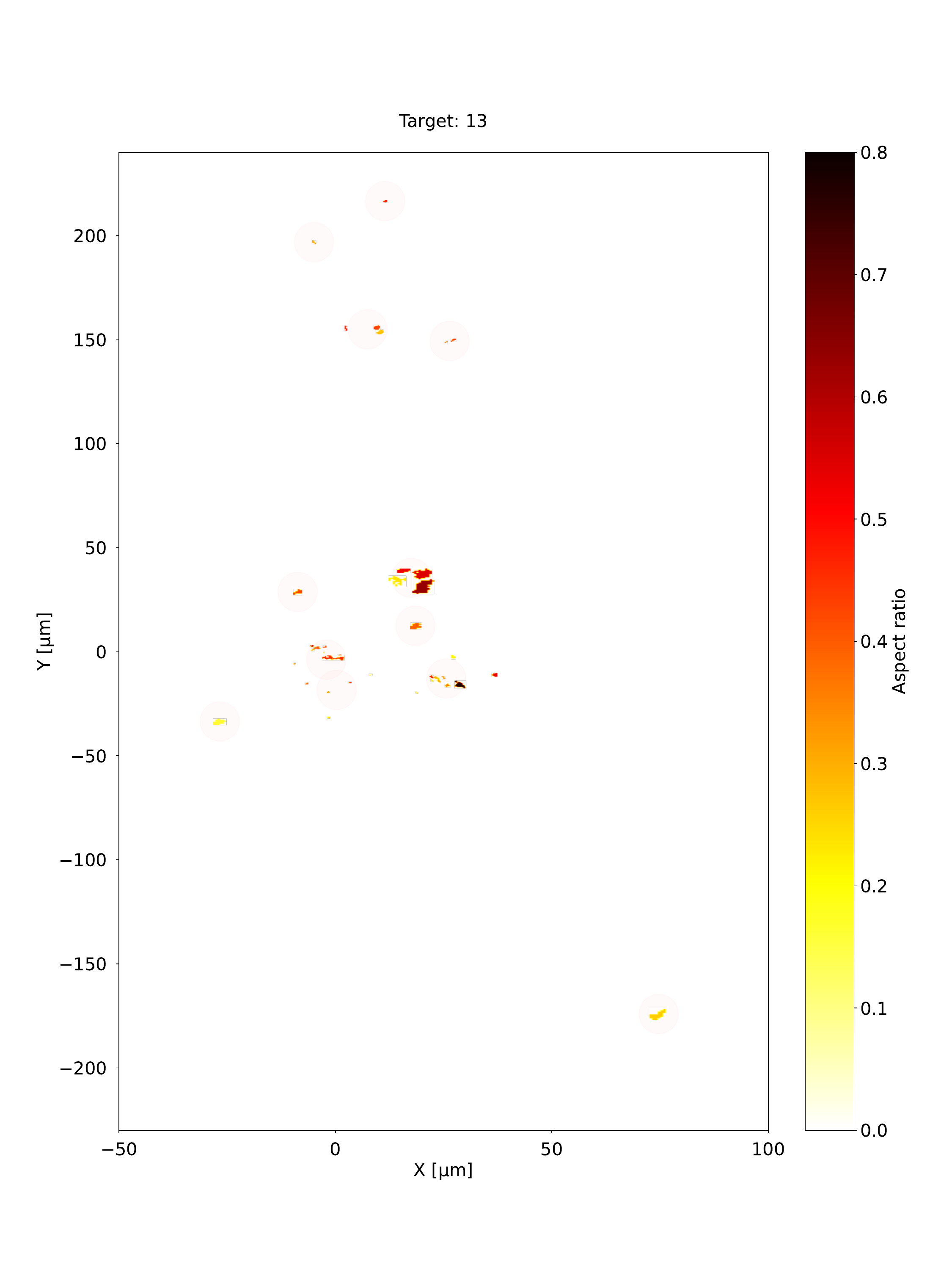}
\caption{Map showing the aspect ratio distribution of dust particles and the particle clustering of target 13. The pink circles represent an approximate size estimation of individual MIDAS clusters.}
\label{fig:2D_dust_clustering_map_12_aspect_ratio}
\end{figure*}

\begin{figure*}
\centering
\includegraphics[width=17cm, height=24cm]{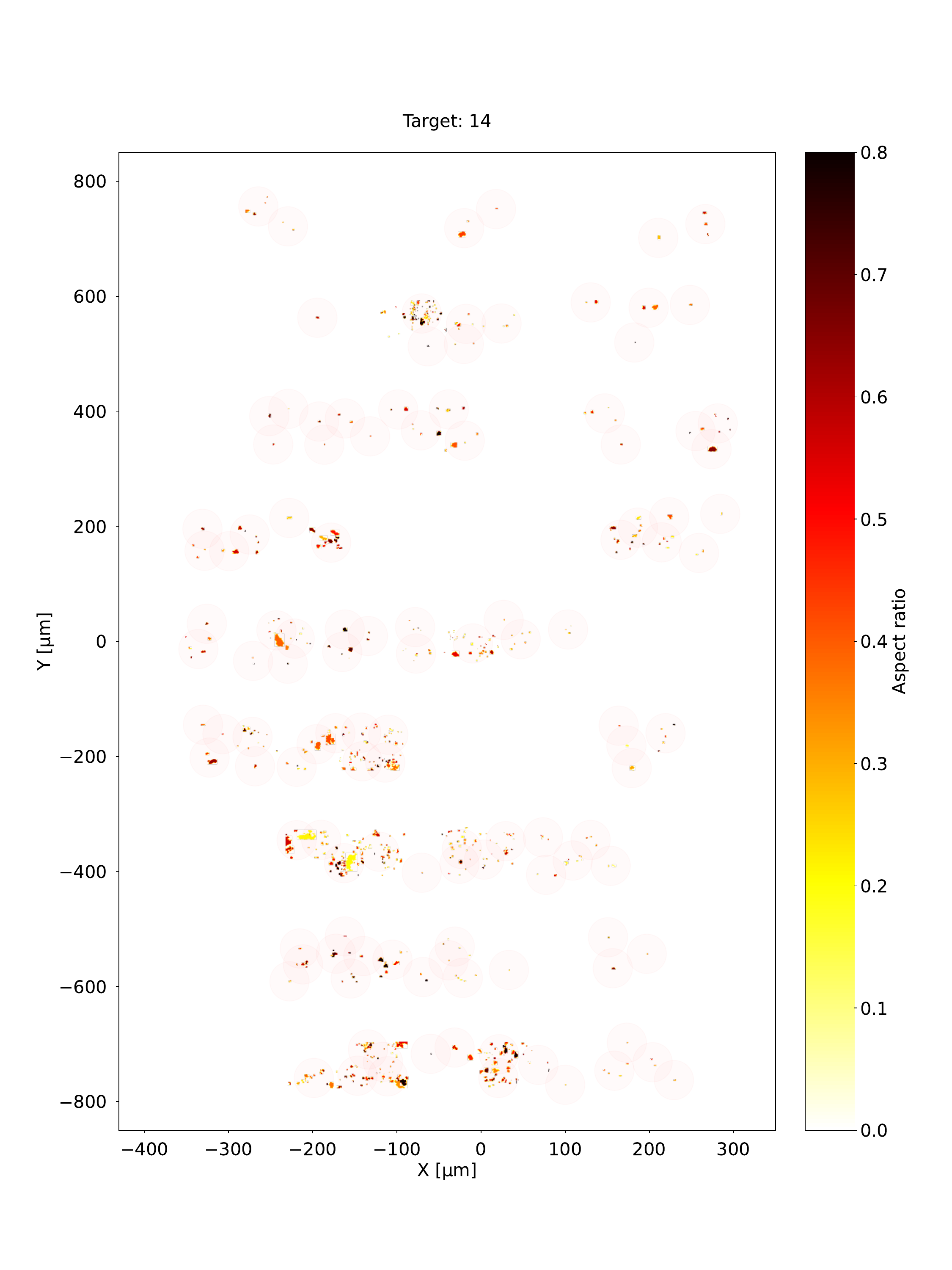}
\caption{Map showing the aspect ratio distribution of dust particles and the particle clustering of target 14. The pink circles represent an approximate size estimation of individual MIDAS clusters.}
\label{fig:2D_dust_clustering_map_13_aspect_ratio}
\end{figure*}

\begin{figure*}
\centering
\includegraphics[width=15cm, height=9.5cm]{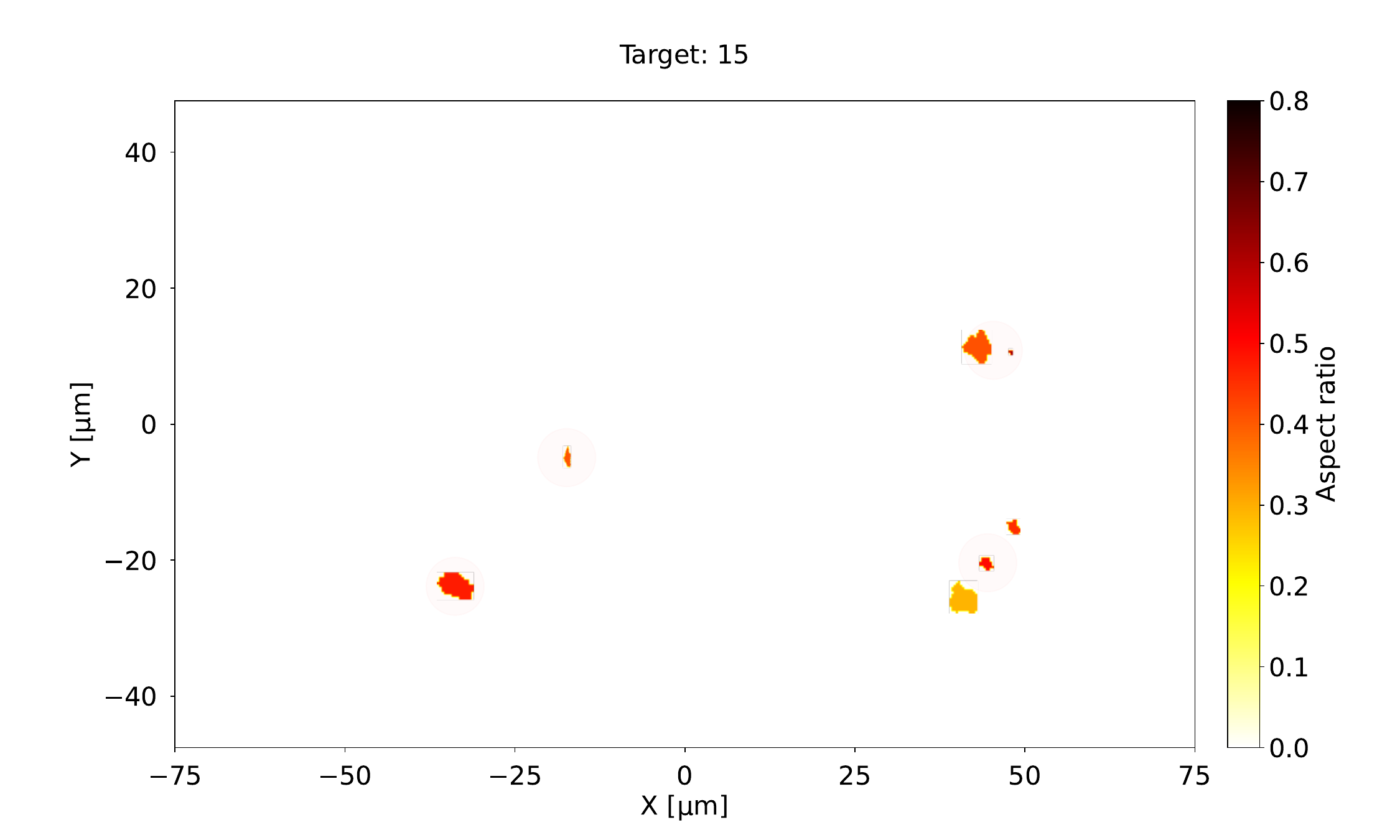}
\caption{Map showing the aspect ratio distribution of dust particles and the particle clustering of target 15. The pink circles represent an approximate size estimation of individual MIDAS clusters.}
\label{fig:2D_dust_clustering_map_14_aspect_ratio}
\end{figure*}

\subsection{MIDAS shape descriptor maps - elongation}\label{appendix_shape_descriptor_maps_elongation}

\begin{figure*}
\centering
\includegraphics[width=17cm, height=23cm]{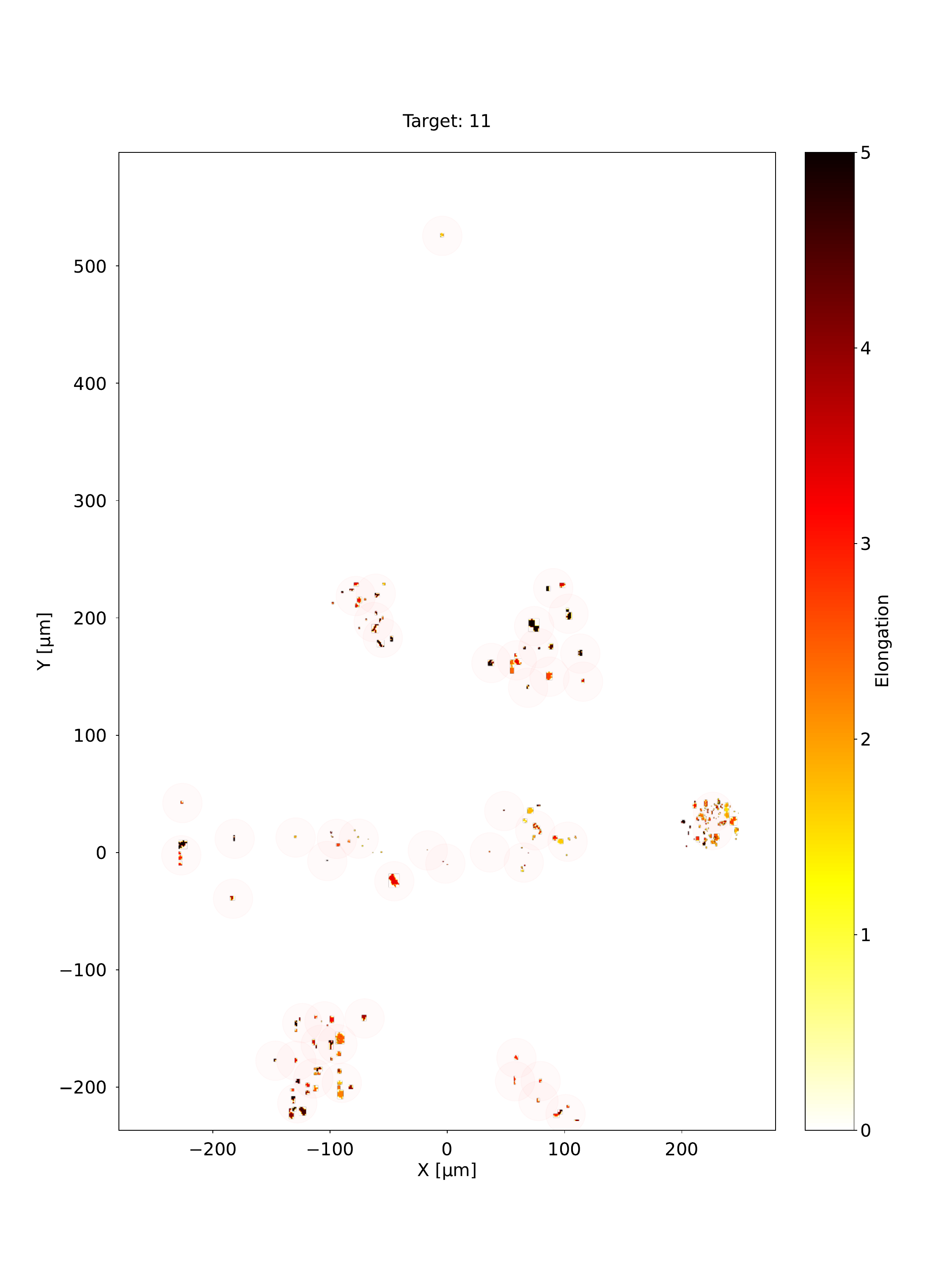}
\caption{Map showing the elongation distribution of dust particles and the particle clustering of target 11. The pink circles represent an approximate size estimation of individual MIDAS clusters.}
\label{fig:2D_dust_clustering_map_10}
\end{figure*}

\begin{figure*}
\centering
\includegraphics[width=13.3cm, height=20cm]{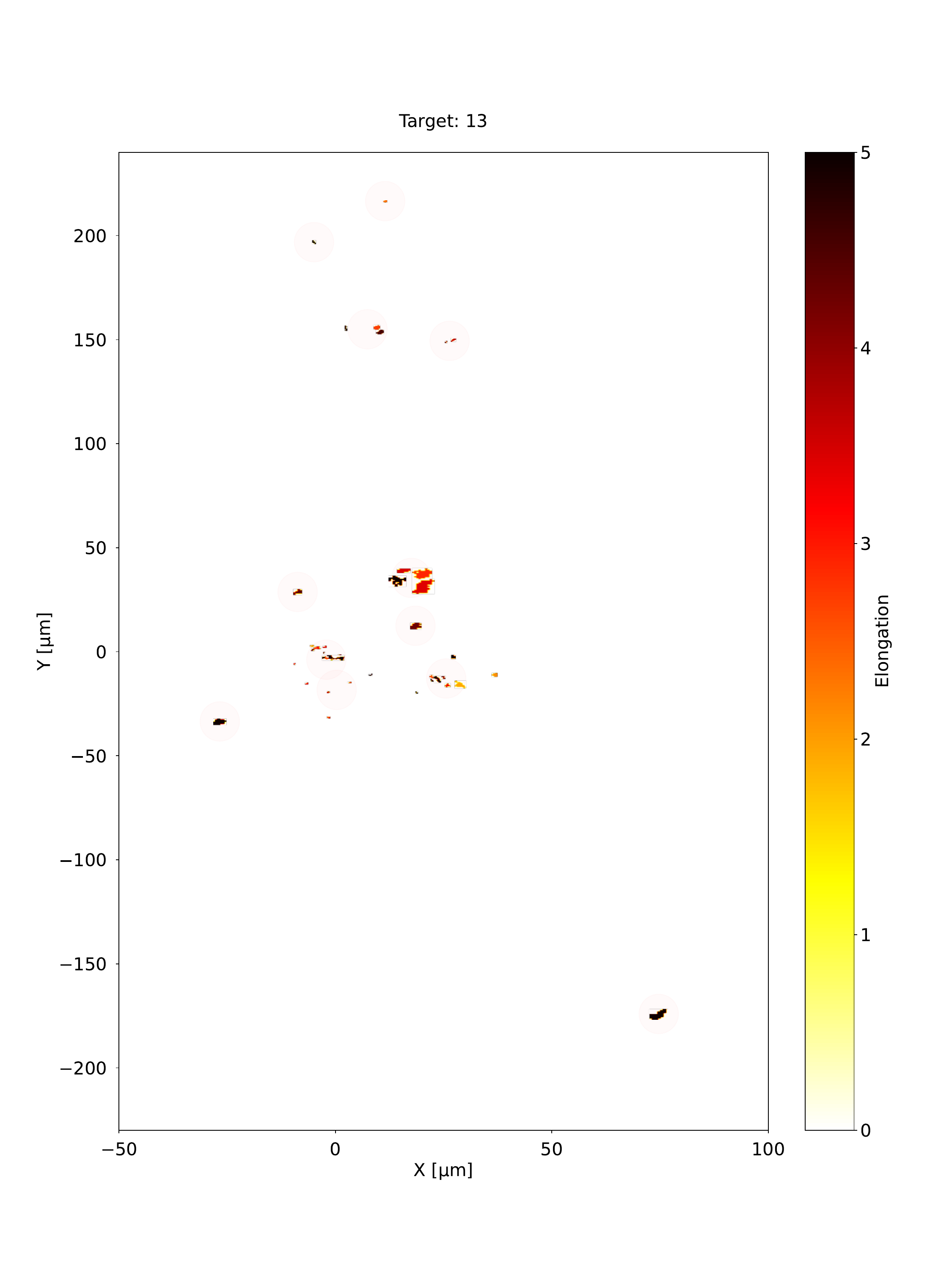}
\caption{Map showing the elongation distribution of dust particles and the particle clustering of target 13. The pink circles represent an approximate size estimation of individual MIDAS clusters.}
\label{fig:2D_dust_clustering_map_12}
\end{figure*}

\begin{figure*}
\centering
\includegraphics[width=17cm, height=24cm]{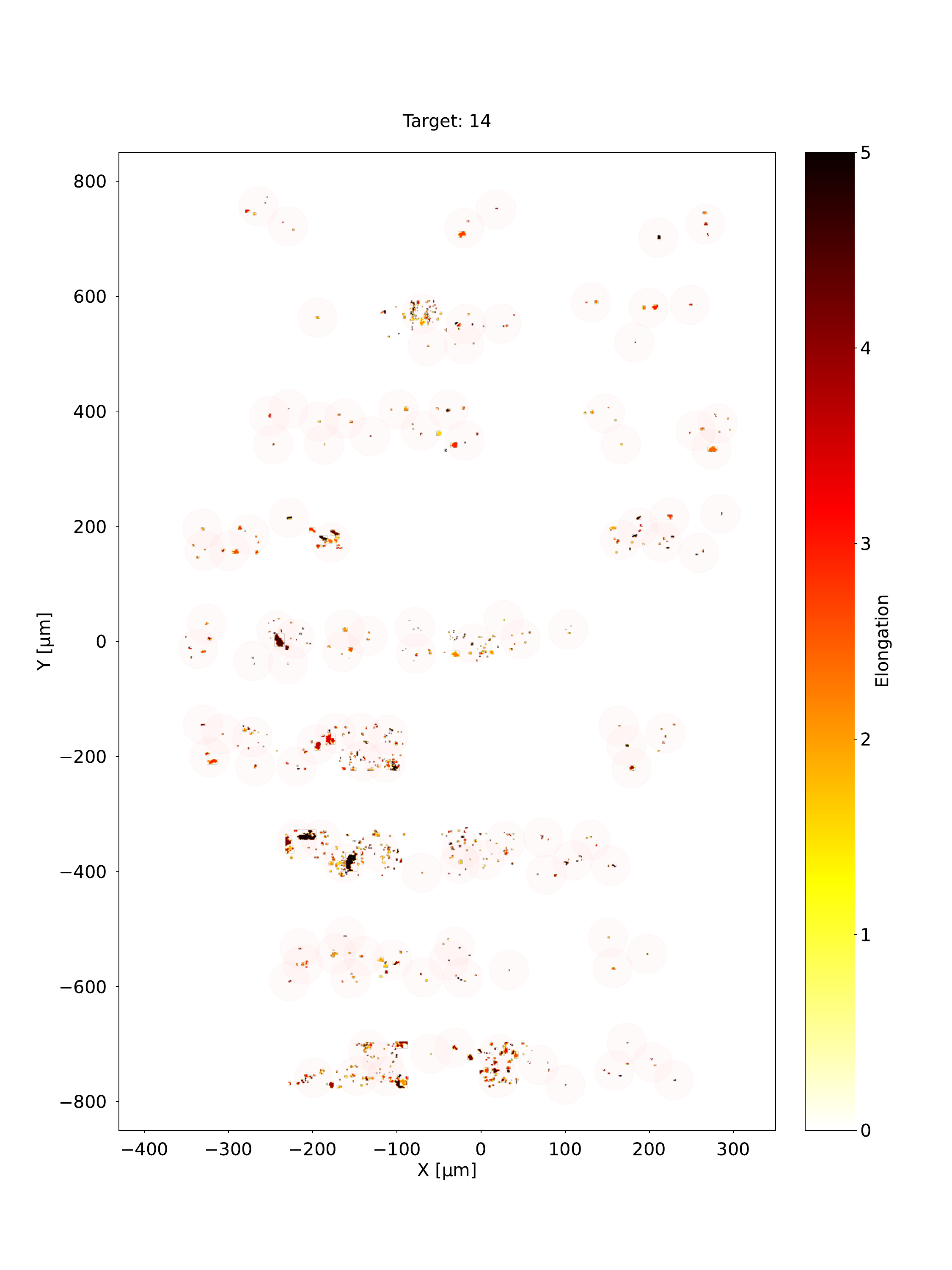}
\caption{Map showing the elongation distribution of dust particles and the particle clustering of target 14. The pink circles represent an approximate size estimation of individual MIDAS clusters.}
\label{fig:2D_dust_clustering_map_13}
\end{figure*}

\begin{figure*}
\centering
\includegraphics[width=15cm, height=9.5cm]{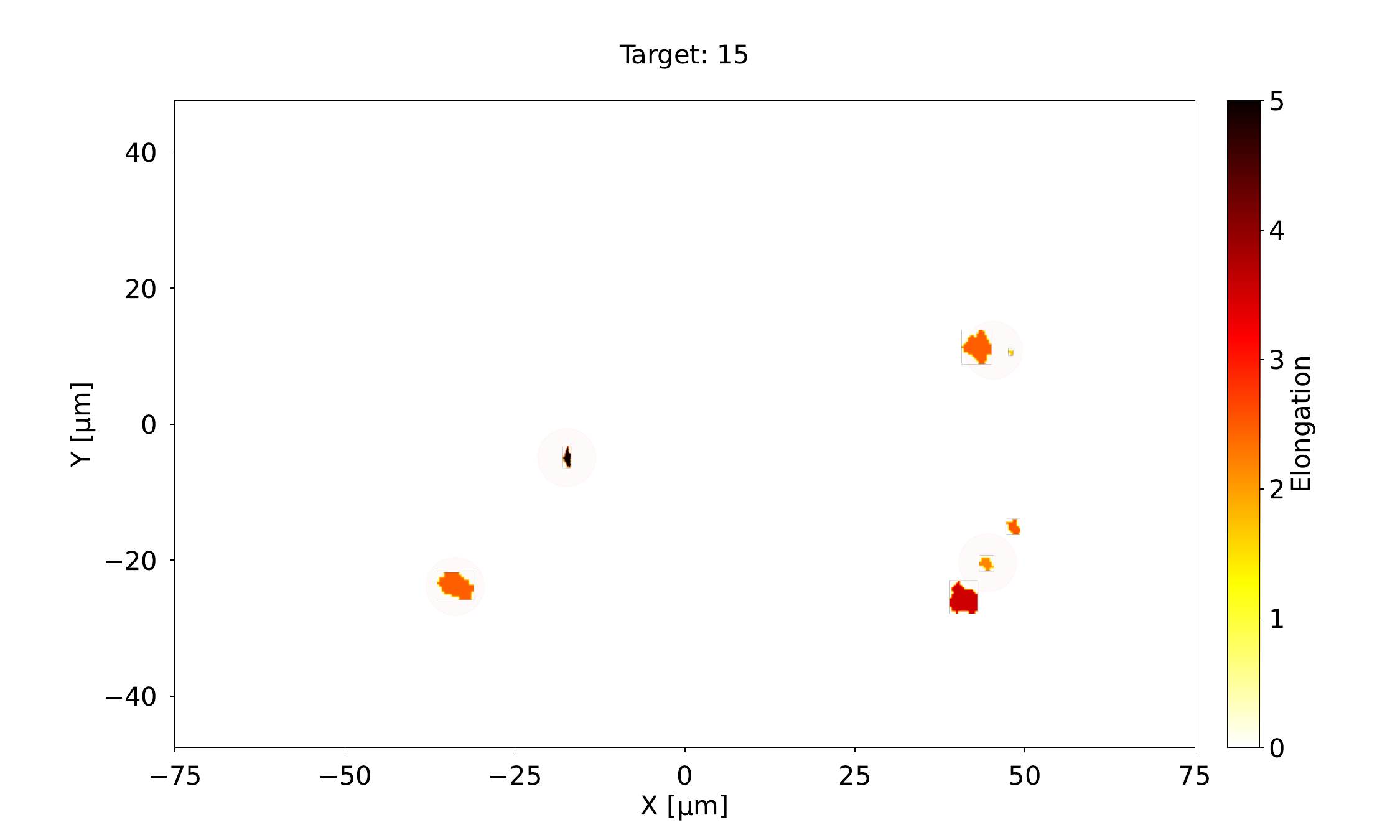}
\caption{Map showing the elongation distribution of dust particles and the particle clustering of target 15. The pink circles represent an approximate size estimation of individual MIDAS clusters.}
\label{fig:2D_dust_clustering_map_14}
\end{figure*}

\subsection{MIDAS shape descriptor maps - circularity}\label{appendix_shape_descriptor_maps_circularity}

\begin{figure*}
\centering
\includegraphics[width=17cm, height=23cm]{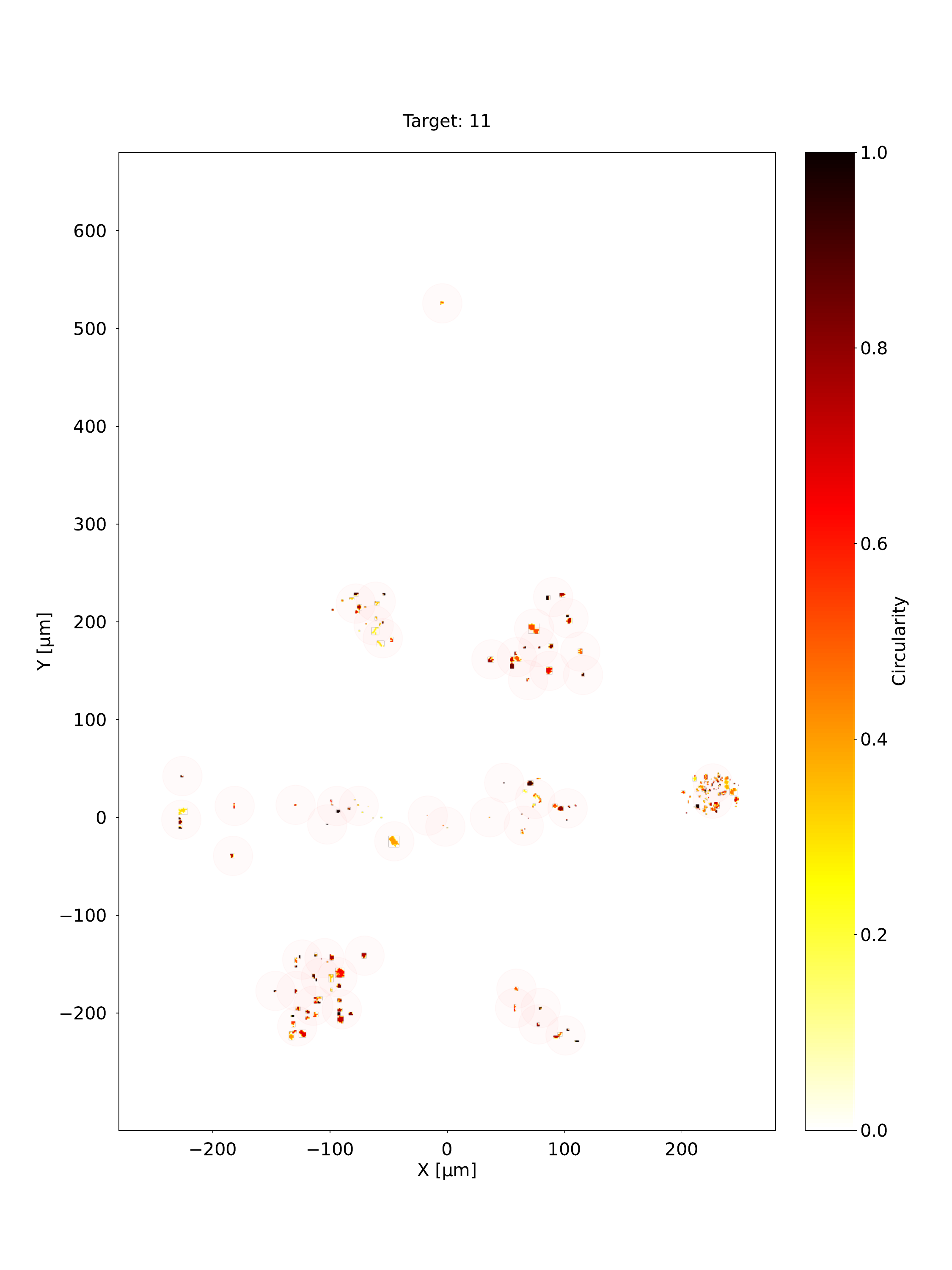}
\caption{Map showing the circularity distribution of dust particles and the particle clustering of target 11. The pink circles represent an approximate size estimation of individual MIDAS clusters.}
\label{fig:2D_dust_clustering_map_10}
\end{figure*}

\begin{figure*}
\centering
\includegraphics[width=13.3cm, height=20cm]{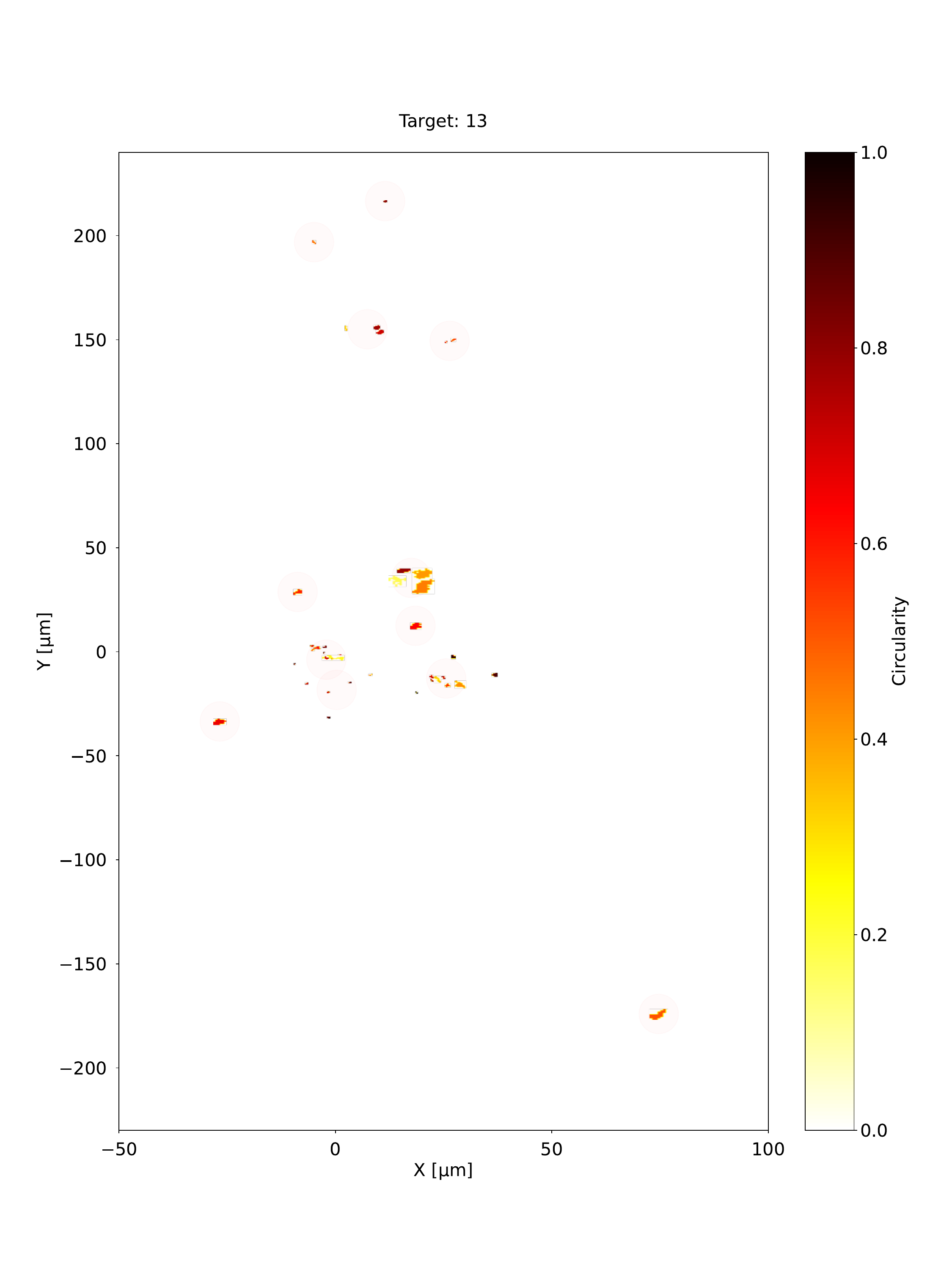}
\caption{Map showing the circularity distribution of dust particles and the particle clustering of target 13. The pink circles represent an approximate size estimation of individual MIDAS clusters.}
\label{fig:2D_dust_clustering_map_12}
\end{figure*}

\begin{figure*}
\centering
\includegraphics[width=17cm, height=24cm]{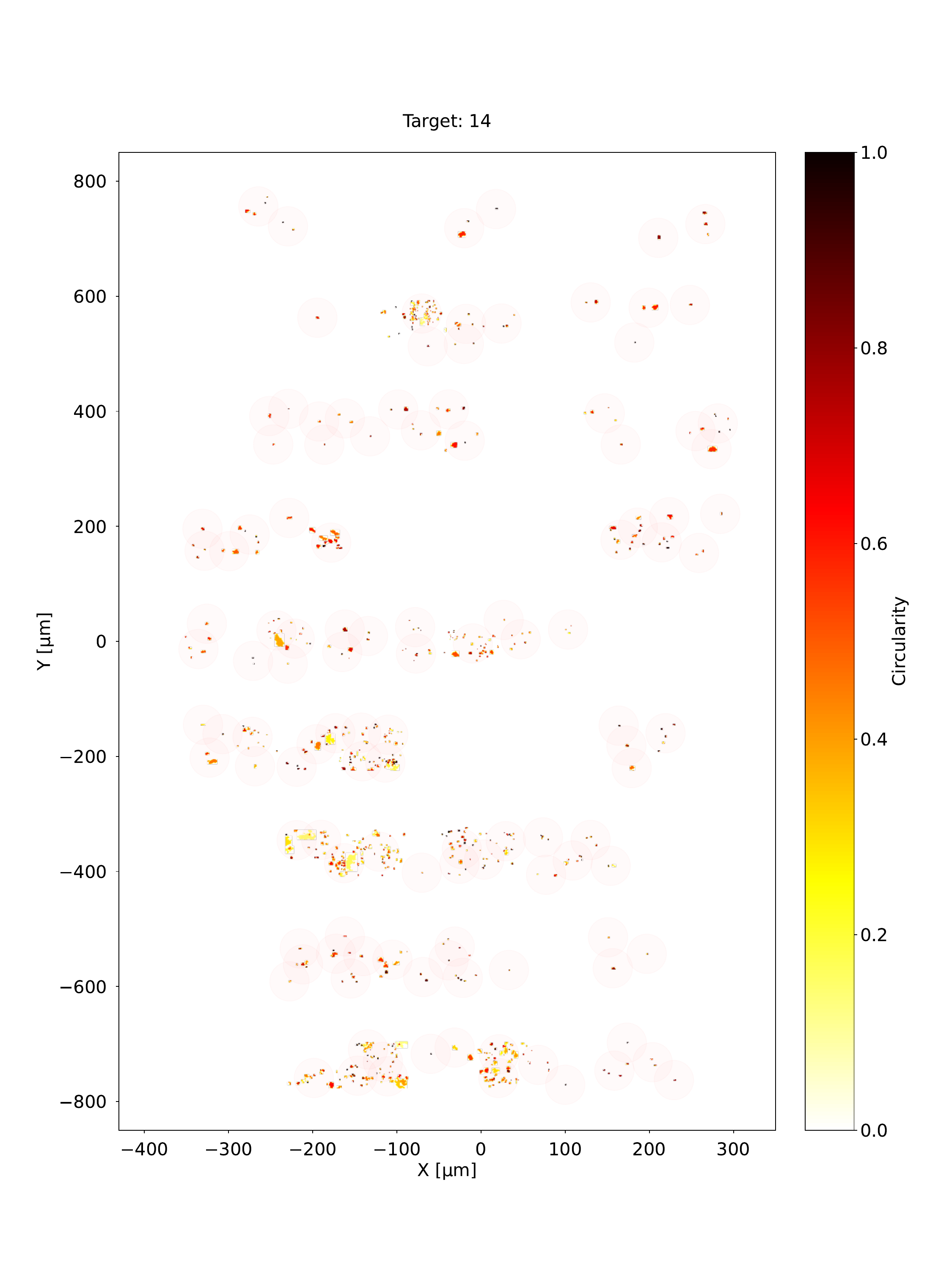}
\caption{Map showing the circularity distribution of dust particles and the particle clustering of target 14. The pink circles represent an approximate size estimation of individual MIDAS clusters.}
\label{fig:2D_dust_clustering_map_13}
\end{figure*}

\begin{figure*}
\centering
\includegraphics[width=15cm, height=9.5cm]{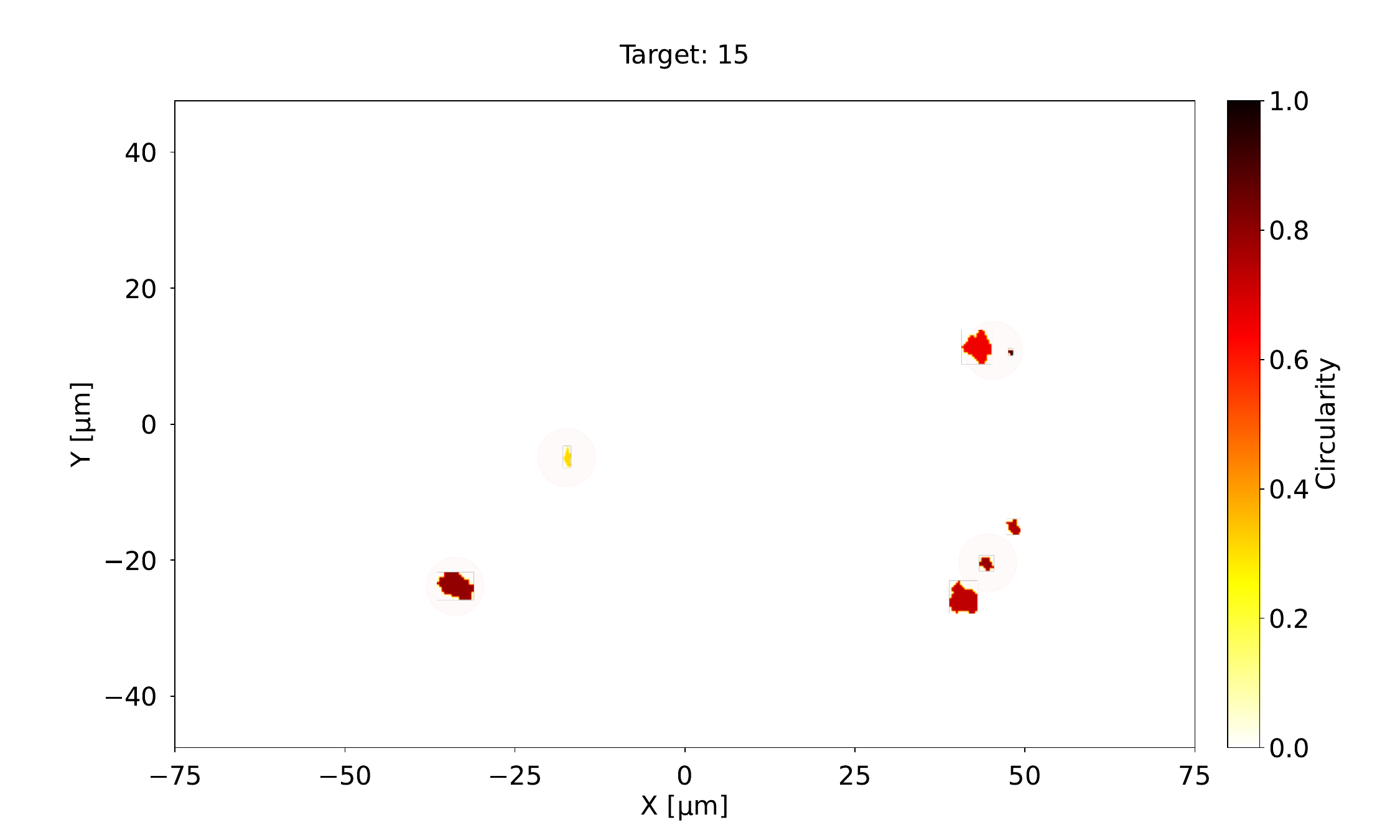}
\caption{Map showing the circularity distribution of dust particles and the particle clustering of target 15. The pink circles represent an approximate size estimation of individual MIDAS clusters.}
\label{fig:2D_dust_clustering_map_14}
\end{figure*}

\subsection{MIDAS shape descriptor maps - convexity}\label{appendix_shape_descriptor_maps_convexity}

\begin{figure*}
\centering
\includegraphics[width=17cm, height=23cm]{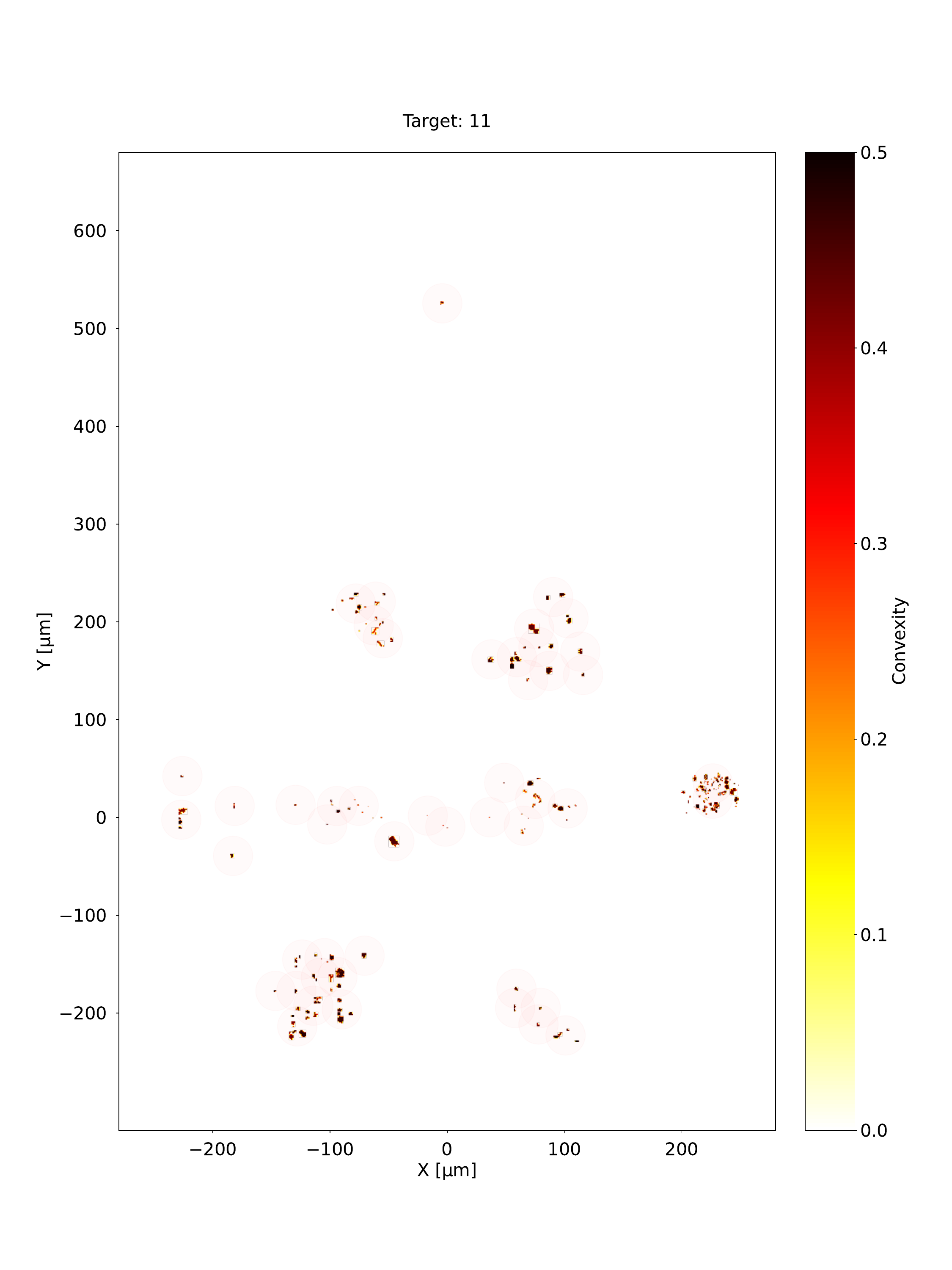}
\caption{Map showing the convexity distribution of dust particles and the particle clustering of target 11. The pink circles represent an approximate size estimation of individual MIDAS clusters.}
\label{fig:2D_dust_clustering_map_10}
\end{figure*}

\begin{figure*}
\centering
\includegraphics[width=13.3cm, height=20cm]{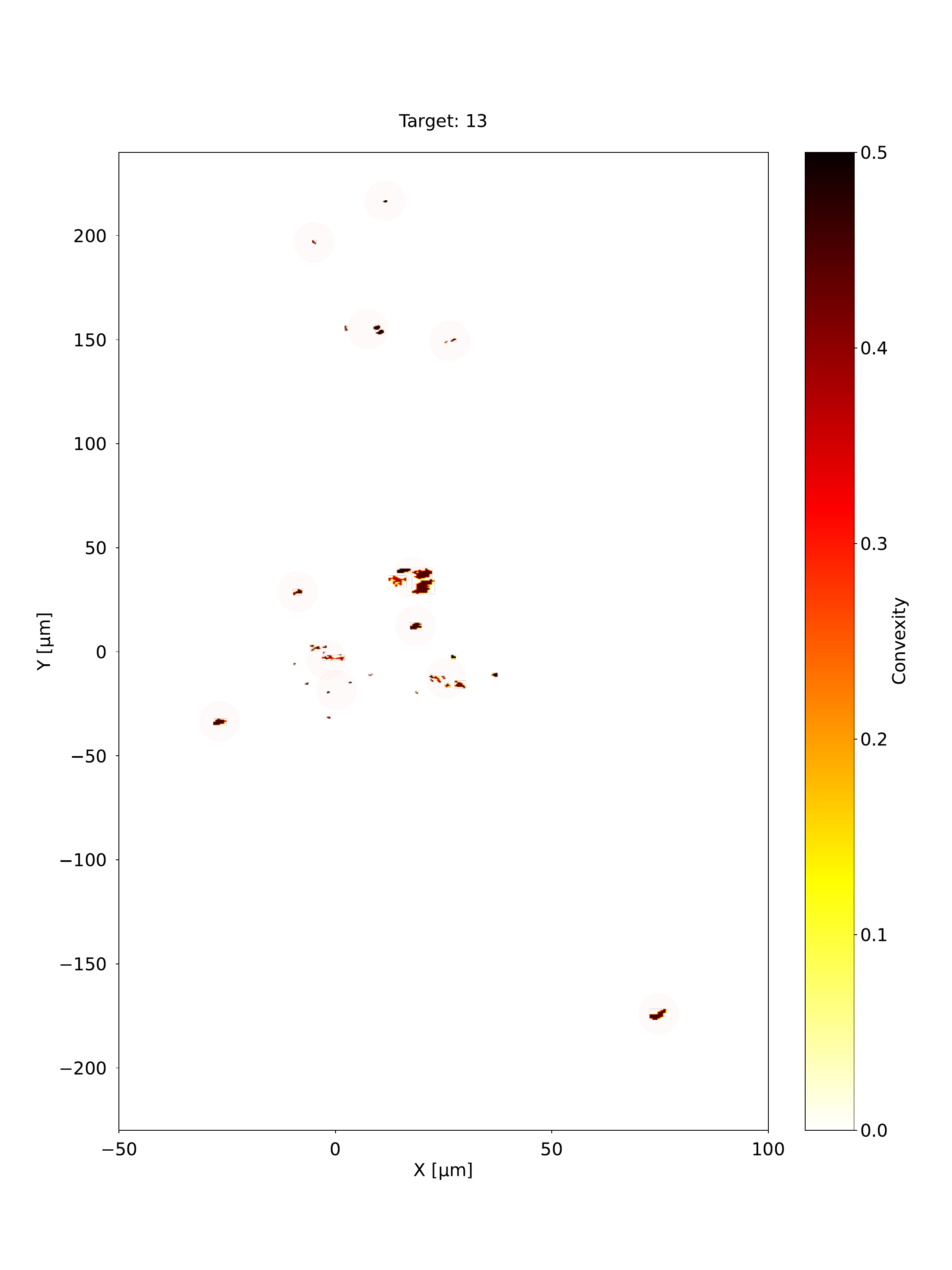}
\caption{Map showing the convexity distribution of dust particles and the particle clustering of target 13. The pink circles represent an approximate size estimation of individual MIDAS clusters.}
\label{fig:2D_dust_clustering_map_12}
\end{figure*}

\begin{figure*}
\centering
\includegraphics[width=17cm, height=24cm]{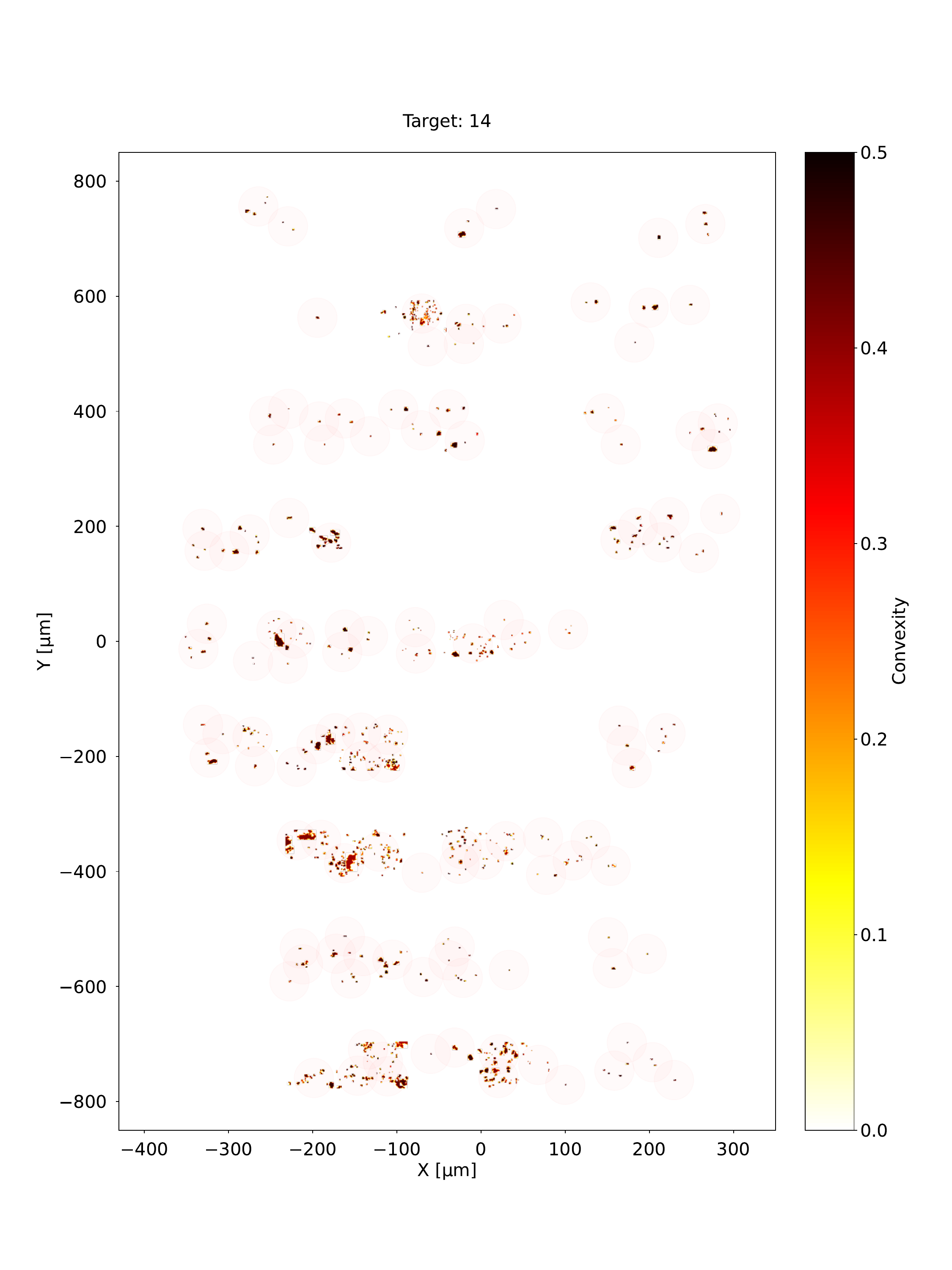}
\caption{Map showing the convexity distribution of dust particles and the particle clustering of target 14. The pink circles represent an approximate size estimation of individual MIDAS clusters.}
\label{fig:2D_dust_clustering_map_13}
\end{figure*}

\begin{figure*}
\centering
\includegraphics[width=15cm, height=9.5cm]{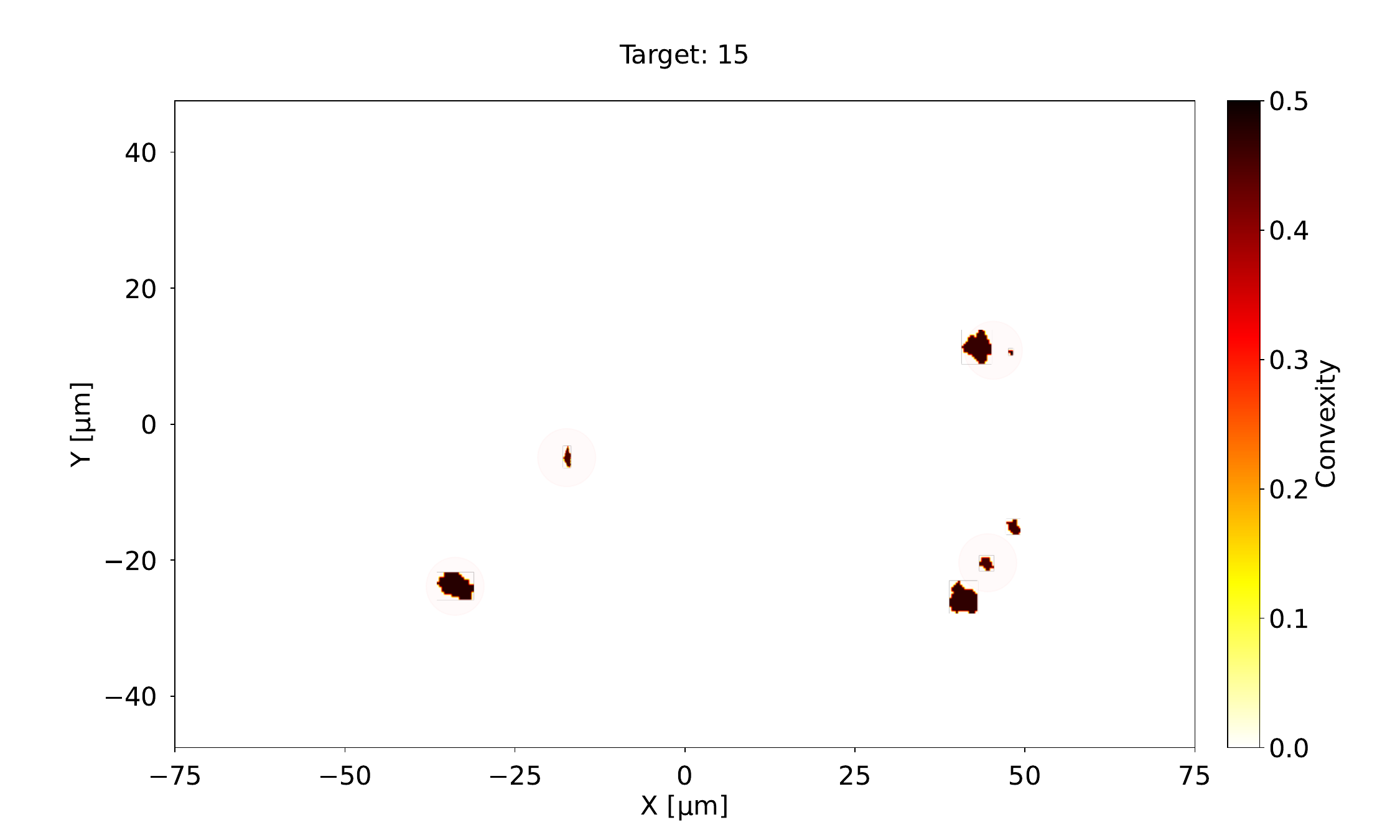}
\caption{Map showing the convexity distribution of dust particles and the particle clustering of target 15. The pink circles represent an approximate size estimation of individual MIDAS clusters.}
\label{fig:2D_dust_clustering_map_14}
\end{figure*}

\subsection{Pristineness of MIDAS particles}\label{appendix_pristineness_of_MIDAS_particles}

A summary of characteristics and shape descriptors of 19 selected particles rated 4 and 5 are listed.

\begin{landscape}
\begin{table}
\label{table: MIDAS_particle_pristineness}
\renewcommand{\arraystretch}{1.3}
\caption{Summary of characteristics and shape descriptors of 19 most pristine MIDAS particles, i.e., pristineness score rated 4 and 5 (marked in boldface).}
\centering
\begin{tabular}{llllllllll}
\textbf{Particle ID} & \textbf{Area} & \textbf{Radius} & \textbf{Height} & \textbf{Aspect Ratio} & \textbf{Elongation} & \textbf{Ellerbroek} & \textbf{COSIMA}   & \textbf{Particle surface and } & \textbf{Pristineness}  \\ 
&\textbf{[$\mu$m$^2$]} & \textbf{[$\mu$m]}& \textbf{[$\mu$m]}& & &\textbf{classification} &\textbf{classification}& \textbf{volume distribution} & \textbf{score}\\
\toprule
2014-11-18T032520\_P01\_T11          & 1.99                & 0.8                  & 0.87                 & 0.62                  & 1.9                 & Single              & Compact           & MIDAS Single                           & 4                          \\
2015-12-13T120938\_P14\_T11          & 0.29                & 0.3                  & 0.47                 & 0.87                  & 1.47                & Footprint           & Shattered cluster & MIDAS Single                           & 4                          \\
2016-04-17T082628\_P07\_T14          & 1.36                & 0.66                 & 0.78                 & 0.88                  & 1.82                & Footprint           & Shattered cluster & MIDAS Single                           & 4                          \\
2016-06-03T062540\_P06\_T14          & 25.27               & 2.84                 & 3.29                 & 0.91                  & 1.68                & Footprint           & Shattered cluster & MIDAS Single                           & 4                          \\
2016-06-03T062540\_P09\_T14          & 34.69               & 3.32                 & 3.72                 & 0.88                  & 1.89                & Footprint           & Shattered cluster & MIDAS Single                           & 4                          \\
2016-06-06T053711\_P03\_T14          & 7.6                 & 1.56                 & 1.67                 & 0.61                  & 1.93                & Single              & Compact           & MIDAS Single                           & 4                          \\
2016-06-19T161200\_P06\_T14          & 12.39               & 1.99                 & 2.34                 & 0.66                  & 2                   & Single              & Compact           & MIDAS Single                           & 4                          \\
2016-06-24T072745\_P32\_T14          & 7.43                & 1.54                 & 1.93                 & 0.98                  & 1.6                 & Footprint           & Shattered cluster & MIDAS Single                           & 4                          \\
2016-07-06T073351\_P14\_T14          & 0.33                & 0.32                 & 0.4                  & 0.95                  & 1.58                & Footprint           & Shattered cluster & MIDAS Single                           & 4                          \\
2016-07-08T002124\_P13\_T14          & 5.95                & 1.38                 & 2.46                 & 1.01                  & 1.78                & Pyramid             & Large Rubble pile & MIDAS Single                           & 4                          \\
2016-08-15T223355\_P03\_T14          & 7.43                & 1.54                 & 1.6                  & 0.88                  & 1.85                & Footprint           & Shattered cluster & MIDAS Single                           & 4                          \\
2016-08-18T123122\_P01\_T14          & 11.89               & 1.95                 & 2.1                  & 0.61                  & 2.04                & Single              & Compact           & MIDAS Single                           & 4                          \\
2016-08-18T220423\_P01\_T14          & 19.82               & 2.51                 & 2.39                 & 0.54                  & 2.29                & Single              & Compact           & MIDAS Single                           & 4                          \\
2016-08-21T211728\_P02\_T14          & 33.7                & 3.28                 & 2.68                 & 0.72                  & 2.67                & Single              & Shattered cluster & MIDAS Single                           & 4                          \\
\textbf{2016-08-21T211728\_P03\_T14} & \textbf{30.72}      & \textbf{3.13}        & \textbf{3.32}        & \textbf{0.93}         & \textbf{2.09}       & \textbf{Single}     & \textbf{Compact}  & \textbf{MIDAS Single}                  & \textbf{5}                 \\
2016-09-08T122335\_P12\_T14          & 0.34                & 0.33                 & 1.83                 & 3.15                  & 1.21                & Footprint           & Shattered cluster & MIDAS Single                           & 4                          \\
2016-09-08T122335\_P23\_T14          & 0.41                & 0.36                 & 1.75                 & 2.74                  & 1.49                & Footprint           & Shattered cluster & MIDAS Single                           & 4                          \\
2016-09-08T122335\_P24\_T14          & 0.21                & 0.26                 & 1.69                 & 3.66                  & 1.7                 & Footprint           & Shattered cluster & MIDAS Single                           & 4                          \\
2016-09-12T200834\_P06\_T14          & 1.9                 & 0.78                 & 2.32                 & 1.68                  & 1.34                & Footprint           & Shattered cluster & MIDAS Single                           & 4                          \\
\cline{1-6}
Mean value & 10.72    & 1.52     & 1.98     & 1.23      & 1.81    &  &  &  & \\  
\bottomrule
\end{tabular}
\end{table}
\end{landscape}

\subsection{MIDAS pristineness maps - pristineness evalution}\label{appendix_pristineness_maps}

MIDAS dust maps combining pristineness scores are in general 2D images (e.g., white: lower value, red: intermediate value, and black: higher value) showing their pristineness score at the respective locations on the individual target. Furthermore, particle clustering with a sophisticated clustering algorithm  (i.e., mean shift method) is described to identify fragments with a common origin (\citealp{Kim_Mannel_MIDAS_catalog}).

\begin{figure*}
\centering
\includegraphics[width=17cm, height=23cm]{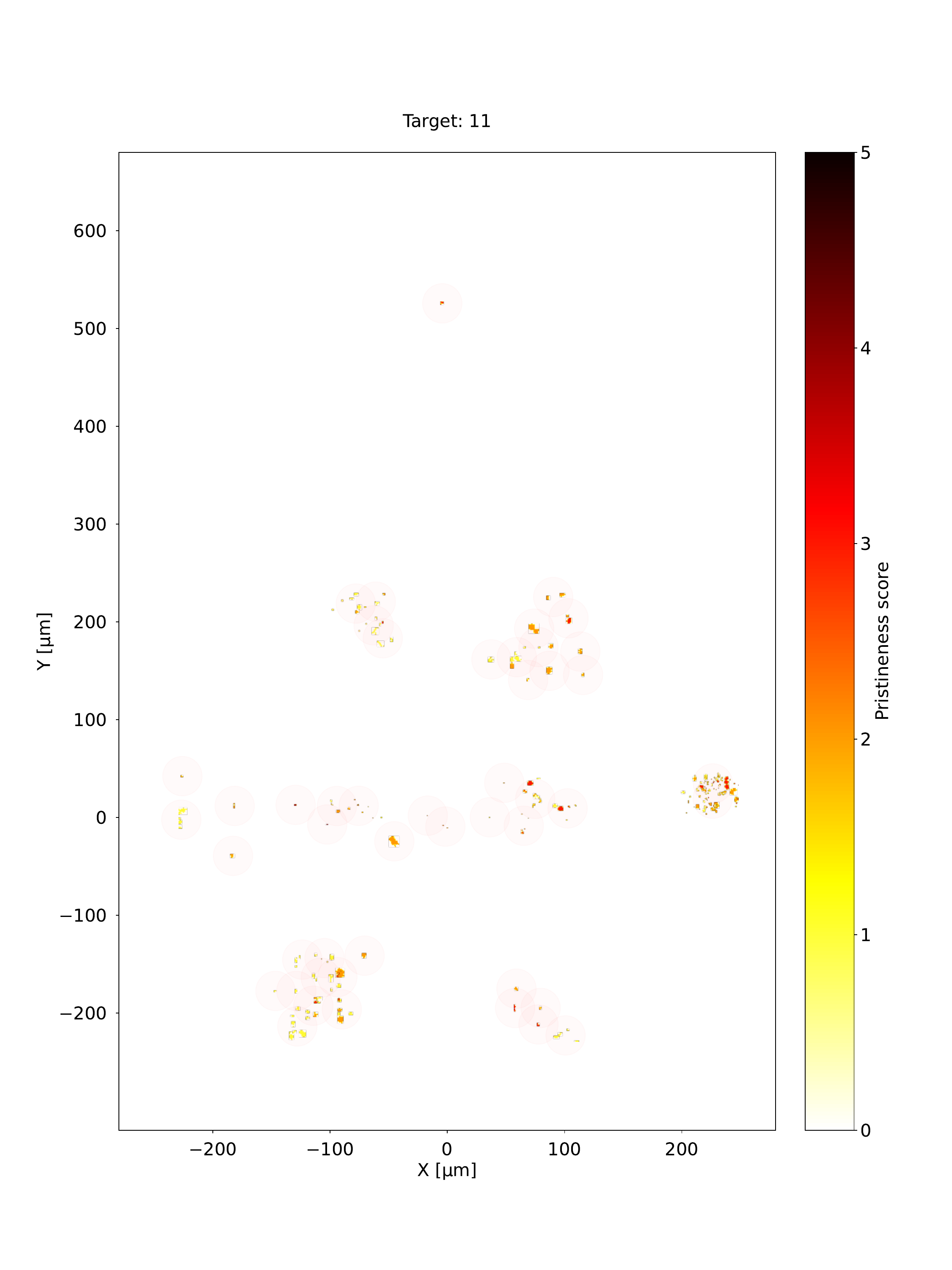}
\caption{Pristineness map combining dust clustering of target 11. The pink circles represent an approximate size estimation of individual MIDAS clusters.}
\label{fig:2D_dust_clustering_map_10}
\end{figure*}

\begin{figure*}
\centering
\includegraphics[width=13.3cm, height=20cm]{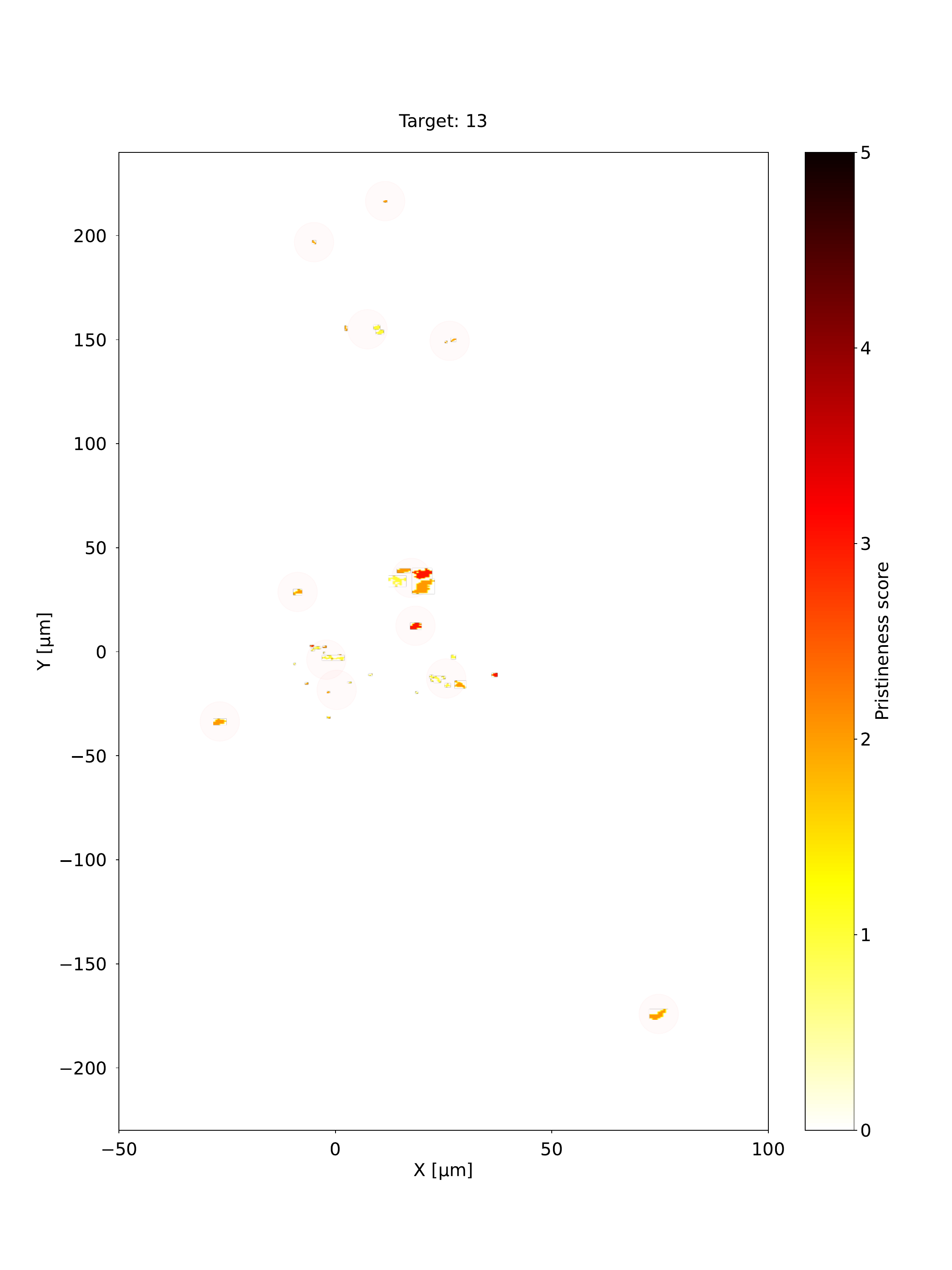}
\caption{Pristineness map combining dust clustering of target 13. The pink circles represent an approximate size estimation of individual MIDAS clusters.}
\label{fig:2D_dust_clustering_map_12}
\end{figure*}

\begin{figure*}
\centering
\includegraphics[width=17cm, height=24cm]{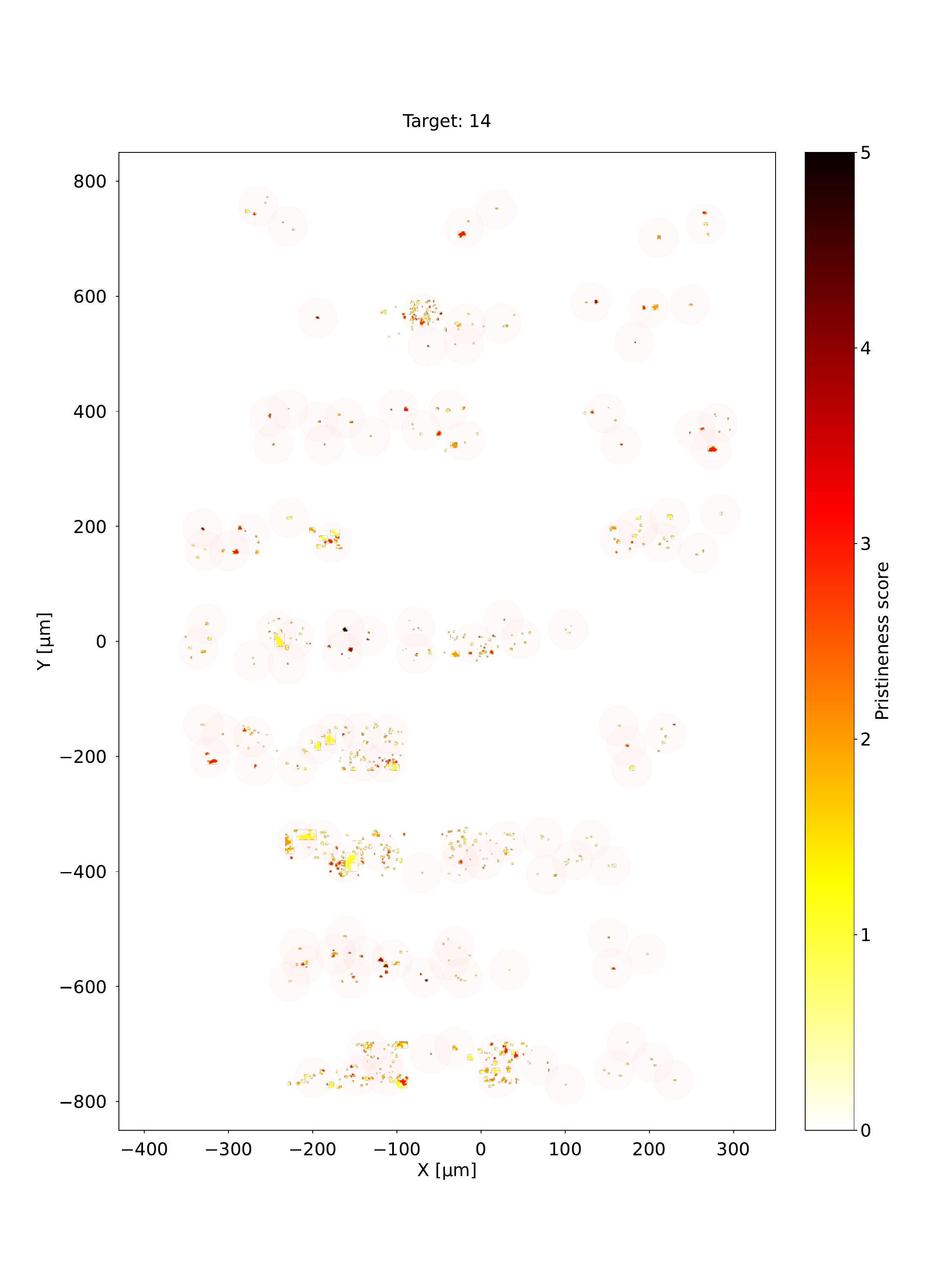}
\caption{Pristineness map combining dust clustering of target 14. The pink circles represent an approximate size estimation of individual MIDAS clusters.}
\label{fig:2D_dust_clustering_map_13}
\end{figure*}

\begin{figure*}
\centering
\includegraphics[width=15cm, height=9.5cm]{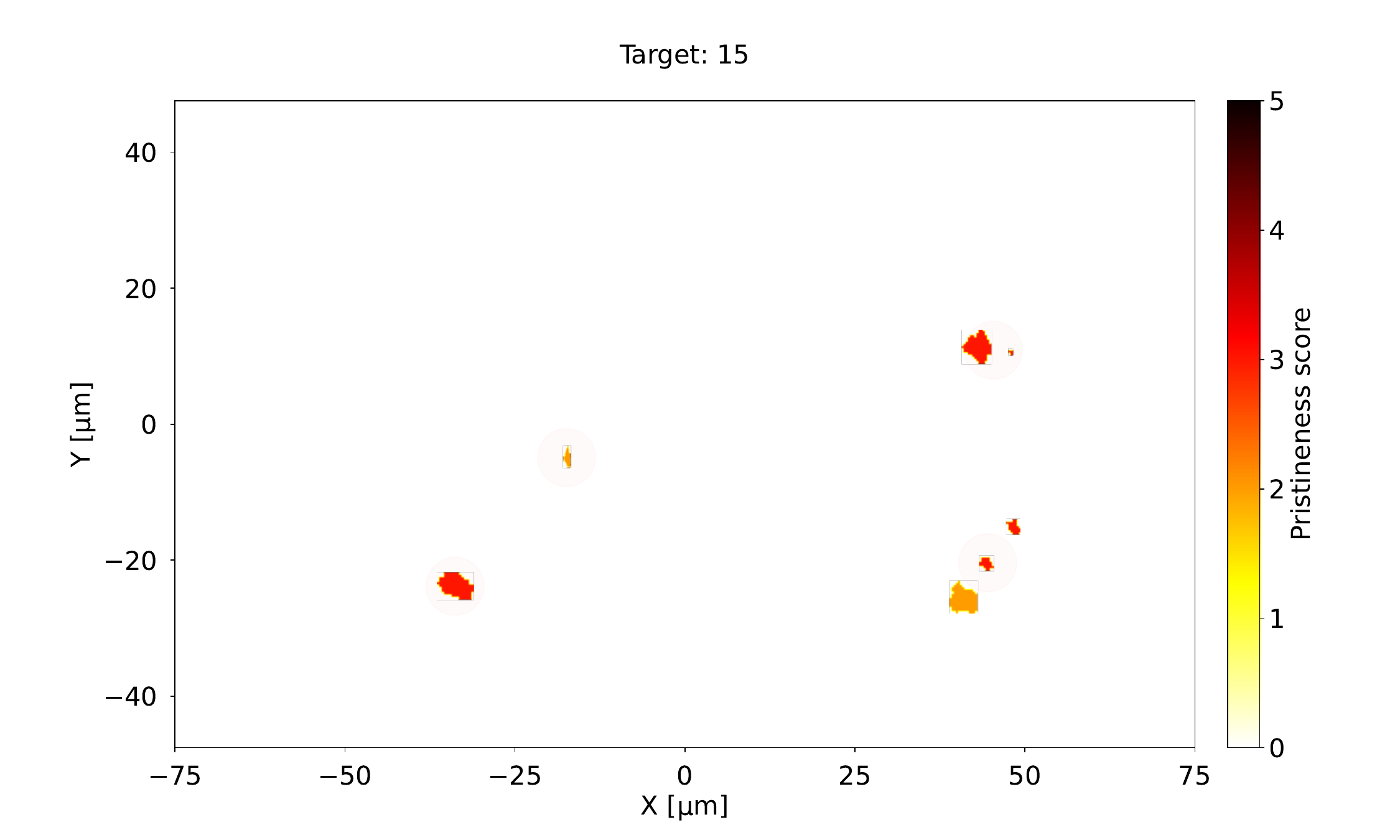}
\caption{Pristineness map combining dust clustering of target 15. The pink circles represent an approximate size estimation of individual MIDAS clusters.}
\label{fig:2D_dust_clustering_map_14}
\end{figure*}

\end{appendix}
	
\end{document}